\documentclass[10pt,aps]{revtex4}
\usepackage{graphics}
\newcommand{\ea}{{\em et al.}}
\newcommand{\eg}{{\em e.~g.}}
\newcommand{\ie}{{\em i.~e.}}
\newcommand{\cf}{{\em cf.}}
\newcommand{\viz}{{\em viz.}}
\newcommand{\wrt}{{w.~r.~t.}}
\begin{document}

\title{Electronic correlation in the quantum Hall regime}

\author{Marcus Kasner}
\email{mkasner@hrzpub.tu-darmstadt.de} 
\affiliation{Institut f\"ur Theoretische Physik, Otto-von-Guericke-Universit\"at 
   Magdeburg, \\ PF 4120, D-39106 Magdeburg, Germany}
\begin{abstract}
Two-dimensional interacting electron systems become strongly correlated if 
the electrons are subject to a perpendicular high magnetic field.  
After introducing the physics of the quantum Hall regime 
the incompressible many-particle ground state and its 
excitations are studied in detail at fractional filling factors 
for spin-polarized electrons. The spin degree of freedom whose importance was 
shown in recent experiments is considered by studying the thermodynamics 
at filling factor one and near one.  

\end{abstract}
\maketitle

\section{Introduction}   
\label{IN}

	\subsection{Overview}
	\label{subsect:overview}

Two-dimensional electron systems (2DES) of high-mobility can be experimentally realized 
in modern semiconductor devices near interfaces. In particular, the 
limit of a strong magnetic field normal to the interface offers the 
opportunity to study an electronic system, which is unique in many respects.   
Due to the strong field, all electrons can be accommodated in the lowest 
orbital Landau level of quenched kinetic energy resulting in a macroscopic degeneracy 
of the non-interacting ground state. Thus, the only relevant energy scale of the system is 
set by the interaction energy, which turns the system into a strongly correlated one.

The existence of energy gaps at certain filling factors is one of the reasons for remarkable 
anomalies in magneto-transport, the quantum Hall effects. These exhibit 
features that are very different from those in lower magnetic field experiments.
In the case of small occupancy of the lowest Landau level, the existence of gaps 
can be traced back to the occurrence of an incompressible many-particle ground state. 

In this Introduction, we review the most important experimental and 
theoretical developments of the physics in the quantum Hall regime 
and derive the appropriate many-particle Hamiltonian.
Then, we discuss the problem of two interacting particles of 
equal and opposite charge, which frequently appears in the many-particle treatment. 

In the second part, we study in detail the properties of the many-particle 
ground state and its elementary charged excitations, the 
quasiparticles, in the limit of an infinitely strong magnetic field neglecting the spin 
degree of freedom. This is explicitly done on the basis of a special model 
and by means of numerical methods in the case that only one third of the 
available one-particle states in the lowest Landau level is occupied, 
\ie~at filling factor $1/3$. 
Furthermore, we discuss the generalizations to filling factors of the 
form $p/q$, where $p,q$ are integers and $q$ odd. 

The third part is devoted to the consideration of the spin degree of freedom, which 
recently received a lot of attention from experiment.
It is shown that at certain filling 
factors the ground state exhibits spontaneous spin magnetization, which defines 
the system as a quantum Hall ferromagnet. 
After the discussion of the ground state properties near filling one, where novel 
charged spin texture excitations appear, a many-particle theory is developed 
to determine thermodynamic properties at exactly filling factor $\nu=1$. 
The theory accounts for the spin-wave like 
excitations above a completely spin-polarized ground state. 
The many-particle theory improving 
the inadequate Hartree-Fock theory predicts the temperature dependence of the  
spin magnetization. The results are compared with recent 
experimental data as well as with different theoretical approaches. 

Our considerations are based on a microscopic electronic 
Hamiltonian. It becomes clear that this strongly correlated 2DES shares many difficulties with 
theories of metallic ferromagnetism, yet it is free from the consequences of 
a complex bandstructure, which have confounded  comparison with experiment. 

Eventually, we speculate about possible routes of future research 
in this lively field that has offered surprises for more than twenty years. 

	\subsection{Two-dimensional electron systems in a magnetic field}
	\label{subsect:2d}
 
Traditionally, any microscopic description in solid state physics starts from 
simplified models trying to capture the essential features before systematically 
extending the model in order to reconcile it with the often much more complex 
experimental results.
Pursuing such an approach, we describe electrons of spin $S=1/2$ and charge $q=-e<0$ 
moving in a three-dimensional 
crystal, which are subject to a spatially constant and time independent magnetic 
field ${\bf B}=B {\bf e}_{z} \equiv B_{\perp} {\bf e}_{z}$ ($B>0$),  
in a first approximation as independent particles in a continuum. Due to the separation 
of the motion in the $x$-$y$-plane and the $z$-direction, the Hamiltonian can be 
written in Cartesian coordinates as
\begin{equation}
H_{0} =
\frac{1}{2 m_{e}}({\bf p} - q {\bf A})^2 + \frac{p_{z}^2 }{2 m_{e}}
- g_{e} \mu_{B}{\bf B}{\bf S} \ .
\label{Hnot}
\end{equation}
Here, the kinematical momentum in the plane is given by ${\bf p}=((\hbar/i)\partial_{x}-q A_{x}, 
(\hbar/i)\partial_{y}-q A_{y})$, whereas the momentum $p_{z}=(\hbar/i)\partial_{z}$ describes 
the free motion 
in the $z$-direction. The gauge freedom of the vector potential allows to write it as 
${\bf A}({\bf r})=B(-\alpha y, (1-\alpha) x,0)$ with the  parameter $\alpha \in R$. 
The gauge parameter $\alpha$ covers for $\alpha=0$ and $\alpha=1/2$ the two mostly
used gauges, \viz~the Landau gauge and the symmetric gauge. The last term is the 
Zeeman energy and is due to the coupling of the dimensionless electron's spin ${\bf S}$ 
to the magnetic field ($g_{e}$ -- gyromagnetic factor, $\mu_B$ -- Bohr magneton of 
the electron). The natural length scale is set by the cyclotron or magnetic length 
$\ell_c=\sqrt{\hbar/|q B|}$, where $\ell_c [nm] = 25.65 /\sqrt{B[T]}$, which exceeds for 
accessible fields up to $B \simeq 20\ T$ the lattice constant by more than a factor of 
10 and justifies the application of a continuum model. 

The solution of Eq.~(\ref{Hnot}) separates in three simple energy eigenvalue problems: first, 
the motion in $z$-direction for finite extension $w$ with energies 
$E_{z,i}=\hbar^2 k_{z,i}^2/(2 m_e)$, where $k_{z,i} \propto 1/w$ 
as long as there is no in-plane component of the magnetic field, second, the spin degree 
can be simply described by the two eigenvalues of $S_{z}=\pm 1/2$, which are denoted 
by the majority spin direction $\uparrow$ for $\sigma=+1$   
and the minority spin direction $\downarrow$ for $\sigma=-1$ 
as long there is no spin-orbit coupling, and third, the 
solution of electrons in the plane that can be reduced to 
a harmonic oscillator problem, whose energy scale is $\hbar \omega_{c}$, and where 
$\omega_{c}=|qB|/m_{e}$ is the cyclotron frequency. 

The problem becomes effectively two-dimensional if one is able to create well distinguished 
energy levels, so-called subbands, by requiring that 
the temperature $T$ is smaller than the energy difference $\Delta E_{01}$ between the 
first excited and the lowest subband energy, \ie~the following inequality holds
\begin{equation}
k_B T <  \Delta E_{01} = \frac{2 \pi^2 \hbar^2}{(m_e w^2)} \; .
\end{equation}
The estimate is based on the assumption of a rectangular quantum well of 
infinite height in $z$-direction. In order to satisfy such a condition 
for, say, temperatures below $200\ K$, a width smaller than at least $120\ $\r{A} is necessary. 
Only the advent of modern semiconductor technology made the practical realization 
of devices possible, where 
potential changes within such a short distance could be achieved. 
By means of appropriate doping, it is possible to create a two-dimensional electron system (2DES) at the 
semiconductor-oxide interface whose Fermi energy can be controlled by a gate voltage. 
Since the end of the seventies, technological progress made the usage of a new class 
of semiconductor devices for such experiments possible, \viz~$GaAs$-$Al_{x}Ga_{1-x}As$ 
heterostructures. They are based on the appropriate doping of stacked $GaAs$ and 
$Al_{x}Ga_{1-x}As$ layers, which form either heterojunctions in case of one interface or 
one and many quantum wells, respectively. In particular, the growth of heterostructures 
by means of molecular beam epitaxy allows a much more controlled manufacturing, which leads to 
a reduced number of impurities and, hence, to an increasing mobility of the samples.

\begin{figure}[h]
\centerline{\resizebox{6cm}{5cm}{\includegraphics{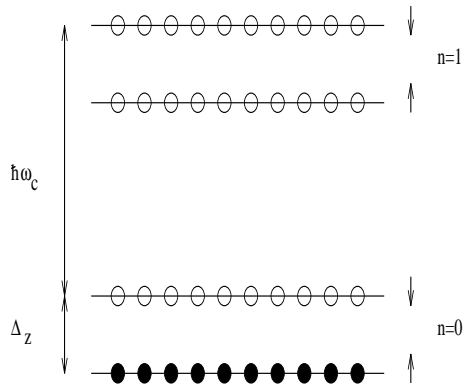}}}
\caption[]{\small
The two lowest spin split orbital Landau levels as in a $GaAs$ heterostructure, where
$0 < \Delta_{z} \ll \hbar \omega_{c}$. The degenerate one-particle states
are denoted by circles. The ground state of non-interacting electrons with majority spin direction  
$\uparrow$ for $\nu=1$ is indicated by filled circles.
}
\label{energy_levels}
\end{figure}

Assuming that the electrons occupy at sufficiently small temperatures 
only the lowest electronic subband, the one-particle problem neglecting 
the subband energy contribution has degenerate eigenvalues 
\begin{equation}
E_{n,\sigma}=\hbar \omega_c \left( n + \frac{1}{2} \right) - \frac{\sigma \Delta_z}{2} \ ,
\label{2d-eigenvalues}
\end{equation}
which are labeled by the orbital quantum number $n=0,1,2,\ldots$ denoting the Landau levels   
and the eigenvalue of the $z$-component of the spin operator, while   
$\hbar \omega_c$ determines the distance between two adjacent Landau levels \cite{Rem1}.  
Note that the situation becomes different in a real 2DES, since one has to account for the effective 
values of the mass, gyromagnetic factor, and dielectric constant of the host semiconductor. 
For $GaAs$, where the effective mass is $m = 0.067 m_{e}$, the gyromagnetic
factor $g=-0.44$ \cite{WH77}, and the dielectric constant $\epsilon = 12.7 \epsilon_{0}$,  
the energy spectrum is schematically depicted in Fig.~\ref{energy_levels}.
The explicit form of the eigenfunctions, belonging to the eigenvalues (\ref{2d-eigenvalues}), 
depends on the gauge. In symmetric gauge, the energy eigenvalue problem 
is reduced to the solution of a two-dimensional isotropic harmonic oscillator problem, 
and in Landau gauge, it describes a free particle subject to periodic
boundary conditions in the $y$-direction and a
one-dimensional harmonic oscillator in the $x$-direction. In either case, each of
the Landau levels is degenerate 
with respect to a quantum number determined by the respective gauge.
For the symmetric gauge, these are the eigenvalues of the
$z$-component of the angular momentum $L_{z}=(\hbar/i) \partial_{\varphi}$, which can be labeled
for each Landau level by  $m=-n,-n+1, \ldots $, and in the Landau gauge by the
momenta in the y-direction
$k_{y,l}=2\pi l/L_{y}$ ($L_{y}$ -- width of a rectangle in the $y$-direction,
$l=0,\pm 1,\pm 2,\ldots$).
To be definite, let us quote the eigenfunctions of the lowest orbital Landau level (LLL) $n=0$
with $m=0,1,2,\ldots$ as a function of the dimensionless complex coordinate 
$z=x-iy = (r_{x}-i r_{y})/\ell_c$ (note $sign(qB)=-1$)  
\begin{equation}
\varphi_{n=0,m}(z) \equiv \varphi_{m}(z) =
\frac{1}{\sqrt{2 \pi 2^m m!}} z^{m}  e^{-\frac{|z|^2}{4}} \ .
\label{sym-n=0}
\end{equation}

The orbital degeneracy $N_{\Phi}=\Phi/\Phi_0$ of each Landau level counts how many 
elementary flux quanta $\Phi_0=h/e$ are contained in the
total flux $\Phi=B A$ through the area $A$ of the plane, {\it i.~e.}~
$N_{\Phi}=A/(2 \pi \ell_c^2)$. This implies
for a system of finite extension in the plane that the degeneracy of
a Landau level is finite too, {\em e.~g.~}on a circle of radius $R$, the degeneracy is
$R^2/(2 \ell_{c}^2)$. In case of the LLL, the eigenfunction $\varphi_{m}(z) \propto z^{m}$ 
covers an area of $2 \pi (m+1) \ell_{c}^2$ and encloses therefore $(m+1)$ elementary 
flux quanta. In other words, the power of $z$, which is also the number of zeros 
of the eigenfunction, equals approximately the number of flux quanta covering the area of the 
system. An important quantity is the  
filling factor $\nu$, which is defined as the ratio of the particle number $N$ to the orbital
degeneracy $N_{\Phi}$, {\it i.~e.~}
\begin{equation}
\nu=N/N_{\Phi}=2 \pi \ell_{c}^2 n\;,
\label{nu}
\end{equation}
where $n=N/A$ is the particle density.\footnote{For finite-size system, we will encounter 
a slightly different definition of the filling factor, which, however, agrees in the thermodynamic
limit with that given here.}
We note that the degeneracy
is proportional to the strength of the magnetic field $B$. This allows to accommodate
an increasing number of electrons in each orbital Landau levels  when the magnetic field
becomes stronger, or for a fixed particle number, the Fermi energy is swept downwards with increasing magnetic
field. If the field is so strong that at zero temperature all electrons reside in the
lowest orbital Landau level, we will loosely speak about the {\it strong field limit} \cite{Rem6}. 

Besides the symmetries, which are related to a specific gauge, and which are satisfied
even for finite systems, there exists
a continuous symmetry of the Hamiltonian Eq.~(\ref{Hnot}) for an infinitely extended 
system since the generator ${\bf t}$ of the magnetic translation operator 
commutes with the Hamiltonian $H_{0}$, 
\ie~$[{\bf t},H_{0}]=0$.
The generator of the magnetic translation in the plane is the canonical momentum 
${\bf t} = {\bf p}+q {\bf A}$, and the magnetic translation operator is hence given by
\cite{transl} 
\begin{equation}
T_{{\bf r_{0}}} = e^{-\frac{i}{\hbar}{\bf r}_{0}({\bf p}+q {\bf A})} \ .
\label{translation}
\end{equation}

Before closing, let us emphasize that such a model of independent electrons ignores at least two 
important ingredients, \viz~disorder and interaction between the electrons, which 
become important when discussing the strong field limit.   

	\subsection{A short historical account of the 2DES in a strong magnetic field}
	\label{subsect:account}

In this Subsection we review some of the remarkable experiments of a 2DES in a strong magnetic field,  
which have made the physics of strong magnetic fields such a lively field of research. 
Up to the discovery of the integer quantum Hall effect (IQHE), the physics of a 2DES did not 
seem particularly spectacular despite the new opportunities due to the 
reduced dimensionality.  For example, the system behaves 
like an ordinary Fermi liquid without magnetic field \cite{AFS82}.

The measurement of transport properties belongs to those methods that allow the phenomenological 
characterization of solids. Hall effect measurements serve as the experimental standard 
method to determine the sign and concentration of the charge carriers, and they are explained in the 
framework of the classical Hall effect. 
It is evident that already at the classical level an independent electron model without disorder 
is insufficient to describe transport. Sources of disorder in a semiconductor 2DES are charged 
impurities, \eg, donors, near the 
2DES or the roughness of the insulator-semiconductor interface in metal oxide semiconductor field effect 
transistors (MOSFETs). 
In the Drude theory of the Hall effect, carriers of charge $q$ move in the 
electric field ${\bf E}=E_{x} {\bf e}_{x}$ and cause a 
current density ${\bf j}= \hat \sigma {\bf E}$ ($\hat \sigma$ - conductivity tensor),
which can be determined by inversion of the resistivity tensor $\hat \sigma = \hat \rho^{-1}$. 
Scattering of the charge carriers on impurities leads to 
elastic scattering events with 
elastic scattering time $\tau$, and the Drude theory gives for the components of the resistivity 
tensor $\hat \rho$ 
\begin{eqnarray}
\rho_{xx}  & = & \rho_{0}   =   \frac{m}{n q^2 \tau} \nonumber \\
\rho_{xy} & =  & - \frac{B}{nq}   =   - \frac{1}{\nu} \frac{h}{q^2} \ ,
\label{drude}
\end{eqnarray}
where $\rho_{0}$ is the resistivity of the Drude theory without magnetic field.
Such a model is often sufficient to explain experiments in fields up to $0.1\ T$.

\begin{figure}[h]
\centerline{\resizebox{6cm}{6cm}{\includegraphics{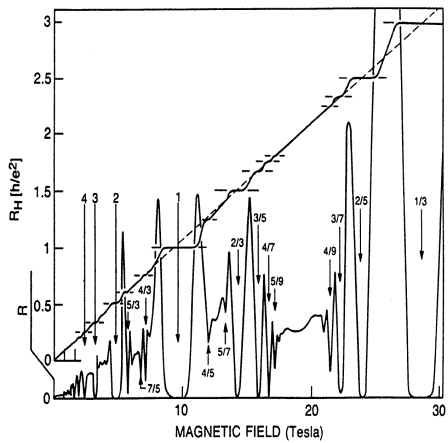}}}
\caption[]{\small
Results of transport measurements in a high-mobility 
$GaAs$-heterostructure  ($\mu \sim 1.3 \times 10^{6} cm^2/Vs$) at low temperature. 
The Hall resistance $R_{H}=\rho_{xy}$ exhibits plateaus at integer (IQHE) and fractional (FQHE)
filling factors $\nu$, while the longitudinal resistance $R=\rho_{xx} (L_{x}/L_{y})$ 
shows deep dips in the plateau region indicating the vanishing of $\rho_{xx}$ 
when approaching zero temperature.
From \cite{WEST87}.
}
\label{qhe-experiment}
\end{figure}

An unprecedented development started with the observation of the IQHE in 1980 
by von Klitzing when he studied the Hall resistance $R_{H}=|\rho_{xy}|$ and the longitudinal 
resistance $R \propto \rho_{xx}$ of a silicon MOSFET with a, at that time, comparably high mobility at 
temperature $1.5\ K$ in high magnetic fields \cite{vKDP80}. 
The Hall resistance exhibited plateau values $R_{H}=(h/e^2)/\nu$ 
around integer filling factors $\nu=1,2,\ldots$,  while the longitudinal resistance started to vanish in 
the middle of the plateaus showing deep dips, \cf~Fig.~\ref{qhe-experiment}.
These features become most pronounced at very low temperatures, where the Hall resistance 
exhibits a staircase like structure.  Simultaneously, the longitudinal resistance vanishes except 
for peak-like structures around 
$\nu=(n+1/2)$, where the transition from one Hall plateau to the next one occurs. 
Particularly remarkable is the accuracy of the Hall resistance in the plateau region, whose values  
are integer fractions of the von Klitzing constant $R_{K}=h/e^2$. Currently, the  
relative standard uncertainty has a value of $3.7 \ 10^{-9}$ 
\cite{CODATA98}, and the IQHE serves as the metrological standard for the resistance. 
These results  are clearly different from those of the Drude theory, since the Drude result 
for the Hall resistance 
merely agrees with the IQHE result at those $\nu$, where the linear $\rho_{xy}$ vs.~$B$ 
curve (\ref{drude}) crosses the staircase structure. 
This happens at those values for the magnetic field, where $\nu=h n/(|eB|)$ attains an 
{\em integer} value, \cf~Fig.~\ref{qhe-experiment}.

When in 1982 Tsui \ea~performed the same experiment with a $GaAs$-sample 
of higher mobility at lower temperatures in stronger fields, they found 
similar features in the Hall and longitudinal resistance 
around certain fractional filling factors of the form 
$\nu=p/q$ with odd $q$, \eg, for $\nu=1/3,\ 2/3,\ 2/5$ to name the most prominent ones, 
see Fig.~\ref{qhe-experiment}. Hence, the name fractional quantum Hall 
effect (FQHE) was coined \cite{TSG82}.  
From a phenomenological viewpoint, these quantum Hall effects are quite similar, although 
the energy scales involved significantly differ. For the understanding of the quantum Hall 
effects (QHEs) it is important to note that 
at those $\nu$, where the effects appear, the disorder-free system exhibits an energy gap above 
the ground state. In the FQHE, this is a gap above the many-particle ground state caused 
by the dominating electron-electron interaction, \cf~Sect.~\ref{sect:polar}, whereas in the 
IQHE, the gap has its origin in the energy gap between adjacent Landau levels at integer filling. 
Since the quantum-mechanical treatment of a disorder free system shows a linear $B$ 
dependence as the Drude theory, a certain amount of disorder is necessary in order to create 
plateaus. 
On the other hand, sufficiently large disorder lets the QHEs disappear, at first, the 
FQHE, at larger disorder, the IQHE too. This can be explained by the different magnitudes of the energy 
gaps: the energy gap in the IQHE is approximately one order of magnitude greater than that 
in the FQHE. 
Hence higher mobility samples than for the IQHE are needed to observe the FQHE as larger 
disorder washes the energy gap out. A similar effect can be observed by increasing 
the temperature, which causes the disappearance of the FQHE, and at higher temperatures 
that of the IQHE \cite{Rem5}.

The theoretical explanation of the IQHE is currently based on a one-particle 
model with quenched disorder. In a two-dimensional system in a strong 
magnetic field, disorder causes the broadening of the Landau levels forming a 
continuum of states \cite{Weg83}. Most of the states 
are localized and only those states exactly at the energy of the originally 
discrete Landau level (LL) energies seem to be extended \cite{Huck95}.  
Such a system is an extensively studied example of a localization-delocalization transition in 
two dimensions. The formation of a mobility gap due to localized states has a similar 
impact on transport as a charge gap in a pure system \cite{Mac95}.  
There are various ways to make this statement more precise. For example, 
using a Kubo formula for the Hall conductivity $\sigma_{xy}$, one can show that only 
extended states below the Fermi energy contribute \cite{AA81}. Hence, 
increasing the chemical potential up to the center of the LLL $\mu=\hbar \omega_c/2$ 
keeps $\sigma_{xy}=0$. 
When reaching the half-filled LLL, the conductivity jumps to the value $e^2/h$ and remains again constant up 
to a filling with a Fermi energy at the center of the next Landau level. 
Other explanations are based on a gauge argument and the role of the edge of the system \cite{Lau81,Halp82}. 
However, we are still far from a microscopic transport theory of the IQHE 
\cite{Pru90} that makes quantitative predictions about the influence of disorder and temperature  
on the conductivities. Nonetheless, there is the belief that this explanation meets the essence of the effect. 
In the FQHE, the energy gap is not the result of a one-particle effect but 
is caused by electron correlation. Its origin can be traced back to the specific properties 
of the many-particle ground state at the so-called magic filling factors $p/q$. 
In particular, Laughlin's proposal for a ground state wavefunction at $\nu=1/q$ paved the way 
for a deeper understanding \cite{Lau83b}.  
One of our main subjects is to explain the existence of these gaps.  
The importance of electron-electron interaction is supported by the experimental fact that 
the FQHE features become more pronounced when increasing $B$, but keeping $\nu$ constant, and 
hence diminishing the influence of higher LLs. 
We see that, contrary to the IQHE, even the study of the ground state and the low 
lying excitations of interacting electrons in strong fields is a difficult subject on its 
own. This is also one of the reasons why a lot of work was devoted to this issue, whereas 
the transport properties of the FQHE are usually explained by 
adopting arguments from the IQHE to the case of the FQHE \cite{Lau83b}.  

Further inspection of Fig.~\ref{qhe-experiment} shows that $\rho_{xx}$ behaves 
around $\nu=1/2$ quite differently. This is attributed to a Fermi liquid like ground state 
of weakly interacting particles, so-called composite fermions, at $\nu=1/2$.  
In particular, surface acoustic wave measurements gave first evidence 
of this effectively one-particle picture \cite{WPR90,WRWP93,WWP95}.  
Moreover, theoretical work \cite{HLR93} based on a singular gauge transformation 
supported such a viewpoint. 

A third group of experiments covers the very low filling factor region, where one expects a 
behavior typical for a Wigner crystal pinned by disorder \cite{Fer97}.  
Various experiments in the low filling factor region indicate that physical properties 
change when compared with those in the  FQHE region. Transport experiments even 
at $\nu \simeq 1/5$ \cite{JWS90}, 
\eg, show a diverging longitudinal resistivity indicating an insulating ground state 
and reentrant behavior in the typical FQHE region.
So far, however, an experiment giving unequivocally evidence for the existence of such a 
ground state is still missing \cite{Fer97}. 

Early experiments in the QHE region explored almost exclusively dynamical properties examining 
the response of the 2DES to electromagnetic fields. 
This was mostly done by 
studying transport properties. The investigation of the thermodynamic behavior 
was particularly pushed by predictions on novel spin structures, although 
Halperin had already noted very early the importance of the spin degree of freedom 
in $GaAs$-heterostructures \cite{Halp83}.   
The spin magnetization can be investigated, for example, by nuclear magnetic resonance (NMR) 
experiments \cite{BTPW94}, magneto-absorption experiments \cite{AGB96}, 
and photo-luminescence measurements \cite{KKE99}.  

Transport experiments at higher LLs emphasized the importance of the spin, and 
at $\nu=5/2$, the only FQHE with even denominator in a one-layer system was found \cite{WEST87}. 
Recently, there were also unexpected observations in the higher Landau level range. 
Near half filling of higher LLs, \eg, at $\nu=9/2$,  
anisotropies in $\rho_{xx}$ were found. They are caused by ground states  
with a strip like density distribution, which is different from the 
LLL behavior \cite{LCE99}. 

Within the last ten years, further experimental information about the physics 
of the quantum Hall regime was gathered by studying double layer systems,  
where two parallel quantum wells are brought close together \cite{Eis97,GM97}.  
In such systems, the phenomena become much richer due to the loss of symmetry of 
the Coulomb interaction and the possible tunneling of charge carriers 
from one layer to the other one. 

Hallmarks of the theoretical understanding of the QH-regime are Laughlin's seminal proposal of 
a ground state wavefunction at $\nu=1/q$, the introduction of the 
quasiparticle concept and the construction 
of a hierarchical theory to explain the occurrence of the FQHE at $\nu=p/q$ by 
Halperin and Haldane and Jain's composite fermion approach.  
The construction of excited states that describe gapped density like intra Landau level 
excitations above the Laughlin state allowed the prediction of a roton-like minimum \cite{GMP85}. 
 
The construction of a field-theoretical approach was initiated by the deep observation 
of Girvin and MacDonald that the application of a special singular gauge transformation 
to the Laughlin state at $1/q$ by attaching $q$ flux quanta per electron leads 
to a bosonic state \cite{GM87}.  
Most remarkable, the resulting bosonic wavefunction exhibits in contrast to the original fermionic 
Laughlin state an algebraic off-diagonal long-range order (ODLRO) of the one-particle 
density matrix. This observation fostered attempts to search for an 
order parameter describing the unique Laughlin state \cite{RH88,Rea89a} and the derivation of a 
phenomenological Landau-Ginsburg theory using the same singular gauge 
transformation \cite{ZHK89a,Zha92}.  
Alternatively, one can also apply the statistical transmutation in order to keep the 
fermionic character by attaching an even number of flux quanta, 
which results in a Chern-Simons problem of interacting electrons at integer filling 
factors \cite{LF91,LF98}.  
These attempts are guided by the desire to derive the quite successful 
trial ground state wavefunctions for the FQHE filling factors from first principles. Furthermore, 
field theories open the opportunity to calculate various observables in a systematic manner. 
The fermionic transformation was also applied to the case of 
even fractions $\nu=1/2q$ for integer $q$ by Halperin \ea~\cite{HLR93}. In this case, 
attachment of $2q$ flux quanta per electron, which  
point oppositely to the external magnetic field, leaves 
the system in average in zero magnetic field. This observation is the starting point 
for a reasoning that the system behaves effectively similar to 
a Fermi liquid in a vanishing field \cite{Sim98}.  
An attempts to get rid of some inconsistencies in these theories led recently to 
the development of a Hamiltonian formulation theory by Shankar and Murthy \cite{SM97b}. 

	\subsection{Theoretical models of the quantum Hall regime and the lowest Landau level 
	approximation}
	\label{subsect:LLL-model}

Our discussion hitherto emphasized that in the region of filling factors, where the FQHE 
occurs, interaction as well as disorder are necessary ingredients of any 
trustworthy theoretical attempt to understand the transport properties.  
On the other hand, the concomitant treatment of disorder and interaction 
is one of the great challenges that solid-state physics encounters. 
In fact, the difficulties are considerable as the example of the treatment of interaction 
in the problem of the metal-insulator transition in disordered solids shows \cite{Local99}.  
Much worse, in most cases each single problem on its own is not exactly solvable. 

In the FQHE regime, the interaction is much larger than 
the fluctuation of the random potential since otherwise the many-particle gap 
above the ground state would be washed out and the effect would disappear.  
Because of the cleanness of the samples, disorder should not play the decisive 
role as long as transport theories are not considered.  
Nonetheless, disorder will quantitatively influence the results.
In any case, our starting point is an interacting  model {\em without} disorder. 
At strong fields, moreover, an independent particle model is an inappropriate 
starting point at partial filling 
of LLs due to the macroscopic ground state degeneracy.  
This argument does not question the usual treatment of 
transport in the IQHE regime, where one starts from a one-particle model with 
disorder neglecting just the interaction. 
The situation at integer filling factors is different from that at partially 
filled Landau levels as the unperturbed ground state is non-degenerate.
Nevertheless, any consistent theory at integer $\nu$ has also to account for the influence 
of the interaction, which has the same order of magnitude as the cyclotron energy 
under typical experimental conditions 
as we will see below. There are some recent attempts to incorporate electron-electron interaction 
in connection with the study of the diverging localization length in the center of the LLL 
indicating a localization-delocalization transition in a random potential \cite{LW96,HB99}.  
Eventually, this whole discussion gets an additional interesting turn due to approaches 
that relate the system at a partially filled LLL $\nu=p/q$ to a model of weakly interacting 
new composite particles at higher integer filling factor $p$. 

Based on these assumptions, we can write down a simplified second quantized 
Hamiltonian describing 
interacting electrons, which move in a plane and are subject 
to a perpendicular magnetic field ${\bf B} = B {\bf e}_{z}, B>0$, \cite{Rem2}
\begin{eqnarray}
H & = & \int d^2 r  \left\{ \sum_{\sigma} \frac{1}{2 m} 
   |({\bf p}-q{\bf A}) \Psi_{\sigma}({\bf r})|^{2}  
   - g \mu_B {\bf B} {\bf S}({\bf r}) \right\}  -\mu N  
\nonumber \\
& & + \frac{\lambda}{2} \int d^2 r \int d^2 r' 
:[n({\bf r}) - n_{b}] V({\bf r} - {\bf r}\;^{\prime})
[n({\bf r}\;^{\prime}) - n_{b}]: \ .
\label{ham}
\end{eqnarray}
The first line renders the one-particle Hamiltonian of Eq.~(\ref{Hnot}) using an effective
mass $m$ and gyromagnetic factor $g$ of the host semiconductor. The total spin density 
operator
\begin{equation}
{\bf S}({\bf r})
= \frac{1}{2} \sum_{\sigma, \sigma^{\prime}} \Psi_{\sigma}^{\dagger}({\bf r})
\vec{\tau}_{\sigma,\sigma^{\prime}} \Psi_{\sigma^{'}}({\bf r})
\end{equation}
can be expressed by means of the field operators $\Psi_{\sigma}({\bf r})$, 
$\Psi_{\sigma}^{\dagger}({\bf r})$ for
electrons with spin $\sigma$ and the  Pauli matrices  $\vec{\tau}$;
$\mu$ is the chemical potential, and
$N=\sum_{\sigma} \int d^{2}r \Psi_{\sigma}^{\dagger}({\bf r})\Psi_{\sigma}({\bf r})$
is the particle number operator.
The second line describes the interaction between the electrons, where 
$\lambda = e^2/(4 \pi \epsilon \ell_c)$ is the coupling constant of a Coulomb like 
interaction. The interaction term 
takes into account the two-particle interaction $V({\bf r}-{\bf r}\;^{\prime})$
between the electrons in the plane as well as the interaction of these electrons
with a homogeneous, compensating neutralizing background charge $n_b$ and
the self-interaction of the background. The sum of these terms can be expressed
by means of the normal order symbol $:$ and the electron density operator
$n({\bf r}) = \sum_{\sigma} \Psi_{\sigma}^{\dagger}({\bf r})
\Psi_{\sigma}({\bf r})$. 
The charge neutrality is satisfied via the constraint
$(1/A) \int d^2 r n({\bf r}) = n_{b}$. In the following, we assume an isotropic
particle interaction $V(|{\bf r}- {\bf r'}|)$.

Despite the restriction to the plane, the interaction is three-dimensional as the
lowest subband wavefunction $\phi_{0}(z)$ is extended in the $z$-direction. 
Hence, the Fourier transformation of the interaction $\tilde V({\bf q})$ 
is corrected by a form factor when 
the influence of the finite width $w$ is discussed. 
The constant subband kinetic contribution to the Hamiltonian is omitted.

The purely electronic Hamiltonian Eq.~(\ref{ham}) 
still covers a huge number of physical situations ranging from the two-dimensional
jellium model for vanishing magnetic field \cite{FW71} to the strong magnetic field situation, 
where the QHEs occur.
If we neglect the Zeeman energy for sake of simplicity,  
the ground state properties in the thermodynamic limit without magnetic field 
are characterized by the dimensionless electron gas parameter $r_{s}=r_{0}/a_{B}$,  
where $r_{0}=1/\sqrt{\pi n}$ is the average particle distance in two dimensions and 
$a_{B}=4 \pi \epsilon \hbar^2/(m e^2)$ is the Bohr radius.
If we switch on a magnetic field, the ratio of Coulomb to cyclotron energy 
\begin{equation}
\frac{e^{2}/(4 \pi \epsilon r_{0})}{\hbar \omega_{c}} = \frac{1}{2} r_{s} \nu
\end{equation}
determines the ground state properties. In the case of an infinitely large magnetic field 
leading to a vanishing interaction energy against the cyclotron energy, it is $r_{s} \to 0$.  
Then, the ground state properties become only a function of $\nu$, which is just 
the situation we encounter in the LLL approximation. 
Low values of $r_{s}$ mean high densities $n$. Consequently, strong magnetic fields 
are needed to reach the low filling factor region. Note that for a typical density of 
$n \simeq 10^{11}\ cm^{-2}$, it is $r_{s} \simeq 1.8$.

Since $\ell_{c}/r_{0}=\sqrt{\nu/2}$, it is appropriate to consider the 
coupling constant $\lambda$  for filling factors $\nu \simeq 1$ 
as the relevant interaction energy scale.
If one is interested in the thermodynamic properties of the model, 
the following three energy scales of an ideal 2DES govern the physics. These are 
the Landau level distance $\hbar \omega_{c}= \hbar |e B|/m$,
the Zeeman energy $\Delta_z=|g \mu_{B} B| = |g B e| \hbar /(2 m)$ and
the Coulomb coupling constant $\lambda = e^{2}/(4 \pi \epsilon \ell_{c})$.
For $GaAs$, the three energy scales expressed as temperatures in 
$K$ are 
\begin{eqnarray}
\hbar \omega_{c}/k_{B} \ [K] & = & 20.05\ B[T]  \nonumber \\
\Delta_{z}/k_{B}\ [K] & = & 0.2953\ B[T]  \nonumber  \\
\lambda/k_{B} \ [K] & =& 51.44\ \sqrt{B[T]}   \ ,
\label{scales}
\end{eqnarray}
when the magnetic field strength $B$ is given in $T$. 
For magnetic fields greater than
$6.58\ T$, the Landau level distance exceeds the interaction energy scale, whereas
the latter one is still larger than the Zeeman energy up to an experimentally non-accessible 
huge magnetic field of $30\;344\ T$.
Although most of the strong magnetic field experiments are done in the range between $7\ T$ and
$20\ T$ and at temperatures below $10\ K$, the omission of higher
Landau level mixing is not automatically justified if the filling factor
is smaller than two.
Nowadays, \eg, samples with small carrier densities down to 
$n \simeq 0.2\times 10^{11}\ cm^{-2}$ are available, and
therefore QH-experiments at $\nu=1$ can be performed in a magnetic field of 
$2.5\ T$ \cite{SEPW95}. Then, the Coulomb energy exceeds the cyclotron gap,
and the LLL approximation has non-negligible corrections.
However, we restrict ourselves to the lowest orbital Landau level throughout this work, 
since the strong magnetic field effects become more pronounced at stronger fields.  
Furthermore, there are some technical simplifications as the 
form of the wavefunctions in the LLL that suggest such an approximation.
	
The Hamiltonian Eq.~(\ref{ham}) can be restricted for $0 < \nu \le 2$ 
to the Hilbert space of the LLL orbital states $n=0$, when the condition 
\begin{equation}
\Delta_{z},  \lambda \ll \hbar \omega_{c} \ 
\label{lll-spin}
\end{equation}
holds, and when the experimentally relevant temperature range satisfies the inequality 
\begin{equation}
k_{B} T \ll \hbar \omega_{c} \ .
\label{temperature}
\end{equation}
In first quantization, the Hamiltonian reads without a positive background term 
in the case of spinless electrons ($0 < \nu \le 1$) \cite{FPA79} 
\begin{equation}
H   =  \frac{\hbar \omega_{c}}{2} N  \  + \ 
\frac{\lambda}{2} \int \frac{d^{2} q}{(2 \pi)^2} \tilde{V}({\bf q}) \  
(\bar{n}^{\dagger}_{{\bf q}}\  \bar{n}^{\vphantom{\dagger}}_{{\bf q}} - 
n_{{\bf q}=0} e^{-q^2 \ell_c^2})   \ ,
\label{ham-lll-first}
\end{equation}
where $n_{{\bf q}}=\sum_{j} e^{-i {\bf q} {\bf r}_{j}}$ is the Fourier transformation 
of the one-particle 
density operator,  $\bar{n}_{{\bf q}} = \sum_{j}  e^{-i q \partial_{z_{j}}}
e^{-i q^{*} z_{j}}$ means its projection onto the LLL, and $z=x-iy, q=q_{x}-i q_{y}$ 
are complex coordinates \cite{GMP85,GJ84}. 

Furthermore, we quote two second quantized representations in planar  
geometry. The first is suited 
for the treatment of a finite system on a disk in front of a positively charged 
background 
\begin{eqnarray}
H & = &  \frac{\hbar \omega_{c}}{2} N
- \frac{1}{2}\Delta_z(N_{\uparrow}-N_{\downarrow}) 
+ \frac{\lambda}{2}\; 
\sum\limits_{m_{1},m_{2}, m_{3},\atop m_{4}=0;\;
\sigma, \sigma'}
W_{m_{1},m_{2},m_{3},m_{4}} \: c^{\dagger}_{m_{1},\sigma}c^{\dagger}_{m_{2},\sigma'}
c_{m_{3},\sigma'} c_{m_{4},\sigma} 			\nonumber \\
& &  - \nu\lambda \sum\limits_{m,m'=0,\atop \sigma}
W_{m,m',m',m} \: c^{\dagger}_{m,\sigma}c_{m,\sigma}
 + \frac{\nu^{2}\lambda}{2} \sum\limits_{m,m'=0} W_{m,m',m',m}  \ .
\label{ham-lll-sym}
\end{eqnarray}
Here, $N=N_{\uparrow}+N_{\downarrow}$ is the particle operator, and the 
field operators are represented solely by the LLL eigenfunctions
$\varphi_{m}({\bf r})$
and LLL fermionic field operators $c_{m,\sigma}, c^{\dagger}_{m,\sigma}$, \ie 
\begin{equation}
\Psi_{\sigma}({\bf r}) =
\sum_{m=0}^{\infty} \varphi_{m}({\bf r})\,c_{m,\sigma} \ .
\label{spin-density}
\end{equation}
The interaction matrix element in symmetric gauge for the LLL is given by 
\begin{eqnarray}
W_{m_{1},m_{2},m_{3},m_{4}} & =& \int d^{2} r \int d^{2} r' \varphi^{*}_{m_1}({\bf r})
\varphi^{*}_{m_2}({\bf r}\;') V(|{\bf r}-{\bf r}\;'|) \varphi_{m_3}({\bf r}\;') 
\varphi_{m_4}({\bf r}) \\ \nonumber 
& \equiv & \langle m_{1}, m_{2} | V | m_{4}, m_{3} \rangle \ 
\label{matrixelement-symm-LLL}
\end{eqnarray}
and satisfies for any isotropic interaction potential the angular momentum 
conservation $m_1+m_2=m_3+m_4$.  Usually, we will omit the constant kinetic energy 
$\hbar \omega_c/2$ multiplied by the particle number.
In Eq.~(\ref{ham-lll-sym}), the constant neutralizing homogeneous background 
density $n_b$ 
is represented by the expectation value of the filled orbital LLL 
weighted by the filling factor $\nu$, which ensures charge neutrality
and for models with long-range interaction the finiteness of quantities in the
thermodynamic limit. 

In Landau gauge, the LLL Hamiltonian of Eq.~(\ref{ham-lll-sym}) 
neglecting the kinetic energy term reads 
\begin{equation}
 H  =  - \frac{1}{2} \Delta_{z}(N_{\uparrow}-N_{\downarrow})
 +   \frac{\lambda}{2} \sum\limits_{p,p^{'},q \atop \sigma, \sigma^{'}}
 \tilde{W}(q,p-p^{'})\, 
c^{\dagger}_{p+\frac{q}{2},\sigma^{\vphantom{'}}} 
c^{\dagger}_{p^{'}-\frac{q}{2}, \sigma^{'}} 
c^{\vphantom{\dagger}}_{p^{'}+\frac{q}{2}, \sigma^{'}} 
c^{\vphantom{\dagger}}_{p-\frac{q}{2}, \sigma^{\vphantom{'}}}  \ ,
\label{ham-lll-landau}
\end{equation}
where $\tilde W(q,p-p')$ is the Landau gauge matrix element 
\begin{eqnarray}
 \tilde{W}(q,p-p^{'}) & = & \int \frac{d^2 k}{(2 \pi)^2} \tilde V({\bf k}) e^{-\frac{{\bf {k}}^2 \ell_c^2}{2}}
e^{i k_{x} (p-p') \ell_c^2} \delta_{k_{y},q}    \nonumber \\
& \equiv & \langle p+\frac{q}{2},p'-\frac{q}{2} |V| p-\frac{q}{2},p'+\frac{q}{2} \rangle
\label{matrixelement-landau}
\end{eqnarray}
using the one-particle eigenfunctions of the LLL in Landau gauge 
\begin{equation}
\phi_{n=0, k_{y,l}}({\bf r}) =
\frac{e^{ik_{y,l}y}}{\sqrt{L_{y}\sqrt{\pi}}\;\ell_c}\; e^{-\frac{(x-X_{l})^2}{2 \ell_c^{2}}} 
\end{equation}
with centers $X_{l}=2\pi \ell_c^{2} l/L_{y}$ ($l=0,\pm 1, \pm 2, \ldots$) along the $x$-direction.  
By excluding the $q=0$ contribution from the sum we take into account a positively charged homogeneous background. 
Diagrammatically, this corresponds to the neglect of the tadpole diagrams.

Due to the macroscopic degeneracy of the non-interacting part of the Hamiltonian 
(\ref{ham-lll-first}), zero temperature field theoretical methods cannot be applied  
except for the special case $\nu=1$ \cite{KH84}.  
Therefore, only a thermodynamic Green's function method 
like the Matsubara technique allows a high-temperature 
expansion in $\lambda/(k_{B}T)$ at arbitrary filling factors.  
For later use let us introduce the Matsubara or imaginary time Green's function defined as
\begin{equation}
{\cal G}_{\sigma,\sigma'}(x,x') = - \langle {\cal T} \Psi_{\sigma^{\vphantom{'}}}(x) 
\Psi^{\dagger}_{\sigma'}(x')  \rangle \ , 
\label{matsubara}
\end{equation}
where $x=({\bf r},\tau)$, ${\cal T}$ is the time ordering symbol, 
$\Psi^{(\dagger)}(x)= e^{(H-\mu) \tau/\hbar}\, \Psi^{(\dagger)}({\bf r})\, e^{-(H-\mu) \tau/\hbar}$  
and $\langle ... \rangle$ means averaging with respect to the grand canonical 
ensemble.
The range of the imaginary time $\tau$ is $0 \le \tau \le \hbar \beta$. For time independent 
Hamiltonians the GF becomes a function of $\tau - \tau'$ and is 
antiperiodic with respect to the period $\hbar \beta$ in case of fermions.
In a constant magnetic field the GF is independent of the degenerate quantum number $k$, \ie  
\begin{eqnarray}
{\cal G}_{n,k,\sigma; n',k',\sigma^{'}} (\tau, \tau^{'})
& = &- \langle {\cal T} c_{n,k,\sigma}(\tau) c_{n',k',\sigma'}^{\dagger}(\tau') \rangle \ \nonumber \\
& = & \delta_{n,n'} \delta_{k,k'} \delta_{\sigma,\sigma'} {\cal G}_{n,\sigma} (\tau-\tau^{'}) \ .
\label{magnetic_GF}
\end {eqnarray}
The free electron frequency dependent Matsubara function ${\cal G}^{(0)}_{n,\sigma}(i \nu_{n'})$ is given by 
\begin{eqnarray}
{\cal G}^{(0)}_{n, \sigma}(i \nu_{n'}) = \frac{1}{i\hbar \nu_{n'}-\xi^{(0)}_{n, \sigma}} = 
\frac{1}{\hbar} \int_{0}^{\hbar \beta} d\tau\, e^{i \nu_{n'} \tau} {\cal G}^{(0)}_{n, \sigma}(\tau) 
\label{GF0}
\end{eqnarray}
with fermionic Matsubara frequencies $\nu_{n'}=(2n'+1)\pi/(\hbar \beta)$  and 
electronic energies $\xi^{(0)}_{n,\sigma} = \hbar \omega_c (n+1/2) \mp \Delta_z/2 -\mu$ 
in the LL $n$ with spin $\sigma$.

In case of the projection onto the LLL, the interaction energy scale dominates over 
any other energy scale except for the temperature, 
which makes the free electron problem inappropriate as a starting point for a 
perturbation theory. This problem is typical for the treatment of strongly correlated systems.
Within the LLL approximation, the constant kinetic energy term with the band mass $m$ can be neglected.  
Nevertheless, there exist situations, where the omission is not justified. For example, 
the construction of the current operator makes the inclusion of higher LLs necessary. 
Note that the LLL projection leads to an effectively one-dimensional problem. 
The Hamiltonian shows also some resemblance with lattice models of itinerant electron systems. 
The kinetic energy term can be viewed as the vanishing limit of a narrow band, but it 
will attain a non-zero value due to existing weak disorder. 

The LLL Hamiltonians (\ref{ham-lll-sym}), (\ref{ham-lll-landau}) can be considered as 
a special case of a Hamiltonian describing a double-layer system consisting 
of two parallel quantum wells \cite{Eis97}.  
Such a system is characterized by the distance of the layers $d \simeq 10 nm$, 
which has the order of $\ell_c$ and hence of $r_{0}$,   
and a hopping matrix element $\Delta_{SAS}$ for interwell tunneling. 
For sufficiently narrow wells and strong fields, we can again assume 
that the electrons occupy only the lowest of the subbands in each well and equally only the 
LLL.  Moreover, it is assumed that the electrons are spin-polarized. Then, in the 
unphysical limit of vanishing layer distance $d=0$, but non-vanishing 
$\Delta_{SAS}$, the system can be exactly mapped 
onto the one-layer system in the LLL with spin-degree of freedom and Zeeman term. 
The energy difference $\Delta_{SAS}$ of the symmetric and antisymmetric combination of the 
subband wavefunctions is the analog of the Zeeman energy $\Delta_{z}$. 
This mapping is accomplished by a unitary transformation in the layer-index space introducing 
a fictitious pseudo-spin \cite{Fer89,MPB90}. 
In case of a non-zero distance $d>0$, the $SU(2)$-symmetry of the Coulomb interaction 
is broken to a $U(1)$-symmetry as the inter-layer Coulomb interaction becomes 
smaller than the intra-layer interaction. This leads to a variety of 
zero and finite temperature phase transitions, which, for example, 
were experimentally and theoretically 
studied at total filling factor $\nu=1$ \cite{YMZMGYZ94,MEBPW94,MMY95,YMB96}. 

In this work, we are mostly concerned with ground state properties of the 
$B \to \infty$ limit, where $\lambda \ll \Delta_z, \hbar \omega_c$ at the typical 
FQHE filling factor $\nu=1/3$ as well as with ground states and 
thermodynamics at and near $\nu=1$. In the latter case, the experimentally important condition 
$\Delta_z \ll \lambda < \hbar \omega_c$ holds.  Therefore, we will ignore 
any effects related to higher LLs. We focus in the work on the bulk properties 
of the 2DES and do not touch two other important problems. First, due to the edges of the 
system there exist in contrast to the bulk region gapless excitations, 
whose low-energy excitations can be described as a chiral Luttinger liquid \cite{Wen92,KF97}. 
Second, one can discuss the physics of a system with a finite number of particles,  
which are confined to an approximately parabolic potential of small extent. 
This corresponds to the situation of a so-called quantum dot, whose physics  
is intensively studied without and in a magnetic field. 

Although this work considers only the physics at single points along the line 
of continuous values of the filling factor $\nu$, we will get an impression of the 
abundance of phenomena occurring in this small filling factor range $0 < \nu \le 2$.  

Research in the field of 2DES in a strong magnetic field has been a lively and 
still expanding field for more than twenty years.  
Besides the original papers, there are review articles and books giving an introduction 
for the newcomer as well as valuable information for the active researcher.
A still very useful source of information on 2DES up to the advent of the quantum Hall effects 
is the review by Ando, Fowler, and Stern from 1982 \cite{AFS82}. 
Meanwhile, there are a few textbooks and collections on the integer and the fractional quantum
Hall effect available \cite{CP95,PG90,JVFH94,DP97,Hei98}.
Pedagogical introductions to the field are lectures given by MacDonald
\cite{Mac95} and Girvin \cite{Gir99}.

	\subsection{Two interacting particles in a magnetic field }
	\label{subsect:2particles}

The study of two interacting spin-polarized fermions of equal and 
opposite charge, respectively,  $q_1=\pm q_2$ in a magnetic field is very instructive 
for the understanding of the many-particle problem. 

The problem is further facilitated when the Hamiltonian is projected onto the LL $n$.
Here we focus on the projection onto the LLL $n=0$. But even projection onto orbital 
LLs $n>0$ is quite common when discussing the FQHE at, \eg, $\nu=5/2$ \cite{Mor98} or higher 
LL effects \cite{KFS96}. On the other hand, LL mixing cannot be ignored as was shown 
for $2 \le \nu \le 4$ \cite{RH90}.
\begin{figure}[h]
\centerline{\resizebox{6cm}{6cm}{\includegraphics{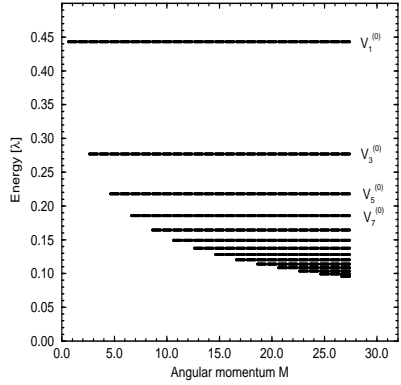}}}
\caption[]{\small
The interaction energy spectrum of two spin-polarized fermions of equal charge in the LLL 
interacting via Coulomb interaction. 
The energy levels are drawn vs.~the total angular momentum $M=m_{-}+m_{+}$. 
Different energy levels indicate eigenfunctions 
with different relative angular momentum $m_{-}$ whose eigenvalues are just the pseudopotential 
coefficients $V_{m_{-}}^{(0)}$ characterizing the Coulomb interaction. 
Due to the antisymmetry of the 
entire wavefunction and the assumed spin-polarization the 
relative angular momentum eigenfunctions exhibit only odd $m_{-}$ values. 
The horizontal degeneracy 
is due to the degeneracy with respect to the center of mass angular momentum eigenvalue.
As long as the total angular momentum $M < m_{max}$, where $m_{max}$ was chosen 
to truncate the Hilbert space of the LLL, the breaking of the translational 
symmetry due to the finite extension of the sample cannot be read off from the spectrum. 
}
\label{two_particle_spectrum}
\end{figure}
The two-particle Hamiltonian can be written as 
\begin{eqnarray}
H & = & H_{0,1} + H_{0,2} + H_{int,1-2} \nonumber \\
& = & H_{0,1} + H_{0,2} \pm \lambda \int \frac{d^{2}q}{(2 \pi)^2} \tilde{V}(q) 
e^{i {\bf q}({\bf r}_1-{\bf r}_2)}  \ , 
\label{two-particles}
\end{eqnarray}
where the sign depends on the repulsion or attraction of the particles' charges.
$\tilde{V}({\bf q})$ is the Fourier transformation of the two-particle potential 
and equals, \eg, $2 \pi/|{\bf q}|$ for the bare Coulomb interaction in two dimensions.
The separation into a free center of mass motion and the relative motion reduces the 
problem to a one-particle problem in the relative coordinate.
In the case of {\em equal} charges and an isotropic interaction, the energetic eigenvalues 
after projection onto the Hilbert space of the $n$th Landau level 
are characterized by $n$ and the relative angular momentum $m_{-}$.
The eigenvalues are given by 
\begin{eqnarray}
E_{m_{-}}^{(n)} & = & \hbar \omega_c (2n+1)  + \lambda V_{m_{-}}^{(n)}  \nonumber \\ 
& = & \hbar \omega_c (2n+1) + \lambda \int \frac{d^2 q}{(2 \pi)^2} 
\tilde{V}(q)e^{-|{\bf q}|^{2} \ell_c^2} [L_{n}(\frac{|{\bf q} \ell_c|^2}{2})]^{2} 
L_{m_{-}}(|{\bf q} \ell_c|^2) \nonumber \\ 
& = & \hbar \omega_c (2n+1) + 
\langle \varphi^{rel}_{n,m_{-}}|H_{int}|\varphi^{rel}_{n,m_{-}} \rangle \ ,
\label{pseudo_n_m}
\end{eqnarray} 
where the $L_{n}(z)$ are Laguerre polynomials and the 
$V_{m_{-}}^{(n)}$ are called pseudopotential coefficients 
characterizing the projection of any isotropic interaction onto the Landau level 
$n$ \cite{Hald83}. The eigenfunctions of the free relative coordinate problem in symmetric 
gauge are given by $\varphi^{rel}_{n,m_{-}}$. For the LLL $n=0$ the explicit energy eigenvalues 
for the Coulomb interaction are 
\begin{equation}
E^{(0)}_{m}
= \hbar \omega_c +\lambda V_{m} 
=\hbar \omega_c + \lambda \  \frac{(2m!) \sqrt{\pi}}{2^{2m+1} (m!)^2} 
\label{pseudo}
\end{equation}
using the notation $V_{m} \equiv V_{m_-}^{(0)} $. These pseudopotential coefficients of the Coulomb 
interaction form a series of monotonically decreasing numbers,  
\cf~also Sect.~\ref{sect:polar}, 
\ie~$V_{0}=\sqrt{\pi}/2, V_{1}=\sqrt{\pi}/4, V_{2}=3\sqrt{\pi}/16, 
V_{3}=5\sqrt{\pi}/32, \ldots $. 
They can be visualized by numerically diagonalizing the LLL two-particle Hamiltonian 
exploiting the conservation of the total angular momentum $M=m_{-}+m_{+}$, 
see Fig.~\ref{two_particle_spectrum}.
In contrast to the discrete energy values in the repulsive case, for {\em attractive} charges 
we get a continuous spectrum from the interaction term in Eq.~(\ref{two-particles}) 
when projecting onto the LL $n$.
This is due to the fact that the exponential in (\ref{two-particles}) can be separated in a part containing 
only harmonic oscillator like Landau level ladder operators and a part that 
depends on the total magnetic translation operator for two particles $\bf T$. However, the latter 
is the projection of the generator of the translation operator $\bf P$ of the center of mass coordinate 
without magnetic field 
onto an arbitrary LL $n$. Since the translation operator has a continuous spectrum  described 
by the two-dimensional vector $\bf k$ the energy spectrum in the LL $n$ reads
\begin{eqnarray}
E^{(n)}({\bf k}) & = & \hbar \omega_c (2n+1) - 
\lambda \int \frac{d^2q}{(2 \pi)^2} \tilde{V}({\bf q}) 
     e^{-\frac{|{\bf q}|^2 \ell_c^2}{2}} [L_{n}(\frac{|{\bf q} \ell_c|^2 }{2})]^{2} 
     e^{i {\bf q} ({\bf e}_z \times {\bf k}) \ell_c^2}  \nonumber \\
&  \equiv & \hbar \omega_c (2n+1) -  \lambda \tilde{a}^{(n)}({\bf k}) \ , 
\label{energy-attractive}
\end{eqnarray}
where $\tilde a^{(n)}({\bf k})$ describes the binding energy of a 
magneto-exciton in the LL $n$. 
In the LLL, this energy becomes 
\begin{eqnarray}
\tilde{a}({\bf k}) \equiv  \tilde{a}^{(0)}({\bf k})   
 & = & \int \frac{d^2q}{(2 \pi)^2} \tilde{V}({\bf q})
   e^{-\frac{q^2 \ell_c^2}{2}} e^{i {\bf q} ( {\bf e}_z \times {\bf k}) \ell_c^2} \nonumber \\
& = & \int \frac{d^2 r}{2 \pi} V({\bf r} - ({\bf e}_z \times {\bf k}) \ell_c^2) 
e^{- r^2/2 \ell_c^2} \ .
\label{attraction-two-particles}
\end{eqnarray}
\begin{figure}[h]
\centerline{\resizebox{6cm}{6cm}{\includegraphics{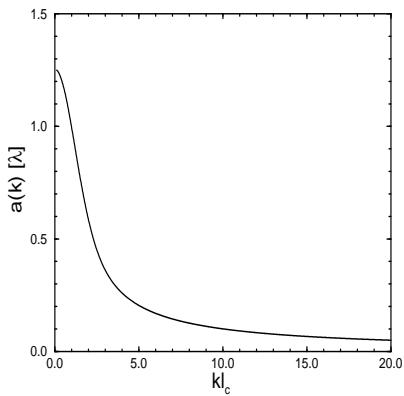}}}
\caption[]{\small
The binding energy $a(\bf k)$ Eq.~(\ref{ak}) of two fermions of opposite charges in the LLL 
for the unscreened Coulomb interaction. Asymptotically, it is $a(k) \sim 1/k$.}
\label{a_k}
\end{figure}
The second line shows that for $k \ell_c > 1$ the quantity 
$\tilde a({\bf k})$ describes 
the interaction between two well separated single particles of distance $k \ell_c^2$.  
For an unscreened Coulomb interaction (no tilde) the isotropic binding energy is depicted in Fig.~\ref{a_k} 
and equals
\begin{equation}
a({\bf k}) = 
\sqrt{\frac{\pi}{2}} e^{-\frac{k^2 \ell_c^2}{4}} I_{0}(\frac{k^2 \ell_c^2}{4}) \ .
\label{ak}
\end{equation}
Here, $I_{0}(z)$ is a modified Bessel function. The value $a(0) = \sqrt{\pi/2}$ 
can be identified as the exchange energy as comparison with 
the value of the Fock diagram shows \cite{KH84,BIE81}. 

\section{Ground states and quasiparticles of spin-polarized electrons in the lowest Landau level}
\label{sect:polar}

	\subsection{The Laughlin function as a quantum liquid ground state at filling 
	factor $1/q$ ($q$ -- odd)}
	\label{subsect:laughlin}

This Section is mainly concerned with the ground state and its charged elementary 
excitations, the so-called quasiparticles (QPs) 
at filling factor $\nu=1/q$. We restrict ourselves to the strongest FQHE filling factor 
$\nu=1/3$. We review the ground state properties of the Laughlin wavefunction 
and develop the notion of the quasiparticles. We illustrate these results on the 
basis of a generic model for the electron-electron interaction, the so-called hard-core model, 
which we will frequently use in modified form throughout this paper. 
In particular, we study the properties of the quasielectrons and discuss 
the merits of various explicit trial wavefunctions proposed for the quasielectron. 
Two theories that try to explain the occurrence of ``weaker'' FQHEs at filling factors 
$\nu=p/q$ with $p>1$ are discussed: the older hierarchical theory and 
Jain's composite fermion theory.
We focus on the disk geometry because direct comparison with the Laughlin theory  
originally formulated in the symmetric gauge is 
possible and experiments are done in a planar geometry. On the other hand,  
results for finite systems on a disk are affected by the edge and show large finite-size corrections.
These are disadvantages when one is interested in bulk properties in the thermodynamic limit, 
since numerical calculations in the spherical geometry don't suffer from edge effects.  
However, we think that it is worth to elucidate the problem in the disk geometry. 

Let us start with the main theoretical arguments supporting the correctness of 
the Laughlin theory for interacting electrons in a strong magnetic field at $\nu=1/q$. 
We assume in this Section that $\lambda \ll \Delta_z, \hbar \omega_c$\,, \ie~the electrons 
are restricted to the subspace of spin-polarized electrons in the LLL. This corresponds 
to the $B \to \infty$ limit in Eq.~(\ref{ham-lll-sym}) and allows the omission of the first 
two terms. Such an assumption is not a priori justified 
when comparing with typical experimental field strengths in Eq.~(\ref{scales}) since 
the inequality (\ref{lll-spin}) is not satisfied.  
But the discussion of the spin-degree of freedom following in Section \ref{sect:spin} will show that 
such an assumption is justified, \eg, at $\nu=1/3$. 
Therefore, in such a limit the only energy scale of the Hamiltonian remains the interaction strength $\lambda$.

The theory of the FQHE put forward by Laughlin \cite{Lau83b,Lau83a} is based on 
proposals for the ground state wavefunction of the interaction dependent part of the 
Hamiltonian (\ref{ham-lll-sym}) at special filling factor $\nu=1/q$ and for the elementary 
charged excitations, the quasiparticles, at this $\nu$. 
At first glance, the search for an appropriate wavefunction seems to be hopeless 
as we encounter here a non-perturbative situation, where standard methods fail. 
However, Laughlin succeeded in finding a ground state wavefunction at the FQHE filling 
factor $\nu=1/q$ ($q$ -- odd) with some remarkable properties.  
It can be shown that
\begin{enumerate}
\item
the Laughlin wavefunction describes a many-particle quantum liquid-like state, 
\item
the elementary charged excitations have a fractional 
charge of magnitude $\pm e/q$, and the sum of their energies defining the 
quasiparticle gap $\Delta_{qe-qh}$ remains positive in the thermodynamic limit. 
\item
The resulting incompressibility ensures at exactly $\nu=1/q$ the position 
of the chemical potential in the many-particle gap above the ground state, 
which is a necessary condition for the peculiar transport properties. 
\end{enumerate}
Laughlin's wavefunction for a finite number $N$ of particles at $\nu=1/q$ approximately 
describes the ground state for a large class of model interactions. 
It is given by
\begin{equation}
\Psi _{1/q}(z_{1},\ldots,z_{N})= \prod_{i>j=1}^{N}(z_{i}-
z_{j})^{q}e^{-\frac{1}{4}\sum_{i=1}^{N}|z_{i}|^{2}} \ .
\label{Laughlin}
\end{equation}
The density corresponds to a circular droplet about the origin of radius $R=\sqrt{2 N_{\Phi}} 
\ell_c$. The Laughlin wavefunction has total angular momentum $M_{N}(q)=qN(N-1)/2$. 
The unnormalized wavefunction is written in first quantization and ensures 
antisymmetry for fermions due to the occurrence of only odd $q$. 
The wavefunction is built out of the symmetric-gauge single-particle states of the LLL $n=0$,
\cf~Eq.~(\ref{sym-n=0}). Therefore, only powers of $z_{i}$, 
but not $z_{i}^{*}$, occur in the non-exponential part of (\ref{Laughlin}). 
Note that analogous wavefunctions can be found for the Landau gauge in the 
torus geometry \cite{Hald85} and for electrons moving in the field of a magnetic monopole on 
the surface of a sphere \cite{Hald83}.
For a finite particle number, the filling factor definition $\nu=N/N_{\Phi}$ 
for the thermodynamic limit Eq.~(\ref{nu}) needs to be corrected 
by a constant $K(\nu)$ forming the ``magic table'' in \cite{Hald83,GM90} so that 
\begin{equation}
N_{\Phi} = \nu^{-1} N + K(\nu)
\label{finite-nu}
\end{equation}
for any Laughlin filling factor 
$\nu=1/q$. Starting with $q=1$ and $K(\nu=1)=0$, we get for a new filling 
factor $(\nu\ ')^{-1} = \nu^{-1} + 2$ a correction $K(\nu')=K(\nu)+2$. Thus, it is 
$K(\nu=1/q)=-(q-1)$, so that $N_{\Phi} = q N - (q-1)$ is the 
corrected finite particle number definition. In fact, applying this scheme to 
(\ref{Laughlin}), we have the following finite-size filling factor definition for the 
Laughlin wavefunction  
\begin{equation}
\nu=(N-1)/(N_{\Phi}-1) = (N-1)/m_{max} = (N-1)/(q(N-1)) = 1/q \ .
\label{finite-nu-q}
\end{equation}
Here, $m_{max}$ 
denotes the maximum one-particle angular momentum in $\Psi_{1/q}$ measuring the 
spatial extension of the electrons or, mathematically,  
the highest power in any of the $z_{i}$ as the polynomial part of the Laughlin function is homogeneous. 
Simple power counting in (\ref{Laughlin}) shows indeed that the wavefunction 
has $\nu=1/q$.
The relation (\ref{finite-nu}) is not only 
useful in order to identify the FQHE states for finite particle numbers at $1/q$, but also 
to identify states as particle-hole symmetry FQHE states or states at filling factor $\nu=p/q$. 

We see that if we fix, \eg, the coordinates of the particles $2,\ldots, N$, the particle 
$1$ is very efficient in keeping the other particles apart due to the $q$th power 
of the relative coordinates $(z_{1}-z_{i})$. This essentially lowers the interaction energy. 
However, this qualitative argument depends on the filling factor and cannot be 
universal, since we expect at very small filling factors $\nu$ or large $q$ that a Wigner crystal and 
not the Laughlin wavefunction is the ground state.

We note that for $q=1$ and any $N$ the Laughlin wavefunction is {\em exact} and 
{\em independent} of the kind of interaction, since all single-particle states 
of the lowest orbital 
Landau level are occupied. Therefore, the only possible Slater determinant agrees for 
the non-exponential part with the Vandermonde determinant in the coordinates $z_{i}$ 
$\Phi_{1}(z_{1},\ldots,z_{N}) = \prod_{i>j} (z_{i}-z_{j})$ and can be written 
either in the form of Eq.~(\ref{Laughlin}) or in second quantization as 
\begin{equation}
|\Psi_{1},N \rangle  = |N-1,N-2,\ldots,1,0 \rangle = 
c_{N-1}^{\dagger} c_{N-2}^{\dagger} \ldots c_{1}^{\dagger} 
c_{0}^{\dagger}\;|0 \rangle  \; .
\end{equation}
Additionally, in the case of two particles requiring $m_{max}=q$ at $\nu=1/q$,  
one is inevitably led to 
$\Psi_{1/q}(z_{1},z_{2}) = (z_{1} - z_{2})^{q} e^{-(|z_{1}|^2 + |z_{2}|^2)/4}$ 
{\it independently} of the interaction. Rewriting this wavefunction as a product of 
a free particle wavefunction in the center of mass coordinate with angular momentum $m_{+}=0$ and  
a LLL wavefunction $z_{-}^{q} e^{-|z_-|^2/8}$ with $q=m_{-}$ shows the direct relation to the two-particle 
spectrum for $n=0$ in Eq.~(\ref{pseudo}).

However, for $q > 1$ and arbitrary particle number this function 
remarkably withstands any analytical translation into the language of second quantization 
for larger $N$ despite its apparent simplicity. This 
leads to difficulties to perform the thermodynamic limit $N \to \infty$, 
see \eg~\cite{FL91}, and explains why to date a lot of our conclusions is 
based on finite-size numerical calculations. 
Indeed, the number of Slater determinants built out of 
one-particle states increases dramatically with increasing particle number $N$.
For $N=10$ and $\nu=1/3$, \eg, there are $135\;281$ 
Slater determinants, whereas the dimension of the sub-Hilbert space of total 
angular momentum $M=135$ is $246\;448$\,. This emphasizes the strong 
correlation in the ground state at fractional $\nu=1/3$ in contrast to the case $\nu=1$, where 
only one Slater determinant, the Hartree-Fock state, occurs. 

Evidence that the Laughlin wavefunction contains the essential physics 
comes from various sources. These are mainly 
\begin{enumerate}
\item
the fact that the Laughlin wavefunction is the exact solution of a model, 
the so-called {\it hard-core model}, with a specific interaction  
\cite{Hald83,TK85,PT85},  
\item
the comparison of the numerically calculated exact ground state wavefunctions for a 
small number of particles $\Psi_{ex}$ with the Laughlin function $\Psi_{1/q}$ in 
various geometries. Here, for small number of particles 
the energy expectation values were compared and the overlap 
\begin{equation}
\frac{ \langle \Psi_{ex}|\Psi_{1/q} \rangle }{\| \Psi_{ex}\|\; \|\Psi_{1/q}\|}
\label{overlap}
\end{equation}
was studied for Coulomb interaction and logarithmic interaction with an  
overlap larger than $0.99$ \cite{Lau83b},  
\item
the analogy with a well studied classical problem, the so-called 
plasma analogy \cite{Lau83b},  
\item
field-theoretical derivations of the Laughlin wavefunction after a singular gauge 
transformation, see below, by studying fluctuations either about a bosonic 
mean-field theory in an average zero magnetic 
field \cite{ZHK89a,Zha92,KKLZ91} or about a fermionic mean-field theory with an average magnetic 
field of the filled lowest Landau level \cite{LF98,LF93}.  
Note that these field theories are justified by their capability to 
derive the already known ground state wavefunctions. 
\end{enumerate}
The first argument provides some firm ground one can build on. 
Based on our considerations of the two-particle problem in Subsect.~\ref{subsect:2particles},  
the many-particle Hamiltonian for an isotropic two-particle interaction can be written in 
first quantization as 
\begin{equation}
H_{int} = \lambda \sum_{i>j=1}^{N} \sum_{n=0, \atop m=-n}^{\infty} 
V_{m}^{(n)} |n,m \rangle _{ij} \langle n,m|_{ij} \ , 
\label{ham-first}
\end{equation}
where the pseudopotential coefficients $V_{m}^{(n)}$ are given by 
Eq.~(\ref{pseudo_n_m}), while the quantity $|n,m \rangle _{ij} \langle n,m|_{ij}$ 
acts as a projection operator onto states with relative angular momenta $m$ in Landau level $n$ 
between particles $i$ and $j$. 
Projection onto the LLL in (\ref{ham-first}) yields the real space representation 
\begin{equation}
H^{(0)}_{int}({\bf r}_1, \ldots, {\bf r}_N) = \lambda \sum_{i>j=1}^{N} 4 \pi \sum_{m=0}^{\infty}
V_{m} {\bf \nabla}^{2 m} \delta^{2}({\bf r}_{i} - {\bf r}_{j})  \ ,
\label{ham-lll}
\end{equation}
where $\delta^{2}({\bf r})$ is the two-dimensional delta function. 

\begin{figure}[h]
\centerline{\resizebox{6cm}{6cm}{\includegraphics{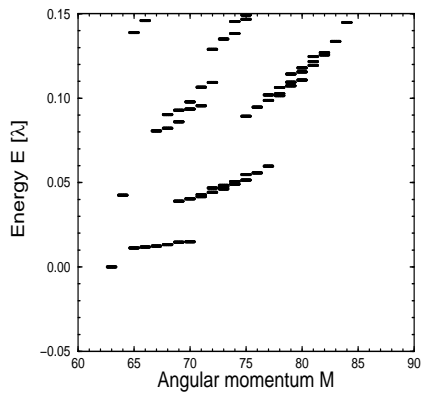}}}
\caption[]{\small
The low-lying energy eigenvalues for the $V_{1}$ model at $\nu=1/3$ and $N=7$. 
The non-degenerate ground state of zero energy at $M_{N=7}(q=3)=63$ can be identified with 
the Laughlin state Eq.~(\ref{Laughlin}). From \cite{KA94}.
}
\label{Laughlin_V1}
\end{figure}

Besides the Coulomb interaction, a whole class of Coulomb like interactions can be constructed 
by truncating the series of pseudopotential coefficients $V_{m}$ 
at relative angular momentum $k$ so that 
the coefficients $V_{0}, V_{1}, \ldots, V_{k}$  are positive, 
but of arbitrary strength. In particular, they can 
agree with the values one gets for the Coulomb interaction, 
while $V_{m} = 0$ if $m > k$. As already mentioned,  
even pseudopotential coefficients do not contribute to the energy expectation 
value, but they become important when the spin is taken into account, see Sect.~\ref{sect:spin}. 
In general, we call this model the {\it hard-core} model or more precisely in dependence on 
the truncation the $V_{k}$ model. 
It is remarkable that the Laughlin wavefunction $\Psi_{1/q}$ is the {\em exact} zero 
energy eigenfunction of the $V_{q-2}$ model ($q$ odd) for any finite $N$ when neglecting the 
neutralizing background, 
{\it i.~e.~}the relation $\langle \Psi_{1/q}|H^{(0)}_{V_{q-2}}|\Psi_{1/q} \rangle =0$ holds.  
In the disk geometry, this eigenstate is unique and 
can be identified in the many-particle spectrum from finite-size diagonalizations, see 
Fig.~\ref{Laughlin_V1} \cite{Rem7}.  
The reason is that only relative angular momentum 
components not smaller than $q$ occur in the Laughlin function and therefore do 
not contribute to the energy as $V_{m}=0$ for $m \ge q$. 
To put it in other words, if the wavefunction 
vanishes at least like $(z_{i}-z_{j})^q$ in each of the relative coordinates of two 
particles, the total energy is zero in the $V_{q-2}$ model. 
Reversely, if a relative angular momentum lower or equal to $(q-2)$ occurs in the wavefunction, 
which is the case at $\nu=1/q$ for all many-particle eigenfunctions except for the 
Laughlin function,  the energy becomes larger than zero due to the positive contributions from 
$V_{1},\ldots, V_{q-2}$. 
Therefore, the Laughlin wavefunction is for a given $q$ among all possible zero-energy 
wavefunctions of the $V_{q-2}$ model that with the smallest 
maximum one-particle angular momentum $m_{max}$. Hence, for $\nu < 1/q$, zero energy 
eigenstates, which are necessarily the ground states, are possible too. 

In particular, if we choose at $\nu=1/3$ as generic model the $V_{1}$ model with 
$V_{1} = \sqrt{\pi}/4$ and 
$V_{m}=0$ for $m \ge 3$, the two-particle interaction is 
\begin{equation}
H_{int}^{(0)}({\bf r}_1 - {\bf r}_2) = 
4 \pi V_{1} \lambda  {\bf \nabla}^{2} \delta^{2}({\bf r}_1 - {\bf r}_2) \;.
\end{equation}
A whole class of eigenfunctions with zero energy at filling factors $\nu \le 1/q$ 
can be constructed by multiplying the Laughlin state with a homogeneous 
symmetric polynomial $S(z_{1},\ldots,z_{N})$ so that the new wavefunctions become  
\begin{equation}
\Psi_{\nu \le 1/q}(z_{1},\ldots,z_{N})  =  
S(z_{1},\ldots,z_{N}) \Psi_{1/q}(z_{1}, \ldots,z_{N}) \;.
\label{symmetric}
\end{equation}
Reversely, any wavefunction 
with filling factor larger than $1/q$ has a positive energy because in any case 
a relative momentum $(q-2)$ occurs. 

In reality, we cannot expect an interaction of such a kind to occur. 
Already in the case of the Coulomb interaction, the Laughlin wavefunction is not 
the exact ground state. This is due to the occurrence of non-zero pair 
amplitudes of relative angular momenta $m \ge 3$ ($m$ -- odd) in the numerically 
determined ground state wavefunction, \cf~\cite{Hald90}. 
Since for the Coulomb interaction all pseudopotential coefficients 
$V_{1}, V_{3}, V_{5}, \ldots > 0$ occur, they contribute to the energy expectation value, 
which is different from the $V_{1}$ model. 
Nevertheless, one can argue that the coefficients $V_{q}$ with 
$q \ge 3$ are weak perturbations of the short-range model introducing additional 
smaller energy scales besides 
the energy scale set by $V_{1} \lambda$. This should not lead 
to significant differences 
between the exact eigenfunction found numerically and the Laughlin function. 
Thus far, this argument was never analytically pursued, but it is strongly corroborated by 
extensive numerical calculations for finite-size systems. To check this, the overlap 
Eq.~(\ref{overlap}) was 
studied in various geometries by many authors. We mention for Coulomb interaction an 
overlap of 
0.979 for $N=4$ in the disk geometry \cite{Lau83b,Lau86}, an overlap of 0.99406 for nine 
particles on the sphere \cite{FOC86}, and of 0.9881 for six particles in a planar 
geometry with periodic boundary conditions \cite{Hald85}.  
 
The robustness of the Laughlin function with respect to changes in the interaction potential 
was shown for some typical interactions. However, there is also strong numerical evidence 
that the Laughlin wavefunction is {\em not always} a good choice for the 
ground state. Such a result is due to Haldane and Rezayi \cite{HR85a} 
who separated the LLL interaction Hamiltonian (\ref{ham-lll}) in the $V_{1}$-part and 
the residual part containing the influence of the pseudopotential coefficients $V_{m>1}$, 
\ie~$H_{int}(\gamma)=H_{V_{1}} + \gamma H_{V_{m>1}}$. 
The numerical study of the low-lying spectrum varying the parameter $\gamma$ in 
$H_{int}(\gamma)$, 
which comprises for $\gamma=1$ the Coulomb interaction and for $\gamma=0$ the $V_{1}$ model,  
revealed different behavior for $\gamma \le 1.25$ and $\gamma > 1.25$. 
In the former case, the ground state has a large overlap with the Laughlin function and 
is gapped, whereas in the latter case, the gap disappears and the Laughlin state 
is only found in the excitation spectrum. 
This observation emphasizes in spite of the artificial character of such interactions  
that ground states of repelling electrons in the LLL 
are not so universal like, for example, the BCS ground state in the theory of 
superconductivity, where already an attractive interaction between electrons leads to a 
new ground state. In summary, the occurrence of the Laughlin state as ground state 
depends on parameters like the filling factor and the pseudopotential 
coefficients characterizing the interaction. These conditions can be satisfied in experiments 
at certain fractional $\nu$.

Now, we turn to some of the remarkable properties of the Laughlin 
ground state, which can be found upon studying the two-point distribution function 
\begin{eqnarray}
n^{(2)}(z, z') & = & \sum_{m_{1},m_{1}',\atop m_{2},m_{2}'=0}
\varphi_{m_{1}}^{*}(z) \varphi_{m_{2}}^{*}(z') \varphi_{m_{2}'}(z') 
\varphi_{m_{1}'}(z) 
\frac{\langle \Psi_{1/q}|c_{m_{1}}^{\dagger} c_{m_{2}}^{\dagger} 
c_{m_{2}'} c_{m_{1}'}|\Psi_{1/q} \rangle} {||\Psi_{1/q}||^2} \nonumber  \\
& = & \lim_{N \to \infty} N(N-1)  \;
 \frac{ \int d^{2} z_{3} \ldots \int d^{2} z_{N} |\Psi_{1/q}(z, z',z_{3}, \ldots, z_{N})|^{2}}
{  \int d^{2} z_{1} \ldots \int d^{2} z_{N} |\Psi_{1/q}(z_{1}, \ldots, z_{N})|^{2} }\ .
\label{two-part-density}
\end{eqnarray}
and the one-particle density 
\begin{eqnarray}
n^{(1)}(z) & = &  \frac{1}{(N-1)} \int dz' n^{(2)}(z,z') = 
\sum_{m_1,m_2 \atop =0} \varphi_{m_1}^{*}(z) \varphi_{m_2}(z) 
\frac{\langle \Psi_{1/q}|c^{\dagger}_{m_{1}} c_{m_2}|\Psi_{1/q} \rangle }{||\Psi_{1/q}||^2}
\nonumber \\
& = & \lim_{N \to \infty} N \; \frac{ \int d^{2}z_{2} \ldots d^{2}z_{N} |\Psi_{1/q}(z,z_{2},\ldots,z_{N})|^2} 
{  \int d^{2} z_{1} \ldots \int d^{2} z_{N} |\Psi_{1/q}(z_{1}, \ldots, z_{N})|^{2}} 
\label{one-part-density}
\end{eqnarray}
for the unnormalized Laughlin wavefunction $\Psi_{1/q}$. 
The two-point distribution function is related to the pair distribution function 
by $g(z,z')=n^{(2)}(z,z')/n^2$. 

At $\nu=1$, the exact two-point correlation function in the thermodynamic limit 
\begin{equation}
n^{(2)}(z,z')=\frac{1}{(2\pi)^2} (1-e^{-|z-z'|^{2}/2}) 
\end{equation}
exhibits translational invariance and isotropy and results in the homogeneous   
one-particle density $n^{(1)}(z)=1/(2 \pi )$.  
For $q>1$, the homogeneity of the Laughlin wavefunction, \ie~$n^{(1)}(z)=\nu/(2 \pi )$, 
can be inferred from a relation of $\Psi_{1/q}$ to a well investigated model of classical 
statistical mechanics. By rewriting the  squared modulus of the Laughlin function 
as the partition function we encounter the Hamilton function of the 
two-dimensional one-component plasma (2DOCP) \cite{Lau83b}. The plasma parameter 
$\Gamma$, which is proportional to the inverse temperature of the classical problem, can be
identified with $\Gamma=2q$ \cite{CLWH82}. As for  
$q \le 70$ the 2DOCOP describes a fluid, while for $q>70$ a crystal occurs, this 
yields a rough estimate for the transition between a fluid-like ground state with constant density 
and the crystalline behavior identified with the occurrence of the expected Wigner crystal. 
However, direct evaluation of (\ref{one-part-density}) for $q=3$ up to 25 particles in the disk geometry 
shows quite large fluctuations around the expected constant density value $1/q$ \cite{MM93}.  
In contrast, the ground state at magic filling factors in the spherical geometry has 
even for finite $N$ a constant density distribution on the surface of the sphere due to its 
identification with a state of zero angular momentum $L=0$ \cite{Hald83}.  

The radial pair distribution function $g(r=|z-z'|)$ for odd $q>1$ was 
analytically calculated by determining the 
coefficients of a power-series in $|z-z'|^{2}$ using charge-neutrality, perfect-screening 
and the compressibility sum rules as constraints \cite{Gir84b}.  
These results agree quite well with data from 
Monte Carlo (MC) calculations for $q=3,5$ up to 144 particles \cite{LWM84,MH86}. 
The radial pair distribution function $g(r)$ shows the 
correlation hole at $r=0$ and tends to one for $r \to \infty$, which is typical 
for a fluid in contrast to the oscillating behavior of $g(r)$ for a crystal exhibiting long-range 
spatial correlations. 

The knowledge of the distribution function and the radial pair-distribution function $g(r)$, respectively,  
allows to calculate the ground state energy per particle $\epsilon(\nu)$ 
in the thermodynamic limit
\begin{eqnarray}
\epsilon (\nu)  =  \frac{\lambda}{2N} \int d^{2}z\, d^{2}z'\, V(z,z')\,(n^{(2)}(z,z')-n^{2}) 
 =  \frac{\lambda \nu}{2} \int_{0}^{\infty}dr\, r\, V(r)\, (g(r)-1) \;. 
\end{eqnarray}
At $\nu=1$, we find for the Coulomb interaction and the $V_{1}$ model, respectively,
\begin{eqnarray} 
 \epsilon_{Coul}(1)/\lambda & = & - \frac{1}{2} \sqrt{\frac{\pi}{2}} \simeq -0.6267    ,   \\
 \epsilon_{V_1}(1)/\lambda & = & 2 V_{1} = \frac{ \sqrt{\pi}}{2}  \ .
\label{exact-polar}
\end{eqnarray}
Similar calculations at $q>1$ show a lower energy of the 
Laughlin wavefunction when compared with 
other trial wavefunctions like, {\it e.~g.~}, charge density waves as long  as $q \le 9$ \cite{LWM84,KJ97}. 
The calculation of the pair-distribution function of the 2DOCP at $\Gamma=6$, \ie~$q=3$, allows 
to determine 
$\epsilon_{Coul}(1/3)/\lambda = -0.4100 \pm 0.0001$ \cite{LWM84}, which is almost 
identical with the value found from an MC evaluation of the Laughlin wavefunction in the 
disk geometry 
$\epsilon_{Coul}(1/3)/\lambda = -0.410 \pm 0.001$ \cite{MH86}.  
Moreover, these results can be compared with finite particle number 
data from exact diagonalization studies at $\nu=1/3$ 
in the torus geometry \cite{YHL83,Yos84b}, with the extrapolation value $-0.409510$  from diagonalizations 
in the disk geometry \cite{KA94} , 
with the value $-0.415 \pm 0.005$ \cite{HR85a,FOC86} in the spherical geometry, and with 
the result $-0.4056$ \cite{CP95} based on the hypernetted-chain method of the 2DOCP.
Similar results can be found for $\nu=1/5$. Although these results show 
the superiority of the Laughlin wavefunction compared with other wavefunctions 
describing a crystalline state, this does not ultimately rule out 
other states with lower energy in the thermodynamic limit. 

Because of the restriction to the spin-polarized LLL, particle-hole 
symmetry holds, which relates in the thermodynamic limit the 
ground state energies per particle at filling $\nu$ and $(1-\nu)$ by the relation 
\begin{equation} 
(1-\nu)\, \epsilon(1-\nu) - \nu\, \epsilon(\nu) = (1-2\nu)\, \epsilon(1) 
\label{particle-hole-polar}
\end{equation}
regardless if a neutralizing background potential occurs.
\begin{figure}[h]
\centerline{\resizebox{6cm}{6cm}{\includegraphics{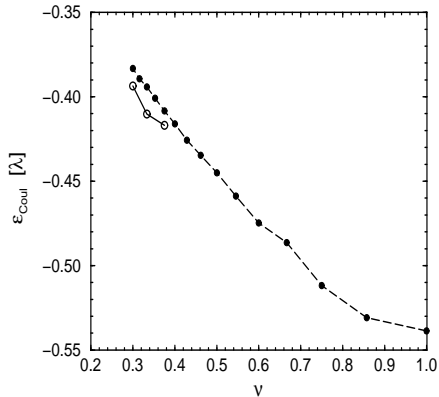}}}
\caption[]{\small
The ground state energy per particle vs.~filling factor for the Coulomb model with 
neutralizing background in the disk geometry using data for six particles. 
Here, the filling factor definition $\nu=N/N_{\Phi}$ is employed.   
The calculated points are shown as filled circles and the long-dashed line serves 
as a guide to the eye. No precursor of a cusp is seen at $\nu=1/3$ for this finite 
particle number. However, 
extrapolations $N \to \infty$ for $1/3$ and two adjacent values might indicate 
the expected cusp when the extrapolated data (open circles) are connected by a solid line.
Note the large finite-size corrections of $\sim 13$\% at $\nu=1$, where 
$\epsilon_{Coul}(1)=-0.6267 \lambda$.
}
\label{ground-coulomb}
\end{figure}

	\subsection{Elementary excitations: quasiholes and quasielectrons as quasiparticles and 
		    the incompressibility of the Laughlin state}  
	\label{subsect:quasiparticles}

\begin{figure}[h]
\centerline{\resizebox{6cm}{6cm}{\includegraphics{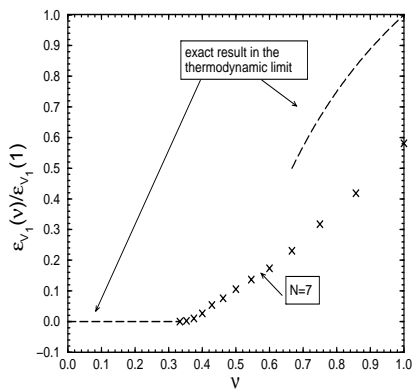}}}
\caption[]{\small
The ground state energy per particle $\epsilon_{V_1}$ 
vs.~filling factor $\nu$ for the $V_{1}$ model without  
background. For $0 \le \nu  \le 1/3$, the energy equals zero (dashed curve). 
At $2/3 \le \nu \le 1$, we find $\epsilon_{V_1}(\nu)=(2-1/\nu) \epsilon_{V_1}(1)$ 
(dashed curve) from (\ref{particle-hole-polar}), where $\epsilon_{V_1}(1)=2 V_{1} \lambda$ 
is the ground state 
energy for the filled LLL. In the range $1/3 \le \nu 
\le 1$, the crosses are the result of finite-size diagonalizations 
for seven particles on a disk indicating strong finite-size corrections when compared  
with the exact results in the range $2/3 \le \nu \le1$. In order to improve the 
results, extrapolations have to be performed for each $\nu$.
The self-consistent Hartree-Fock approximation, see Sect.~\ref{sect:spin}, yields a linear 
dependence $\epsilon_{V_1}(\nu)=2 \nu V_1 \lambda$ ($0 \le \nu \le 1$) and shows 
its limited worth except at $\nu=1$.
Note that the quasielectron energy cannot be estimated from the 
slope at $\nu=1/3$ in contrast to equivalent data from the spherical geometry.  
The energy is given in units of $\epsilon_{V_1}(1)=2 V_{1} \lambda$.
}
\label{ground-hard-core}
\end{figure}

The ground state properties of the Laughlin state are remarkable on its own. 
However, in order to explain the unique transport properties an additional ingredient 
is necessary, \viz~the occurrence of gapped neutral particle-hole excitations 
above the ground state \cite{Lau83b,Lau86,Halp84}. This is equivalent to a discontinuity 
in $\partial \epsilon/\partial \nu$ 
at $\nu=1/q$ creating downward cusps in the $\epsilon(\nu)$ dependence or to 
the incompressibility of the Laughlin state as we will see below.
Cusps of $\epsilon(\nu)$ at, \eg~$\nu=1/3$, as they can be seen in 
finite size diagonalizations, \cf~Figs.~\ref{ground-coulomb} and \ref{ground-hard-core}, 
are of limited value and do not prove the existence of a gap above the many-particle 
ground state. In order to show this, we employ the quasiparticle construction by taking properly 
the thermodynamic limit into account. 
The quasiparticles (QPs) at $\nu=1/q$ are related to the ground state on the left and 
right side, respectively, infinitesimally close to $1/q$, and they can be viewed 
as the {\em elementary charged excitations} of a quantum Hall system \cite{Lau83b}. 
There are three ways to create such QPs by slightly changing 
the filling factor $\nu=(N-1)/(B A/\Phi_{0}-1) = 1/q$ to $1/q \pm \delta \nu$ 
by changing either the area $A$ ({\em neutral} QPs), the particle number $N$ ({\em gross} QPs) 
or the magnetic field $B$ ({\em proper} QPs). 

Within the construction of {\em neutral} QPs, the 
system area is either increased or decreased by an amount of 
$2 \pi \ell_c^2$ altering the number of elementary flux quanta $N_{\Phi}=BA/\Phi_{0}$ 
by one elementary flux quantum $\Phi_{0}=h/e$ while keeping $N$ and $B$ fixed. 
The new ground state when compared with the homogeneous Laughlin state exhibits 
either a localized charge deficiency (one flux quantum added) 
or a localized accumulation of charge 
(one flux quantum removed) of total charge $|e|/q$ whose location is parameterized 
by the coordinate $\xi$. 
The former QP is called the neutral quasihole (QH, index $-$) and the 
latter the neutral quasielectron (QE, index $+$).\footnote{Some authors denote the quasielectron as quasiparticle, but we use 
the term quasiparticle only in the comprehensive way.} 
The total energy difference between the new ground state and the ground state at 
the QHE filling factor defines the neutral quasihole energy $\epsilon_{-}^{n}(\nu)$ 
and the quasielectron energy $\epsilon_{+}^{n}(\nu)$, respectively.  
These energies are given in a finite system and in the thermodynamic limit, respectively, by
\begin{eqnarray}
\epsilon_{\mp}^{n}(N,N_{\Phi}) = E_{0}(N,N_{\Phi} \pm 1) - E_{0}(N,N_{\Phi})  \nonumber \\
\epsilon_{\mp}^{n}(\nu) = 
 \mp \nu^2 \left( \frac{\partial \epsilon}{\partial \nu} \right)_{\nu^{\mp}} \ .
\label{qp-energies}
\end{eqnarray} 
There are two other ways to create QPs, \viz~{\em gross} QPs with energy 
$\epsilon_{\pm}(\nu)$ and 
{\em proper} QPs with energy $\tilde \epsilon_{\pm}(\nu)$. In the first case, the number of particles 
is changed by one and leads to the formation of $q$ independent gross QPs in order to 
conserve the total charge, while 
the latter case is based on the change of the magnetic field and therefore 
altering $N_{\Phi}$ by $\pm1$, but keeping $N$ and $A$ fixed. 
In each of the three constructions, the relative charge is increased or 
decreased \cite{Lau83b,Halp83,MH86, MG86b}.  
One should note that these QPs are different from quasiparticles used in Landau's 
theory of Fermi liquids,  
where they are introduced to describe an interacting fermionic system approximately in terms of 
non-interacting new particles, whose life-time diverges like $T^{-2}$ when approaching the 
Fermi surface \cite{LL9}. 

One can easily show that in the thermodynamic limit the various energies are related by 
the ground state energy per particle $\epsilon(\nu)$, 
\cite{MH86, MG86b}  
\begin{equation} 
\epsilon_{\pm}^{n}(\nu)= \epsilon_{\pm}(\nu) \mp \nu \epsilon(\nu) 
= \tilde{\epsilon}_{\pm}(\nu) \pm \frac{1}{2} \nu \epsilon(\nu) \ .
\label{qp-relation}
\end{equation}
Furthermore, in case that the single QPs are identified in a QHE system, we can 
construct neutral excitations {\em above} the QHE ground state consisting of 
a QE and a QH far apart and hence non-interacting. The energy of such a particle-hole pair 
is given by the sum of their energies regardless of the kind of construction 
as Eq.~(\ref{qp-relation}) shows 
\begin{equation}
\Delta_{qe-qh}(\nu)= \epsilon_{+}^{n}(\nu) + \epsilon_{-}^{n}(\nu)
= \nu^{2} \left( \left( \frac{\partial \epsilon}{\partial \nu} \right)_{\nu^{+}}
- \left( \frac{\partial \epsilon}{\partial \nu} \right)_{\nu^{-}} \right) \ .
\label{eh-gap}
\end{equation}
Note that the positive definiteness of the gap implies downward cusps of the $\epsilon(\nu)$ 
curve as suggested in Fig.~\ref{ground-coulomb} because of 
$\partial \epsilon/\partial \nu <0$. 
An instructive example of gross QPs can be found for {\em non-interacting} electrons at 
$\nu=1$, see Fig.~\ref{energy_levels}. The addition of one electron leads to a ground state 
with one electron promoted to the minority spin $n=0$ Landau level, \ie~the gross quasielectron 
energy is $\epsilon_{+}(1)= \hbar \omega_{c}/2 + \Delta_z$. 
On the other hand, the removal of one electron from the ground 
state in Fig.~\ref{energy_levels} causes only a change of the kinetic 
energy so that the gross quasihole energy is $\epsilon_{-}(1)=-\hbar \omega_{c}/2$. 
Therefore, the quasielectron-quasihole gap equals 
$\Delta_{qe-qh}(\nu=1)=\Delta_z$, which is indeed the energy of the lowest-lying 
particle-hole excitation. 
Note that the non-interacting system is for non-integer $\nu$ compressible 
since $\Delta_{qe-qh}(\nu < 1)=0$. 
But this is altered at FQHE filling factors when interaction is taken into account.

\begin{figure}[h]
\centerline{\resizebox{6cm}{6cm}{\includegraphics{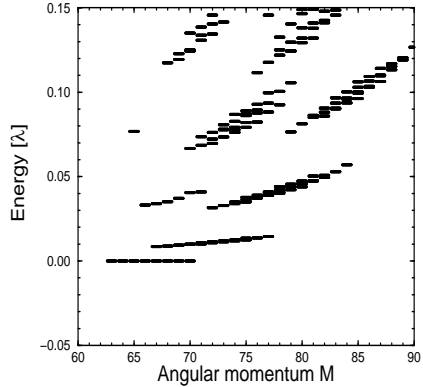}}}
\caption[]{\small
The low-lying excitations in the $V_{1}$ model showing the $\nu=1/3$ QHs for a 
seven particle system. Besides the Laughlin state at $M_{N=7}(q=3)=63$, each of 
the zero energy eigenstates is non-degenerate and is described by the states of 
Eq.~(\ref{qh-laughlin}) with fixed total angular momentum $M$ ranging from 64 to 70.
From \cite{KA94}.
}
\label{spectrum_qh}
\end{figure}

To be definite, let us first turn to the quasiholes of the Laughlin state illustrated 
by numerical results at $\nu=1/3$. 
In order to find an approximate analytical expression for the ground state of the QH  
wavefunction, Laughlin gave a prescription for the construction of such a state \cite{Lau83b}.  
The construction resembles the way a proper QH is created as we increase the magnetic field in 
average by one flux quantum.
Let us assume a one-particle energy eigenstate $z^{m}e^{-|z|^2/4}$. 
When a magnetic solenoid with flux $\Phi_{0}$ and magnetic field 
parallel to the external magnetic field is added to the one-particle Hamiltonian, a new eigenstate 
$z^{m+1}e^{-|z|^2/4}$ of same energy can be found after gauging away the vector potential outside 
the origin.
If such an operation is applied to the non-degenerate Laughlin ground state in symmetric 
gauge $\Psi_{1/q}$ by adiabatically 
switching on a solenoid at the origin, the Laughlin state will evolve approximately in 
a new many-particle state $\Psi_{1/q}^{(-)}$, where the angular momentum of each particle is 
increased by one. 
The resulting many-particle wavefunction 
\begin{equation}
\Psi_{1/q,\xi=0}^{(-)}(z_{1},\ldots,z_{N})=
\prod_{i=1}^{N} z_{i} \Psi_{1/q}(z_{1},\ldots,z_{N}) 
\label{quasihole-origin}
\end{equation}
is particularly efficient in keeping away the electrons from the origin and describes a quasihole at 
this location. 
Comparison of (\ref{quasihole-origin}) with Eq.~(\ref{symmetric}) shows that 
this quasihole is an exact eigenstate of zero energy of the $V_{q-2}$ model due to 
the symmetric prefactor $z_{1} \cdot \ldots \cdot z_{N}$. Moreover, there exist in total 
$N$ of such symmetric polynomials that increase within the neutral 
QH construction the maximum angular momentum by one, \ie~$\nu=(N-1)/N_{\Phi} < 1/q$,  
and which are zero energy eigenstates of the $V_{q-2}$ model. This $N$-fold degeneracy of 
QHs with total angular momentum reaching from $M_{N}(q)+1 \equiv M^{*}+1$ to 
$M_{N}(q)+N=M^{*}+N$ and the Laughlin state itself, 
see Fig.~\ref{spectrum_qh}, can be exploited to construct a QH located at $\xi$ 
by superposing these $N+1$ states with weights $\xi^{m}$ ($0 \le m \le N$) 
\begin{eqnarray}
\Psi_{1/q,\xi}^{(-)}(z_{1},\ldots,z_{N})  
=  \prod_{i=1}^{N}(z_{i}-\xi) \prod_{k > j=1}^{N}(z_{k} - z_{j})^{q}e^{-\frac{1}{4}\sum_{l=1}^{N}|z_{l}|^{2}} 
\; .
\label{qh-laughlin}
\end{eqnarray}
An alternate way to derive the QH wavefunction at $\xi$ is the application of the 
magnetic translation operator (\ref{translation}) to the QH located 
at the origin (\ref{quasihole-origin}). 
This leads to a shift of the QH at $\xi=0$ to the position $\xi$ in front of the homogeneous 
background 
\begin{eqnarray}
\prod_{i=1}^{N} T_{i,\xi}\,\Psi_{1/q,\xi=0}^{(-)}(z_{1},\ldots,z_{N})  =  
\Psi_{1/q,\xi}^{(-)}(z_{1},\ldots,z_{N}) e^{-N |\xi|^2/4} \prod_{i=1}^{N} e^{z_{i}\xi^{*}/2} \ .
\label{qh-translation}
\end{eqnarray}
Laughlin's QH wavefunction can be recovered by neglecting the last exponential, which 
means projection to the correct filling factor \cite{KA94}. 
The charge deficiency is localized around $\xi$ within the magnetic length scale $\ell_c$, 
what can be analytically checked at $\nu=1$ since 
\begin{equation}
n^{(1)}(z) = \frac{1}{2\pi}(1-e^{-|z-\xi|^{2}/2}) \ . 
\label{quasihole-density}
\end{equation}
In total, it amounts for one charge, while far away from $\xi$, the constant 
density of the Laughlin state remains unchanged. 
For $\nu=1/q$, $q>1$, the one-particle density is given by the expression 
(\ref{quasihole-density}) multiplied by $\nu$ with a total charge diminished by $1/q$ 
when comparing with the Laughlin state. 
The correctness of Laughlin's QH function for odd $q$ and Coulomb interaction 
is again corroborated by generalizing 
the plasma analogy mentioned above. Now, additionally to the 2OCP Hamiltonian function 
a term shows up, which indicates an interaction between the plasma charges 
$q$ and a phantom charge of strength $+1$. Similar conclusions can be drawn from 
the overlap of the trial wavefunction 
with results from exact diagonalizations and from MC calculations. Exact diagonalizations  
are particularly successful in the spherical geometry because the 
QH eigenstates can be identified as degenerate states with angular momentum $L=N/2$  
due to their $(N+1)$-fold degeneracy \cite{Hald83}.  

Another trial QH state formulated in second quantization takes the Laughlin construction 
literally \cite{MG86b}. Then, the operators 
$\hat{U}_{m}=c^{\dagger}_{m+1} c_{m} + 1 - n_{m}$ are successively applied 
to the second quantized Laughlin state $|\Psi_{1/q} \rangle$ at $1/q$ with 
$n_{m}=c^{\dagger}_{m} c_{m}$ as angular momentum distribution operator. It 
shifts each occupied one-particle angular momentum state  $|m \rangle  \to |m+1 \rangle $. 
This corresponds to a QH at the origin and reads in second quantization
\begin{equation}
|\phi_{1/q}^{(-)} \rangle =  \hat{U}_{0} \hat{U}_{1} \ldots \hat{U}_{\infty} |\Psi_{1/q} \rangle \ .
\label{qh-girvin}
\end{equation}
It contains the same Slater determinants as the Laughlin QH (\ref{qh-laughlin}), 
but this QH wavefunction differs in its weights of the Slater determinants. 
Despite the disadvantage that this QH state is not an exact eigenstate 
of the hard-core model, it is analytically more accessible and 
its energy expectation value for a more realistic interaction 
needs not to be energetically higher than that of Laughlin's QH state (\ref{qh-laughlin}).  
This is indicated by Tab.~\ref{tab:energies-nu-1/3}, where for 
Coulomb interaction and $V_{1}$ model 
QH energies and QE energies are listed at filling factor $1/3$.

When turning to the QEs, it seems at first glance that 
the concept of QH and QE creation is symmetric \wrt~the stable Laughlin 
state at $\nu=1/q$. However, there exist differences between these two types of QPs, which  
become already apparent when looking for the QE wavefunction in the $V_{q-2}$ model. 
Contrary to the QH, no exact solution for arbitrary particle number is known. Moreover, 
no wavefunction with $m_{max}=q(N-1)-1$ can prevent the occurrence of relative 
angular momentum $(q-2)$, \ie~$\epsilon^{n}_{+}(1/q)>0$ for any finite particle number.
Hence, quite a lot of trial wavefunctions was proposed for the description of the QE 
emphasizing various aspects of the problem. 

Laughlin proposed a QE wavefunction centered at $\xi$\ \cite{Lau83b}, which is given by 
\begin{equation}
\Psi ^{(+)}_{1/q,\xi}=\prod_{i=1}^{N} 
e^{- \frac{1}{4}|z_{i}|^{2}}(2\partial_{z_{i}}-\xi^{*})
\prod_{k>l=1}^{N}(z_{k}-z_{l})^{q}\:.
\label{qe-laughlin}
\end{equation}
The derivative operator is the adjoint of the product operator in (\ref{qh-laughlin}).

MacDonald and Girvin constructed a QE at the origin in analogy to the 
QH construction (\ref{qh-girvin}) 
\cite{MG86b} 
\begin{equation}
|\phi_{1/q}^{(+)}> = \hat{d}_{\infty} \ldots \hat{d_{1}}
(1-\hat{n}_{0})|\Psi _{1/q}> \ .
\label{qe-girvin}
\end{equation}
Here, the operator $\hat{d}_{m}=c_{m-1}^{\dagger}c_{m} +1 -n_{m} $ decreases all angular momenta 
by one except the one for $m=0$, which has to be projected out. 

A third trial wavefunction is due to Jain \cite{Jai89b} and is related to the composite 
fermion (CF) theory, which is discussed in Subsect.~\ref{subsect:composite}. Here, each of the total 
angular momentum components $M^{*}-N+m+1$ ($m=-1,0,\ldots,N-2$) is given by 
\begin{equation}
\chi_{1/q,M^{*}-N+m+1,\xi=0}^{(+)}={\cal P}_{0}\prod_{i>j=1}^{N}
(z_{i}-z_{j})^{q-1}
\left| \begin{array}{ccc}
z_{1}^{m}|z_{1}|^{2} & \ldots &
z_{N}^{m}|z_{N}|^{2} \\
1 & \ldots & 1 \\
z_{1}  & \ldots & z_{N} \\
\vdots & \ddots & \vdots \\
z_{1}^{N-2} & \ldots & z_{N}^{N-2}
\label{qe-jain}
\end{array} \right|
e^{-\frac{1}{4} \sum_{i=1}^{N}|z_{i}|^{2}}  
\end{equation}
where initially a wavefunction is constructed that contains one state from the LL $n=1$ before 
the LLL projection operator ${\cal P}_{0}$ is applied \cite{Rem8}.  
 
A fourth QE wavefunction, which transfers the idea of the magnetic translation 
(\ref{qh-translation}) to the 
QE at the origin, is the disk equivalent \cite{KA93} of a function 
originally proposed by Haldane for the spherical geometry \cite{Hald83}. 
It describes a QE centered at $\xi$
\begin{equation}
\Phi_{1/q,\xi}^{(+)}
=\prod_{i=1}^{N} ((N-1)q-z_{i}\partial_{z_{i}}+\xi \partial_{z_{i}}) \Psi_{1/q} \ .
\label{qe-kasner}
\end{equation}
For completeness, we list a proposal due to Morf and Halperin for a QE at the origin 
motivated by the plasma analogy \cite{MH86}
\begin{equation}
\tilde {\Psi}_{1/q,\xi=0}^{(+)} = {\cal A} \left( \left[ \frac{1}{z_{1}-z_{2}} \right]^2 
\prod_{j=3}^{N} \frac{z_{j}-(z_{1}+z_{2})/2}{(z_{1}-z_{j})(z_{2}-z_{j})} \Psi_{1/q} \right) \ ,
\label{qe-morf}
\end{equation}
where ${\cal A}$ denotes the antisymmetrization operator. 
\begin{figure}[h]
\centerline{\resizebox{6cm}{6cm}{\includegraphics{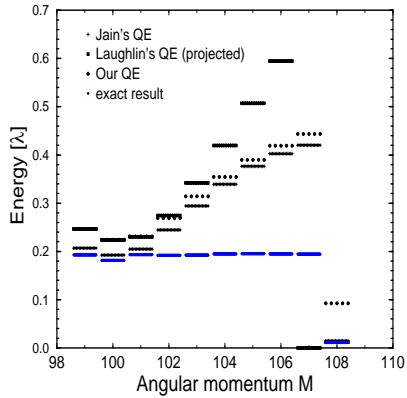}}}
\caption[]{\small
Numerically calculated expectation values for Laughlin's (projected to the correct 
$\nu$), our, and Jain's 
quasielectron trial wave functions arranged with increasing $M$ in comparison 
with exact energies for $N=9$ particles in the hard-core model. Low lying states 
with total angular momentum $M \ge M^{*}$ are attributed to edge excitations.
From \cite{KA94,KA93}.
}
\label{quel_N=9}
\end{figure}
All wavefunctions describe 
a local accumulation of charge $|e|/q$ around the position $\xi$, although 
differences in charge distribution in the core region of the QE exist \cite{HR85a,MG86b}.  
A formal proof for the existence of fractional charges in the QH and QE wavefunctions 
can be performed via a Berry phase \cite{Ber84} calculation based 
on the explicit form of the QP trial wavefunctions. 
Then, the QP charge $e^{*}=\pm |e|/q$ can be inferred from the phase the QP gains when adiabatically 
moving around a closed loop \cite{ASW84}.   

For small particle numbers, exact diagonalization data can be used to check the 
quality of the trial wavefunctions. Moreover, trustworthy trial wavefunctions 
can be evaluated by MC calculations for larger $N$ not accessible to diagonalization 
studies \cite{MH86,MH87,JK97}. 
Fig.~\ref{quel_N=9} shows for nine particles such a comparison assuming 
a $V_{1}$ interaction in the disk geometry. 
While the approximate energetic degeneracy of the numerically calculated low-lying 
states allow their identification with QE states of angular momentum $M$ in the range 
$M^{*}-N$ to $M^{*}-1$, the quality 
of Laughlin's QE wavefunction (\ref{qe-laughlin}), Jain's QE (\ref{qe-jain}) and our QE 
proposal (\ref{qe-kasner}) become worse when moving from the origin to the edge of the system.
The calculations show that among the proposed QE wavefunctions after extrapolation 
to the thermodynamic limit Jain's QE wavefunctions  
come closest to the finite-size diagonalization results found in the disk geometry \cite{KA93}.  
A similar outcome was reported for the spherical geometry, where Jain's QE is energetically 
superior to Haldane's QE wavefunction \cite{GH94}.  
\begin{table}[h]
{\small 
\begin{tabular}{lrrrrrr} 
\\ \hline 
Model, geometry & $\epsilon$ &  $\epsilon_{-}^{n}$ & $\epsilon_{+}^{n}$ & $\epsilon_{-}$ & $\epsilon_{+}$ & \hspace*{0mm}$\Delta_{qe-qh}$  \\
method & & & & & & \hspace*{0mm}
\\ \hline\hline \\[-3mm]
Coul., sphere, diag. \cite{FOC86} &$-0.410$ &0.0947& $0.0089$ & $0.2314$	&$-0.1278$ & $0.1036$ 
\\ 
 &  &  &  & 	& &     \\ \hline  

Coul., disk, analyt. \cite{MG86b} &-0.4098 &0.0959  &$0.017$  &0.2337 &$-0.120 $& $0.114$  \\ 
Eqs.~(\ref{qh-girvin}), (\ref{qe-girvin})& & &$ \pm 0.002$  & 
& $\pm 0.002 $& $\pm 0.002$   	 \\ \hline

Coul., disk, MC \cite{MH86}&$-0.410$ &0.0943  &$0.0046 $  &0.231 	&$-0.132 $& $0.099$  \\ 
Eqs.~(\ref{qh-laughlin}), (\ref{qe-laughlin}) &$ \pm 0.001$ &  & & & & $ \pm 0.009$  \\ \hline

Coul., sphere, VMC \cite{Bon95} & &  &   & & & $0.106(3) $ 	 \\ 
Eqs.~(\ref{qh-laughlin}), (\ref{qe-jain})& &  &  & 	& &     \\ \hline \hline 

$V_{1}$ model, disk, & 0 	 & 0  & 0.1865  & 0   &  0.1865  &0.1865  \\ 
diag. \cite{KA94} &   &  &   &   &    &  \\ \hline

$V_{1}$ model, sphere, & 0 & 0  & 0.1905  & 0  &  0.1905  & 0.1905  \\ 
diag. \cite{GM90} &  &   &   &   &  & \\ \hline
\end{tabular}
}
\caption[]{ 
The ground state energy per particle $\epsilon$, the neutral 
quasihole and quasielectron energies $\epsilon_{-}^{n}$ and $\epsilon_{+}^{n}$, 
the gross quasihole and quasielectron energies $\epsilon_{-}$ and $\epsilon_{+}$ 
and the quasielectron-quasihole gap $\Delta_{qe-qh}$ in units of $\lambda$ at $\nu=1/3$.  
The results encompass data from exact diagonalizations as well as from energy expectation 
values of 
trial wavefunctions quoted by their numbers in the text (QH wavefunction first) 
for two models of spin-polarized electrons in the LLL. Moreover, the geometry of the system 
used and the method applied are listed. 
In one case \cite{Bon95}, only the gap energy $\Delta_{qe-qh}$ is given, since the energy of 
an excited state with a QE and QH far apart was calculated. 
}
\label{tab:energies-nu-1/3}
\end{table}
Eventually, it is worth to note that for electrons on a disk with Coulomb interaction 
the QE degeneracy could not be observed up to nine electrons \cite{KA94} probably due 
to strong finite-size effects, whereas on the sphere the QEs appear as ground states with 
angular momentum $L=N/2$ like the QHs and hence exhibiting an $(N+1)$-fold 
degeneracy \cite{Hald83}.  
The essential quantitative result is that the quasiparticle gap $\Delta_{qe-qh}$ 
at $\nu=1/3$ has a value of approximately $0.1 \lambda \simeq 5 \sqrt{B[T]}\ K$ 
for Coulomb interaction. This is considerably smaller than the quasiparticle gap 
$\hbar \omega_c$ for non-interacting electrons.
Similar calculations yielding smaller quasiparticle gaps were performed at $\nu=1/5, 1/7$, 
which are in accordance with the observation that the FQHE at $1/3$ is the strongest one.
These theoretical results for $\Delta_{qe-qh}$ allow also comparison with the experimentally 
determined activation gap $\Delta(\nu)$. It is found from 
measurements of the temperature dependent longitudinal resistivity $\rho_{xx}$ 
at sufficiently low temperatures $k_{B} T \ll \Delta $ in the plateau region of a 
FQHE filling factor using high-mobility samples \cite{WST88}. 
The activated behavior of the longitudinal resistivity is governed by the relation 
\begin{equation}
\rho_{xx}(T)=\rho_{xx}^{0} e^{-\Delta/2k_{B}T} \ ,
\label{activation}
\end{equation}
where $\rho_{xx}^{0}$ is a constant, which is thought to be temperature 
independent. Assuming a spin-polarized ground state and non-spin-reversed QPs, 
which is the case at 
high enough magnetic fields, and which is supported by a non-zero Zeeman term, 
we identify $\Delta$ with $\Delta_{qe-qh}$ and hence 
a dependence $\Delta_{qe-qh}(1/3) \simeq 0.1 \lambda \propto \sqrt{B}$ is expected to occur.
In experiment, this can roughly be observed, however, the gap value is at least more than 50 \%  
smaller than the calculated ideal value at $10\ T$ \cite{WST88}.  
Effects reducing the activation gap 
are the finite extension of the wavefunction in the $z$-direction, disorder and Landau level mixing as 
we will discuss later in more detail.

Before closing, let us mention two remarkable properties of the QPs. 
First, the QPs are anyons, \ie~they obey fractional particle statistics \cite{Wil90}.  
If in three dimensions two quantum-mechanical particles are interchanged, \ie~one 
particle arounds the other one in a half circle,  the two-particle wavefunction 
acquires a prefactor of $e^{i \pi}=-1$ and $e^{i 2 \pi}=+1$ for fermions and bosons, 
respectively. However, in two dimensions this phase can have continuous values 
$\varphi$ with $0 < \varphi \le 2 \pi$ for topological reasons. 
To evaluate the phase for the case of two QHs, one employs the two-quasihole wavefunction 
$\prod_{i} (z_{i}-\xi_{1})(z_{i}-\xi_{2})\Psi_{1/q}$ and calculates the extra 
Berry phase \cite{Ber84} due to the adiabatic movement of the QP located at $\xi_{1}$ 
around the other one fixed at $\xi_{2}$. 
This results for Laughlin's QP wavefunctions (\ref{qh-laughlin}), (\ref{qe-laughlin}) 
in a phase factor for interchange of $\pi \nu=\pi/q$ \cite{Halp84,ASW84} indicating 
that the QPs are $\pi/q$ anyons. Note that the QHs of the $\nu=1$ state are fermions. 
Unfortunately, this property has no direct experimental consequences. 

The second surprising property is the already mentioned fractional charge $\pm e/q$ of the $1/q$ QPs 
whose experimental detection caused some debate in conjunction with 
Laughlin's gauge argument \cite{Lau81,CHS89,KP89}. 
The most direct evidence for this fractional charge was found recently in two different experiments. 
In the first one, resonant tunneling through an anti-dot was performed, where the 
observed periodic conductance peaks in dependence on magnetic field and gate voltage allow to 
determine the charge of the quasiparticles of the $1/3$ state \cite{GS95}.  
In the second experiment, a {\em non-equilibrium} property was exploited, 
\viz~the shot noise of the backscattered tunneling current 
through a constriction between two  $\nu=1/3$ quantum liquids. The shot noise is proportional 
to the charge of the carriers and experiments showed indeed an effective charge of $e/3$ 
\cite{SGJE97,PRH97}. In order to exclude the possibility that such an experiment measures 
rather the conductivity but not the QP charge, a similar experiment was performed at $\nu=2/5$. 
Here, a QP charge of $|e|/5$ was found \cite{RPT99}, which is 
in accord with the prediction that QPs at $\nu=p/q$ carry charge $\pm e/q$. 
 
Now let us turn to the incompressibility of the Laughlin state. 
The isothermal compressibility can be either microscopically calculated as the long-wavelength limit 
of the density-density correlation function or from its thermodynamic definition 
\begin{equation}
\kappa_{T} = - \frac{1}{V}  \left( \frac{\partial V}{\partial p} \right)_{T,N} \ .
\end{equation}
At zero temperature, this can be related to the ground state energy per particle $\epsilon(\nu)$ via
\begin{equation}
\kappa_{T}^{-1} = \frac{\nu}{2\pi \ell_c^2}\  
\frac{\partial}{\partial \nu} \left(\nu^{2} \frac{\partial \epsilon}{\partial \nu} \right)\ .
\label{inv-comp2}
\end{equation}
Incompressibility is therefore achieved by a discontinuity 
of the first derivative of the ground state energy at a certain filling factor, 
\cf~(\ref{inv-comp2}), which corresponds to the expected 
cusp in an $\epsilon(\nu)$ representation or by a jump of the chemical 
potential $\mu$, which differs in its left and right values 
$\mu_{\pm}=\left( \partial E/\partial N \right)_{\pm} 
= \left( \partial (\nu \epsilon(\nu)) / \partial \nu \right)_{\pm}$. 
In a finite particle system, this jump in the chemical potential 
$\Delta\mu=\mu_{+}-\mu_{-}$ is  
\begin{equation}
\Delta\mu(N) = E(N+1) + E(N-1) -2E(N) \; ,
\label{delta-mu}
\end{equation}
however, the thermodynamic limit is decisive as our discussion of the cusp in 
Fig.~\ref{ground-coulomb} in a finite particle system showed.
From the comparison with (\ref{eh-gap}), it follows that the jump of the 
chemical potential is proportional to the quasiparticle gap  
\begin{equation} 
\Delta\mu(\nu) = \frac{1}{\nu} \Delta_{qe-qh} \ .
\end{equation} 
Therefore, the existence of a positive quasiparticle gap $\Delta_{qe-qh}$ 
means incompressibility. 

In contrast to the compressible gas of non-interacting electrons {\em without} magnetic field,  
the situation changes for non-interacting electrons in a constant magnetic field 
when a Landau level is filled. For arbitrary $\nu$, the 
ground state energy per particle in units of $\hbar \omega_{c}$ depends only on $\nu$,  
even for a finite particle number \cite{symbol} because of  
\begin{equation}
\epsilon(\nu)= \frac{1}{\nu} \left\{ [\nu]\;(\nu-\frac{[\nu]}{2}) + \frac{1}{2}(\nu - [\nu])\right\} \ .
\end{equation}
At integer filling $\nu=n$, the jump of the chemical potential, which is located in 
the charge gap in the middle of the two adjacent LLs, becomes 
$\Delta \mu =1/n$, and therefore it is $\Delta_{qe-qh}=1$. 
Thus the system is incompressible for each integer $\nu$,   
while it is compressible for any non-integer $\nu$. 
Note that the quasiparticle gap is independent of the Landau level {\em and} particle 
number and can also be found from the finite $N$ expression (\ref{delta-mu}).  
Of course, the reason for the incompressibility is the existence of discrete Landau 
levels, and this holds also when the electron's spin is taken into account.
Now let us return to the case of interacting electrons, where we expect also for FQHE $\nu$ 
incompressibility. We showed in the last Subsection that this is the case at 
$\nu=1/q$ by extrapolating diagonalization data 
and trial wavefunction results to the thermodynamic limit. 
In the $V_{1}$ model at $\nu=1/3$, such a quasiparticle gap develops only by the 
non-vanishing contribution from the quasielectron energy. This is illustrated 
in Fig.~\ref{ground-hard-core}.  
On the other hand, this intuitive argument for a finite particle system cannot replace a 
formal proof since the probability 
amplitude of this contribution remains unknown in the thermodynamic limit. 
Up to now, a rigorous analytical proof hasn't been found yet \cite{Mac95,Hald90}.  
When comparing this situation with that for non-interacting electrons, we realize that we know 
neither the exact QE wavefunction nor a lower bound of its energy.
There are other attempts to show the incompressibility at $\nu=1/3$ in the $V_{1}$ model. 
They are based on a diagrammatic high-temperature 
expansion of the grand potential \cite{ZM94}, a direct high-temperature 
expansion of the energy \cite{SA2000}, and a virial expansion \cite{TM97}. 

	\subsection{The microscopic understanding of filling factors $p/q$ }
	\label{subsect:composite}

Besides the strong FQHE at $\nu=1/q$ for $q=3,5$ and their particle-hole analogs at $\nu=1-1/q$ 
in the LLL approximation 
there exist much more $\nu$, where plateaus in $\rho_{xy}$ emerge, \cf~Fig.~\ref{qhe-experiment}. 
The most prominent ones are of the form $\nu=p/(2p + 1) < 1/2$ and the corresponding 
particle-hole symmetric states 
$\nu=1 - p/(2p + 1)= (p+1)/(2p+1)> 1/2$ for $p=2,3,4$. 
So far, we outlined the main features of the Laughlin ground state strongly 
corroborating the existence of an incompressible ground state at $\nu=1/q$. However, 
this does not explain the existence of ground states with similar properties at 
$\nu=p/q$ for $p>1$ like the 2/5-state. 

The hierarchical theory \cite{Halp83,Hald83,Hald90,MH86,Halp84} tries to apply  
the stability condition at $\nu=1/q$ (\ref{finite-nu-q}) between 
the number of degenerate electrons $N$ and the degree of degeneracy $N_{\Phi}$ to the equally 
degenerate boson-like quasiparticles \cite{Hald90} of the parent state at $1/q$. If the quasiparticle 
filling fraction attains the value $1/p$ ($p$ - even integer),  
the hierarchical theory predicts two new second level hierarchy FQHE filling factors 
for the daughter states, either at $\nu=p/(qp+1)=1/(q + 1/p)$ in case of 
quasiholes or at $\nu=p/(qp-1) = 1/(q - 1/p)$ in the case of quasielectrons. 
In a short notation these finite chain fractions  
correspond to a $[q,+p]$ and a $[q,-p]$ state, respectively. 
The denominators remain 
odd since $p$ is even. Within this theory, \eg, the $2/7$ state is the daughter state of the 
quasiholes of the $1/3$ state, and the $2/5$ state is created by the condensation of 
the $1/3$ quasielectrons if $p=2$.

In principle, this mechanism can be applied again and again by constructing 
new states from the quasiparticles of one hierarchy to get a stable state of the 
next hierarchy level unless the new incompressible state becomes unstable against the formation 
of a different kind of state as the Wigner crystal in the low filling factor region.
In any case, the states in the hierarchical theory have filling factors with odd denominators. 

Clearly, there are some shortcomings in this theory \cite{Jai90b}.  
First, the starting point of the construction are states at $1/q$,  
what seems to be rather arbitrary. 
Second, the number of quasiparticles becomes macroscopically large in order to reach 
a new filling factor. The deeper one goes in the hierarchy by consecutive application of the 
hierarchical construction the more questionable the results become. 
Third, the essential prediction of the hierarchical theory is that a daughter state 
is not observed unless all of its parent states are observed. This agrees with experiment 
and shows that the quasiparticle gap $\Delta_{qe-qh}$ of the daughter state is smaller than that 
of the parent states. 
On the other hand, the experimentally observable state at $\nu=3/7$ appears 
in the short notation as the $[3,-2,-2]$ state, while $\nu=5/13$ corresponds to 
$[3,-2,+2]$ and hasn't been yet observed. Naively, one would expect that both 
states have the same degree of stability. Fourth, there are fundamental difficulties to write down 
explicit wavefunctions. There were attempts to introduce wavefunctions using 
quasiparticle coordinates \cite{Halp84}, but there remains the difficulty to 
translate them into electronic wavefunctions. This problem is closely related 
to the occurrence of anyons and their fractional statistics, which the quasiparticles obey \cite{Halp84}. 
Last, but not least, there is also a technical difficulty omitted so far: 
the QPs are not point-like and therefore corrections are needed in order to render 
correctly their short-range interaction \cite{BM91}.

Up to the general acceptance of an alternative and more accessible theory by Jain \cite{Jai89a},  
there were, nevertheless, few 
attempts to put the hierarchical theory on a more quantitative 
footing by determining the interaction of the quasiparticles on a sphere from finite 
size numerical calculations for the case of quasielectrons 
\cite{BM91} and for the quasiholes \cite{ET92}. 
In the first case, the authors succeeded in predicting quite accurate values for the ground 
state energy and the quasiparticle 
gap by determining the QE interaction of the $1/3$ parent state, 
whose QEs condense into the $2/5$ state. 
In the latter case, the authors found that the low-lying excitations of a 
bosonic system with Coulomb interaction 
at quasihole filling factor $\tilde \nu=1/2$ are similar to those of an electronic system 
at $\nu=2/7$, whereas the ground state energy was not determined. 
In another attempt we considered the quasiholes of the $\nu=1$ state in the $V_{1}$ model 
and showed that the ground state energy of the electronic $2/3$ state is identical 
to the ground state of an effective bosonic Hamiltonian of interacting quasiholes 
at filling factor $1/2$ \cite{Kas94}.

In summary, on one hand, the hierarchical theory is at first glance qualitatively appealing, however, 
it is rather clumsy and of limited value in their predictability at general filling factors, 
in particular, when compared with the composite fermion theory. 

The composite fermion theory by Jain \cite{Jai89b,Jai89a,Jai90a} proposes  
explicit electronic trial wavefunctions for the ground state of $N$ interacting 
particles at filling factor $\nu=p/q$ ($q$ -- odd). 
The construction of the wavefunctions at a filling factor of the form 
$\nu=p/(2kp+1)$, $k$ -- integer, is done in the following way: 
accommodate all $N$ particles in the lowest $p$ (spin-polarized) Landau levels. 
This results in the wavefunction $\Phi_{p}$, which is the Slater determinant of 
non-interacting particles at $\nu=p$. Then, $\Phi_{p}$ 
is multiplied by the $(2k)$th power of the Vandermonde determinant 
$\Phi_{1}$ of $N$ particles. Contributions from higher Landau levels, that inevitably occur 
for $p>1$, are projected out by the application of the projection operator ${\cal P}_{0}$ 
leaving us with the wavefunction
\begin{equation}
\chi_{p/(2kp+1)}(z_{1},\ldots,z_{N})= {\cal P}_{0}\Phi_{1}^{2k} \Phi_{p} = 
{\cal P}_{0}  \prod_{i<j}^{N} (z_{i}-z{j})^{2k} \Phi_{p}  \ .
\label{wv-jain}
\end{equation} 
Actually, the number of flux quanta per particle or the inverse filling factor 
is $1/\nu = 1/p + 2k = (1+2kp)/p$. 
This is the so-called D-operation within the formal classification of operations to generate  
wavefunctions at certain FQHE filling factors starting from integer $\nu=p$.  
In order to generate wavefunctions at other $\nu$, Jain introduced additionally 
the C-operation allowing to create a 
wavefunction at $1-\nu$ when the wavefunction at $\nu$ is known because of the particle-hole symmetry. 
Eventually, the L-operation creates 
states at $1+\nu$ from FQHE states at $\nu$. For example, the $2/7$ state appears as the 
result of the D-operation from $2/3$, which again is the result of the C-operation from the 
$1/3$ state \cite{JG92}. 

These trial wavefunctions are meanwhile well accepted since extensive numerical calculations 
show that their overlap with the numerically determined ground state wavefunctions 
is very large. Moreover, the ground state energy per particle found from trial wavefunction 
expectation values and diagonalization results show an astonishing agreement, 
\eg~at $\nu=2/5$ for ten particles,  
the Jain function yields $-0.4693$, whereas the exact value is 
$-0.4694$ \cite{JK98}. Furthermore, QP wavefunctions can 
be constructed by creating first the QPs of filled LLs and then performing the same operations 
as described above. Such a QE wavefunction was already discussed in the disk geometry, where projection 
is even necessary for the QEs of the Laughlin state, \cf~Eq.~(\ref{qe-jain}). 
The important outcome of all these calculations at various $\nu=p/q$ with 
$p>1$ is similar to that of 
the Laughlin wavefunction: the neutral excitation of a QH-QE pair is gapped, and 
it follows the incompressibility at these filling factors.

It is interesting to note that even without projection onto the LLL only a small portion 
of spectral weight resides in higher LLs so that the excess due to kinetic energy is quite 
small \cite{TJ91}. 
Two properties are particularly appealing: first, the Laughlin wavefunctions appear as a 
special case by starting from the filled LLL, \ie~, $p=1,k=1$. Second, the unique wavefunctions 
for integer and fractional $\nu$ are related by a simple operation, and a deeper connection 
between the FQHE and IQHE becomes obvious below. 

As in the case of Laughlin's function, the ultimate reason for the incompressibility 
is poorly understood. While for $\nu=1/q$ the hard-core model with the Laughlin state as 
exact solution gives a hint how incompressibility is caused,  
at $\nu=p/(2kp+1)$ a similar model, where the Jain function appears as the exact ground state,  
is missing.  
However, there exists for filling $\nu=p/(2p+1)$ an interesting mechanism that supports the 
incompressibility property of the Jain function. In case of $p=2$, we choose the somewhat 
artificial model, where the Hilbert space consists of only the two lowest Landau levels 
$n=0,1$ and whose energetic distance vanishes, \ie~this corresponds to the band mass limit 
$m \to \infty$. 
It turns out that in this case the function $\chi_{2/5}$ is an exact solution of 
the $V_{1}$ model, which is also related to a wavefunction for electrons with spin \cite{Halp83}. 
It was shown by numerical calculation that in the limit $m \to 0$,  
which eventually leads to the LLL approximation, the overlap of the Jain function and the 
numerical diagonalization does not vanish and becomes approximately 0.73 \cite{RM91}.

One rather puzzling feature, that we will discuss below,  
is the fact that the starting wavefunction was found for a Hamiltonian,  
whose energy scale is set by $\hbar \omega_c$, while the resulting Jain function describes 
the physics of a Hamiltonian of the interaction energy scale $\lambda$. 

Now let us approximately reverse the above construction in a more physical way. 
We saw that the Jain function $\chi_{p/(2pk+1)}$ provides an extremely accurate description of the 
ground state correlations of the interacting Hamiltonian. 
The statistical transmutation mentioned in Subsect.~\ref{subsect:account} is in fact a singular gauge 
transformation with a parameter $\alpha$ counting the number of flux quanta, which are attached to 
each electron and corresponds to an infinitely thin magnetic solenoid. 
A usual non-singular gauge transformation 
leads to an additional phase factor $e^{i (e/\hbar) f} = e^{2 \pi i f /\Phi_{0}} $ 
in the one-particle wavefunction when the vector potential ${\bf A}$ is changed to 
$\Delta {\bf A}  = {\bf A} + {\bf \nabla} f \equiv {\bf A} + {\bf a}$. 
In a singular gauge transformation, an additional magnetic field 
${\bf b} = {\bf \nabla} \times {\bf a}$ occurs only at  
the location of the particle. In a many-particle formulation, such a transformation can be 
accomplished by adding, \eg~for the particle with index $1$, a vector potential  
\begin{eqnarray}
{\bf a}(z_{1}) & = & \sum_{j=2}^{N}   \nabla_{1j} \frac{\alpha \Phi_{0}}{2 \pi} \Theta_{1j}  = 
\frac{\alpha \Phi_{0}}{2 \pi} 
\int dz' n^{(1)}(z') \frac{ {\bf e}_z \times  (z_{1}- z')} {| z_{1} - z'|^2 }  \ , 
\label{gauge-field}
\end{eqnarray}
where for arbitrary $i$ the condition $i<j$ prevents double counting.
The $\Theta_{ij} \equiv arctg((y_{i}-y_{j})/(x_{i}-x_{j}))$ is the azimuthal angle 
between $z_{i}$ and $z_{j}$ relative to the $x$-axis. The   
solenoid occurs for $z_{i}=z_{j}$ with flux $\alpha \Phi_{0}$, where the sign of $\alpha$ 
determines the direction of the additional magnetic field relative to $B$.
Note that (\ref{gauge-field}) is a constraint between the vector potential 
and the particle density that is responsible for the Chern-Simons term occurring in a 
Lagrangian formulation of the problem \cite{HLR93,ZHK89a}. 
Applying such a singular gauge transformation to the Jain wavefunction with the special choice 
$\alpha = - 2k$ yields a new 
electronic wavefunction obeying the same Hamiltonian except for the additional vector 
potentials ${\bf a}(z_{i})$.  It reads 
\begin{eqnarray}
\tilde {\chi_{p}}(z_{1},\ldots,z_{N})   =  \prod_{i>j=1} e^{-i 2k \Theta_{ij}} \chi_{p/(2kp+1)}
= \prod_{i>j}  \left( \frac{z_{i}-z_{j}}{|z_{i}-z_{j}|} \right)^{-2k} \chi_{p/(2kp+1)} \ .
\label{transmutation}
\end{eqnarray}
This leaves us {\em in average} with a magnetic field $B^{*}$ of strength 
$B^{*} =  B - 2k n \Phi_{0}$,  
where $B=n \Phi_{0}/\nu = n \Phi_{0}(2kp+1)/p = B^{*} (2kp+1)$, and $n$ is the particle density. 
The resulting integer filling factor $\nu^{*}=p$ of an inhomogeneous field satisfies the relation
\begin{equation}
\nu = \frac{\nu^{*}}{2k \nu^{*} + 1} \ . 
\label{CS-filling}
\end{equation} 
We ignore the possibility that the resulting field $B^{*}$ is antiparallel to $B$, 
which would give $\nu=\nu^{*}/(2k \nu^{*} -1)$ in (\ref{CS-filling}). 

It is obvious that the gauge transformation is not exactly the inverse operation $D^{-1}$ 
as can already be seen in the Laughlin case $p=1$: we get neither the wavefunction 
of the filled LLL at $\nu=1$, since we leave this subspace, nor is the magnetic field homogeneous. 
But this transformation is a starting point for field theoretical considerations 
to derive the Laughlin wavefunction \cite{LF91,LF98} by carrying out 
a gauge transformation at $\nu=1/q$ 
with an even number of $(q-1)$ flux quanta. The resulting fermionic field theory is 
evaluated at $\nu=1$ within a random-phase approximation (RPA).  
Also a bosonic formulation is possible, when $\alpha=-q$ is odd \cite{ZHK89a}. 
This is equivalent to the statistical transmutation applied to the Laughlin wavefunction 
that was used to prove algebraic 
off-diagonal long-range order (ODLRO) \cite{GM87}. In the bosonic field theory, the particles 
move in an average zero magnetic field, and an approximate derivation of the Laughlin state 
is also possible. Moreoever, the field theoretical formulation opens the 
opportunity to calculate correlation functions.

Now let us return to Jain's wavefunction, whose product 
$ \prod_{i<j} (z_{i}-z_{j})^{2k} \Phi_p$ can 
be interpreted as weakly interacting ``quasiparticles'' in Landau level $p$,  
to which $2k$ flux quanta are attached despite the difference to the singular gauge 
transformation (\ref{transmutation}). Indeed, the 
prefactors behave like vortices, whose arounding yields just a phase $2k \times 2\pi$. 
Hence, the idea arouse to consider the complicated many particle state of electrons 
at magnetic field $B$ as  weakly interacting composite particles, the so-called 
composite fermions (CF), in a reduced field $B^{*}$. 
In this picture, most of the interaction is compensated by the $2k$th power of the Vandermonde 
prefactor $\Phi_{1}$. 
However, one should keep in mind that the term composite fermion is not only used in 
connection with Jain's construction, but also when Chern-Simons theories are actually meant. 

The idea of the Chern-Simons theory 
can be extended to the so-called ``unquantized Hall effect'' filling factors like 
$1/2$. The fermionic field theory by Halperin, Lee and Read \cite{HLR93} and their refinements, 
\cf~\cite{Sim98,Halp97}, start by applying the gauge transformation with $\alpha=-2$, 
which keeps the fermionic character, and consider the  
interacting particles in a vector potential fluctuating about zero. 
The main difficulty is to take appropriately into account the electron-electron interaction 
as well as the fluctuations of the gauge field, which 
couples to the electron density \cite{Sim98}. The appealing feature of such a theory is the fact 
that the unperturbed ground state describes independent fermions forming a Fermi sea, while the original 
problem has no perturbational starting point. 
This theory predicts a Fermi surface and {\em compressibility} despite the perturbations.  
Hence, the transport behavior is very different from that at QHE filling factors. 
When the Chern-Simons idea is properly translated to Jain's construction of wavefunctions,  
we are led to a $\nu=1/2$ trial wave function of the form 
\begin{equation}
\chi_{1/2} =  {\cal P}_{0}|FS \rangle  \prod_{i<j} (z_{i}-z_{j})^2   \ , 
\end{equation} 
where $|FS \rangle$ describes the Fermi sea of spinless electrons at zero field. The 
overlap of this Rezayi-Read wavefunction with diagonalization results 
is again impressive in the spherical geometry, 
\eg, for nine particles, it is $0.9988$ \cite{RR94}. 
Of course, the existence of a Fermi surface should have clear experimental consequences.  
Experiments at the primary series of filling factors $\nu=p/(2p \pm 1)$,  
whose limit for $p \to \infty$ 
is just $\nu=1/2$, are elucidating. 
According to Jain, the effective magnetic field $B^{*}$ is related to the 
field at $\nu=1/2$ by $\pm B^{*} = B_{p/(2p \pm 1)} - B_{1/2} = \pm n \Phi_{0}/p$. 
Hence, the activation gap energy $\Delta_{qe-qh}$ should scale like $1/p$ or linearly with $B^{*}$ 
as long as the effective mass $m^{*}$ is independent of $B^{*}$ \cite{DST93}.  
Moreover, the symmetric appearance of oscillations in $\rho_{xx}$ around $\nu=1/2$, 
\cf~Fig.~\ref{qhe-experiment}, resembles 
those known from Shubnikov-de Haas oscillations and replicate the features due to the 
IQHE at $\nu=p$. This suggests to interpret these oscillations as those of 
composite fermions in the weaker field $B^{*}$. 
Thus, the FQHE appears as the IQHE of composite fermions \cite{Jai89b,Jai89a}.  

Other peculiarities of the conductivity $\sigma_{xx}$ at and near $\nu=1/2$ can only 
be seen when going beyond d.~c.~transport measurements by exploring length scales 
that are much smaller than the mean free path $\ell^{*}$ 
of the impurity scattered composite fermions, \ie~for $q \ell^{*} \gg 1$. 
Therefore, Willett \cite{WRWP93,Wil98} exploited the surface 
acoustic wave (SAW) technique to get information on $\sigma_{xx}(q,\omega)$ 

via the amplitude damping and the velocity shift $\Delta v_{s}/v_{s}$ of the SAW at sufficiently 
high frequencies ($v_{s}=\omega/q$ - velocity of sound). These quantities can be 
related to the conductivity, which, on the other hand, shows for $ql^{*} \gg 1$ a linear $q$-dependence.  
At sufficiently high frequency (or wavevector $q$), the semi-classical theory in the small 
effective field $B^{*}$ predicts minima in the velocity shift $\Delta v_{s}/v_{s}$ of the SAW, which 
occur at certain values of the product $q R_{c}^{*}$ \cite{HLR93}. Here, $R_{c}^{*}$ is 
the composite fermion cyclotron radius 
$R_{c}^{*}=v_{F}/\omega_{c} = \hbar k_{F}/(|e B^{*}|)$. Hence, the observation of the 
principal resonance (minima) in the velocity shift away from the minimum at $B_{1/2}$ 
allows the determination of the composite fermion Fermi wavevector $k_{F}$, which is found  
to be very close to the value $k_{F}= \sqrt{4 \pi n}$ of a gas of free electrons with one 
spin direction \cite{WRWP93}. This result dramatically supports the CF picture. 
The origin of the SAW resonances is of geometrical nature due to the commensurate 
resonance between the cyclotron radius $R_{c}^{*}$ and the SAW wavelength $2 \pi/q$. 
Another experimental set-up studies the longitudinal resistance 
$R_{xx}$ near $\nu=1/2$ in a regular arrangement of anti-dots, which  
acts like a superlattice with lattice constant $a$. Geometrical resonances  
are observed if a commensurate condition between 
$a$ and the effective cyclotron radius $R_{c}^{*}$ is satisfied \cite{KSP93}. 
These resonances are similar to those found in real low magnetic field experiments \cite{WRM91}. 
Despite all these successes, the Chern-Simons theory has to answer the fundamental question 
of the correct energy scale, which is influenced by $B^{*}$ as well as the CF mass $m^{*}$. 
The CF energy gap $\hbar |e B^{*}|/m^{*} \propto |B|/((2kp+1) m^{*})$ must agree with 
the FQHE activation energy gap for every $p$ in the primary series,  
which scales in the strong field limit like $\sqrt{B}$. 
Thus, the effective mass $m^{*}$ has a $\sqrt{B}$ dependency at fixed $\nu$ and has to 
diverge when approaching $B_{1/2}$ as $p \to \infty$. 
This is difficult to achieve in the diagrammatic treatment by Halperin \ea~. Therefore, the 
effective mass $m^{*}$ was used as a phenomenological parameter, which is fitted to 
the known activation gaps at a certain $\nu$ of the primary series. Furthermore, 
experiments trying to measure $m^{*}$ are ambiguous and depend on the method used to 
extract the effective mass \cite{WWP95,DST94,LNF94}.  
On the other hand, the unrenormalized band mass $m$ cannot be completely abandoned. For example, 
it couples to spatial inhomogeneities of the magnetic field and enters the magnetic moment. 
Therefore, any satisfying theory should distinguish these kinds of masses.

An approach that tries to circumvent these difficulties was recently developed \cite{SM97b,MS98a}. 
After performing the fermionic statistical transmutation,  
Murthy and Shankar enlarge the Hilbert space by imposing additional constraints in order to 
properly consider the physical degrees of freedom. Approximate treatment of this problem and 
back projection onto the physical space yield Jain's and the Rezayi-Read wavefunction  
for $\nu=p/(2p+1)$ and $\nu=1/2$, respectively. To go beyond the derivation of wavefunctions,  
canonical transformations and the approximate treatment in the long-wavelength limit result in 
a mass renormalization to higher values $m^{*}$ and exhibit the desired $\sqrt{B}$ dependency 
of the LLL approximation. 
These features are very pleasing and pave the way to a more unified view of the 
odd and even denominator states in the QH-regime. 

\section{Ground states and thermodynamics of electrons with 
spin degree of freedom in the lowest Landau level }
\label{sect:spin}

	\subsection{The spin degree of freedom}
	\label{subsect:spin degree}

In the last Section, our considerations were restricted to 
spin-polarized electrons.  
However, the underlying assumption of an infinitely strong magnetic field 
is not satisfied in experiment. In contrast, 
the Zeeman energy is much smaller than the interaction scale, but the magnetic field 
is so strong that the inequality (\ref{lll-spin}) holds.  
Hence, we return in this Section to the Hamiltonian of the form 
(\ref{ham-lll-sym}) and (\ref{ham-lll-landau}), respectively, taking explicitly into 
account the Zeeman term. Then, the appropriate parameter to describe 
the various physical regimes is the effective gyromagnetic factor 
$\tilde g = \Delta_z/\lambda \propto \sqrt{B}$.

This fact, however, does not devaluate our previous investigations in the limit 
$\tilde g \to \infty$. The results are correct as long as ground state 
and charged excitations remain spin-polarized,  
even when (\ref{lll-spin}) is not fulfilled. This is possible due 
to the influence of interaction on the spin-orientation. 
Indeed, finite-size diagonalizations indicate that the 
Laughlin ground state at $\nu=1/q$ is even in the absence of a Zeeman term 
spin-polarized in the LLL approximation, 
while the ground state at $\nu=2/5$ yields a spin singlet at vanishing Zeeman term, 
but becomes spin-polarized under the influence of a Zeeman term at accessible 
values for $\tilde g$ in $GaAs$ \cite{CP95}. Therefore, the existence of spin 
polarization in the ground state must be checked in each case. 
Even more intriguing is the question whether both 
charged excitations or only one of them is accompanied by spin reversals. 
Ideally, this problem can be solved by studying the magnetic field dependence 
of the activation gap $\Delta$ in the longitudinal resistivity 
$\rho_{xx} \sim e^{-\beta \Delta/2}$, \cf~Eq.~(\ref{activation}). Then, a  
square root dependence on $B$ indicates purely interaction driven excitations, whereas 
a linear $B$-dependence shows that spin-flip excitations are involved. 
Such different regimes could be observed, \ie, at lower fields a linear dependence on $B$ 
with a cross-over to a square root $B$-dependence when increasing the strength of the field. 
Unfortunately, the influence of disorder and the finite extension of the 2DES 
in the $z$-direction spoil often the experimental results.

Trial wavefunctions at fractional $\nu$ considering the spin degree of freedom were early 
proposed \cite{Halp83}. But only with the extension of the LLL Hilbert space 
up to $\nu=2$, the physics at $\nu=1$ appeared in a new light. While in the spin-polarized case, the 
QE at $\nu=1$ does not exist, now, the QE has to reverse at least one spin. 
Surprisingly, the charged excitations are not necessarily 
one-spin flip quasiparticles changing only locally the homogeneous spin magnetization 
of the ferromagnetic ground state at $\nu=1$. 
At a sufficiently small effective gyromagnetic factor $\tilde g $, 
charged spin texture excitations with a large number of flipped spins are energetically 
preferred \cite{SKKR93}. These excitations are called skyrmions. Their existence 
has experimental consequences for the spin magnetization 
so that a lot of experimental and theoretical activity was sparked.

In this Section, we are mainly concerned with the physics at exactly and near $\nu=1$. 
The methods we apply are in some sense specific to this filling factor. 
This is in accordance with the fact that there exists no many-particle theory in the 
quantum Hall regime comprising the physics of the entire filling factor range. 
Therefore, general attempts are not able to concomitantly reveal the specific features 
of the physics for all $\nu$. Theories trying to explain experiments 
over a large range of filling factors as, \eg, tunneling experiments 
between two parallel layers  contain so strong approximations that they do not incorporate 
features specific to FQHE filling factors \cite{EPW92b,Hau96}.

Now, the particle-hole symmetry for spin-polarized electrons is generalized 
to particle-hole symmetry around $\nu=1$, \cf~Eq.~(\ref{particle-hole-polar}).  
For the ground state energy in the thermodynamic limit, $\epsilon(\nu)$, it holds
\begin{equation}
(2-\nu)\ \epsilon(2-\nu) - \nu\  \epsilon(\nu) = 2(1-\nu) \epsilon(2) \ , 
\label{particle-hole-spin}
\end{equation} 
which is invariant with respect to the substitution $\nu \leftrightarrow (2-\nu)$. 
Since we are mostly interested in magnetic properties due to 
the spin of the electron, let us quote some of the repeatedly used elementary 
relations.
Assuming an external magnetic field in the $z$-direction,  
the temperature dependent extensive 
total spin magnetization $M$ is proportional to the difference of the expectation values  
of the particle numbers $N_{\uparrow}$ for spin-up electrons and $N_{\downarrow}$ 
for spin-down electrons 
\begin{equation}
M(T, \nu) = \frac{|g| \mu_{B}}{2} (N_{\uparrow} - N_{\downarrow}) = 
M_{0} \frac{N_{\uparrow} - N_{\downarrow}}{N_{\uparrow} + N_{\downarrow}}
= M_{0} (\nu_{\uparrow} - \nu_{\downarrow}) \ ,
\label{magn-formula}
\end{equation}
where $M_{0} \equiv N |g| \mu_{B}/2$ is the total magnetization of $N$ 
spin-polarized electrons, and $\nu_{\sigma}$ is the temperature dependent 
filling factor for the 
respective spin direction $\sigma$ with $\nu_{\uparrow}+\nu_{\downarrow}=\nu$. 
The expectation value of the local spin operator (\ref{spin-density}) can be expressed for a 
state $|\Psi \rangle $ in the LLL by 
\begin{equation}
{\bf S}({\bf r}) = \frac{1}{2} \sum_{m,m' \atop \sigma, \sigma'} 
\varphi_{m}^{*}({\bf r}) 
\varphi_{m'}({\bf r}) \langle \Psi|c_{m,\sigma}^{\dagger} 
\vec{\tau}_{\sigma,\sigma^{'}} 
c_{m',\sigma'} |\Psi \rangle  
\end{equation}
($m$ denotes here a general degenerate quantum number in the LLL) 
and defines the dimensionless local 
magnetization ${\bf m}({\bf r})= 2 {\bf S}({\bf r})$.
Consequently, the total spin magnetization is 
\begin{equation}
M = \frac{|g| \mu_{B}}{2} \frac{1}{2 \pi \ell_c^2} \int d^{2} r\, m_{z}({\bf r}) 
= |g| \mu_{B} \frac{1}{2 \pi \ell_c^2} \int d^{2} r\, S_{z}({\bf r}) \ .
\end{equation}
When discussing spin configurations, it is interesting to know the 
total number of reversed spins $K$ in the minority spin band. This number 
can be extracted from $m_{z}({\bf r})$ via the relation 
\begin{equation}
K = \frac{1}{4 \pi \ell_c^2} \int d^2 r\, (1 - m_{z}({\bf r})) - \frac{1}{2}(N_{\Phi}-N) \ .
\label{flipped-spins}
\end{equation}
The interacting Hamiltonian (\ref{ham-lll-landau}) 
exhibits a continuous $SU(2)$-symmetry in the spin-space when the Zeeman term vanishes. 
Hence, the role of any non-zero Zeeman coupling is that of a symmetry 
breaking field. In case that the spin of the ground state for a vanishing 
gyromagnetic factor $g$ is different from $S=0$,  
the system develops {\em spontaneous magnetization}, which is parallel to the direction 
of the external magnetic field. We will see that such a situation can occur for quite a lot 
of filling factors in the LLL. If one compares this phenomenon with ferromagnetism in a 
metal, it is at first glance astonishing that in a QH-system such a situation 
occurs although the electrons are exposed to a 
strong magnetic field when the Hilbert space is restricted to the LLL. 
However, it is important to realize that the role of the symmetry breaking 
field is played 
by the entire Zeeman term, which can be independently tuned by the magnetic field as well as by $g$. 
Thus, the symmetry breaking Zeeman term can even vanish in the strong magnetic field limit. 
This can be realized in experiments under hydrostatic pressure. 
This allows to sweep the 
$g$ factor from its ``natural'' negative value through zero \cite{MPP96}. 
Under such circumstances, the strong magnetic field enters only indirectly through 
a ``band'' of vanishing bandwidth and the LLL interaction matrix elements, which are 
quite different from the unprojected Coulomb matrix elements. 
Hence, a QH-system that exhibits spontaneous magnetization is called a 
{\em quantum Hall ferromagnet} (QHFM).
One should keep in mind that the two-dimensionality of the system and the continuous 
$SU(2)$-symmetry for a vanishing Zeeman term prohibits any 
magnetization at non-zero temperature because of the Mermin-Wagner theorem \cite{MW66}.
Any reasonable theory of the spin magnetization should satisfy this constraint. 

	\subsection{Ground state and charged excitations at and near $\nu=1$ }
	\label{subsect:ground}

We start with the discussion of the ground state of interacting 
electrons with spin in the LLL for filling factors exactly at $\nu=1$. 

The physics at $\nu=1$ is 
comparably well understood. We choose the $V_{0}$ model for the interaction 
and construct a many-particle electronic wavefunction 
of zero energy in the spirit of the derivation of the Laughlin wavefunction at $\nu=1/3$ 
for the $V_{1}$ pseudopotential \cite{MFB96}.  
\begin{figure}[h]
\centerline{\resizebox{6cm}{6cm}{\includegraphics{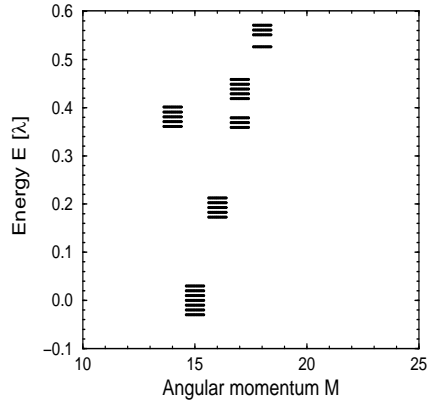}}}
\caption[]{\small
The spectrum at $\nu=1$ for six particles with spin interacting via the $V_{0}$ model 
on a disk ($V_{0}=\sqrt{\pi}/2$). The degeneracy of each 
energy level can be read off from 
the spectrum due to the chosen non-zero value of the Zeeman energy $\Delta_z=0.01 
\lambda$. {\em E.~g.}, the zero energy ground state at $M=15$ is split into 
seven non-degenerate levels, {\em i.~e.~}$S=3$.
}
\label{ground state_spin_v0}
\end{figure}
Note that due to the additional spin degree of freedom denoted by $\chi$ the entire wavefunction 
has still to be antisymmetric, but the symmetry condition for the orbital part is relaxed and 
allows now also relative angular momenta $m$ that are even. 
We know that any $N$-particle eigenfunction of the $V_{0}$ model without 
neutralizing background and Zeeman term of the form 
\begin{equation}
\Psi[z,\chi] = [\prod_{i<j}^{N} (z_{i}-z_{j})] \Psi_{B}[z,\chi] \ ,
\label{zero-state}
\end{equation}
has zero energy because of the occurrence of relative angular momentum one in 
the antisymmetric prefactor. The residual wavefunction $\Psi_{B}[z,\chi]$ 
for $N$ particles with spin $1/2$ has to be symmetric. The maximum power of $z_{i}$ in 
the prefactor is $N-1$ so that for $\nu=1$, {\em i.~e.~}$N=N_{\Phi}$, no further 
orbital dependence of $\Psi_{B}$ as a power of $z$ may occur (except for the exponential part).  
In order to be an eigenfunction of ${\bf S}^2$, $\Psi_{B}[\chi]$ 
must be a symmetric $N$-particle spinor with total spin $S=N/2$. 
The zero energy level is $(N+1)$-fold degenerate with 
spin component $S_{z}$ between $-N/2,-N/2+1,\ldots,N/2$. For $V_{0}>0$, any 
state different from (\ref{zero-state}) has a positive energy, and therefore 
\begin{equation}
|\Psi_{\nu=1} \rangle  = |\uparrow_{1} \uparrow_{2}  \uparrow_{3}  
\ldots \uparrow_{N} \rangle |\Psi_{S} \rangle  
\label{spin-polarized}
\end{equation}
is a zero eigenenergy state with maximum $S_{z}=N/2$, $|\Psi_{S} \rangle$ is the Slater determinant 
of the filled LLL. All the other degenerate ground states  
with spin-eigenvalue $|S,S_{z}\rangle$ can be constructed 
by subsequent application of the ladder operator 
$S^{-}=\sum_{i=1}^{N} S^{-}_{i}$ to (\ref{spin-polarized}) as $[S^{-},{\bf S}^{2}]=0$. 
Any non-zero Zeeman term lifts the degeneracy and selects from the spin multiplet the state 
Eq.~(\ref{spin-polarized}) as the unique ground-state of energy $-N \Delta_z/2$, 
see Fig.~\ref{ground state_spin_v0}. 
In the ground-state, all available orbitals with $m=0,1,\ldots,N_{\Phi}-1$ are occupied, and the 
spin for each orbital state is up $\uparrow$. Hence, the schematic representation of the ground state 
in Fig.~\ref{energy_levels} remains exact for interacting electrons in the $V_{0}$ model.
It is clear that any double occupancy of an orbital costs energy due to the 
occurrence of relative angular momentum zero. 
Actually, the maximum $S_{z}$ ground state is identical with the Laughlin state for $q=1$ 
in the spin-polarized model. There, however, such a state occurs due to the infinitely 
large Zeeman term,  
while in the current case this is a pure interaction effect 
resulting from the spontaneous breaking of the symmetry.
The $V_{0}$ model without neutralizing background at $\nu=1$ plays a role analogous 
to that of the $V_{1}$ model for the spin-polarized case at $\nu=1/3$. 
Its real space representation is in accordance with (\ref{ham-lll}) 
\begin{equation}
H_{int} = 4 \pi V_{0} \lambda \sum_{i<j} \delta^{2}({\bf r}_{i}-{\bf r}_{j}) \ ,
\label{real-space-v0}
\end{equation}
where we adopt for $V_{0}$ its Coulomb value in the LLL, \ie~$V_{0}=\sqrt{\pi}/2$.
The Fourier transformation of (\ref{real-space-v0}) is $\tilde{V}({q}) = 4 \pi V_{0}$  
so that $\tilde a(0)= 2 V_{0}$, \cf~Eq.~(\ref{attraction-two-particles}).  

A translation of Eq.~(\ref{zero-state}) into an equivalent formulation 
within a bosonic language for $\Psi_{B}[z, \chi]$ is 
possible. Then, the bosonic operators $b_{m,\sigma}^{(\dagger)}$ with angular 
momentum $m$ and spin $\sigma$ allow to write the degenerate ground state as 
\begin{equation}
|n_{0\downarrow}=K, n_{0\uparrow}=N-K \rangle  = 
\frac{1}{\sqrt{(N-K)!K!}} \left (b_{0,\downarrow}^{\dagger} \right)^{K} 
\left (b_{0,\uparrow}^{\dagger} \right)^{N-K} |0 \rangle  \ ,
\end{equation}
where $K=0,\ldots,N$ is the number of reversed spins as $S_{z}=N/2-K$, $n_{m\sigma}$ 
is the occupation number for bosons with angular momentum $m$ and spin $\sigma$,
and $|0\rangle$ is 
the bosonic vacuum. The state with $K=0$ corresponds to (\ref{spin-polarized}). 
Hence, the bosonic ground state is completely characterized by the occupation 
numbers $n_{0\sigma}$ of orbital states with $m=0$. 
\begin{figure}[h]
\centerline{\resizebox{6cm}{6cm}{\includegraphics{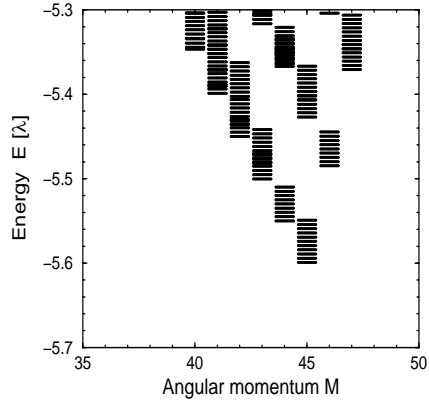}}}
\caption[]{\small
The spectrum at $\nu=1$ for ten particles with spin interacting via the Coulomb interaction 
on a disk with a neutralizing background. The Zeeman energy is $\Delta_z=0.005 \lambda$. 
The ground state has, as expected, total angular momentum $M=45$ and $S=5$.
}
\label{ground state_spin_coul}
\end{figure}
The situation does not change qualitatively when we consider the more realistic 
Coulomb interaction with pseudopotential coefficients monotonically decreasing from 
the maximum value $V_{0}$. 
Although we cannot rigorously infer that the ground state is identical to that of 
the $V_{0}$ model,  
finite-size numerical calculations in different geometries show that, 
see Fig.~\ref{ground state_spin_coul} \cite{MMY95}. 
In physical terms, the existence of aligned spins is favored by the Pauli principle. 
Parallel spins exclude the occupation of the same orbital state and thus reduce 
the overlap between two electrons and the Coulomb energy contribution. Moreover, 
parallel spins lead to a gain in exchange energy and to a decrease of the ground state
energy compared with that of non-interacting particles. 
The mechanism is particularly effective for a flat band, whereas in a system of 
finite band-width a competition between increase of kinetic energy and exchange energy 
gain occurs when all spin are aligned. 
However, even the zero band-width is not sufficient to prevent the breakdown of 
the complete spin-polarization leading to spin-singlet ground states when there is no  
Zeeman coupling, {\em e.~g.~}at $\nu=2/3$ and $\nu=2/5$ \cite{ZC84,MDH86}.  
In the Coulomb model, the ground state energy is the same as for the spin-polarized 
case since only the orbital part contributes, and we recover 
$\epsilon_{Coul}(1)/\lambda= - \tilde a(0)/2= -\sqrt{\pi/8}$, \cf~Eq.~(\ref{exact-polar}).
\begin{figure}[b]
\centerline{\resizebox{6cm}{2cm}{\includegraphics{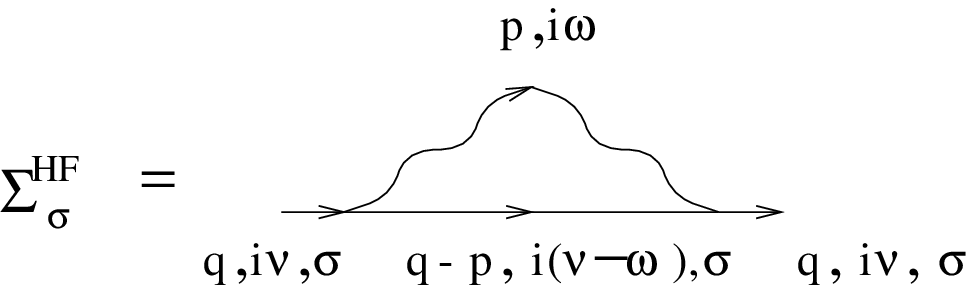}}}
\caption[]{\small 
Proper self-energy diagram in the self-consistent Hartree-Fock approximation.
The propagator line in this diagram must be determined 
self-consistently.  This approximation leads to a frequency
independent self-energy and hence to a Green's function whose spectral
weight consists of a single $\delta-$function.  The first order tadpole 
diagram is absent because of the neutralizing background charge.} 
\label{Fock_diagram}
\end{figure}
In order to introduce the diagrammatic many-particle approach, let us repeat the 
ground state energy determination 
by calculating the thermodynamic Green's function (GF) in a 
self-consistent Hartree-Fock approximation (SHF), \cf~Eq.~(\ref{GF0}), 
\begin{equation}
{\cal G}_{\sigma}^{HF}(i \nu_{n}) = \frac{1}{i\hbar \nu_n - \xi_{\sigma}^{HF}} \; .
\label{GHF}
\end{equation}
The combination of the Dyson equation with the 
temperature and filling factor dependent, but frequency independent self-energy 
$\Sigma_{\sigma}^{HF}$ approximated by the Fock diagram yields 
\begin{equation}
\Sigma_{\sigma}^{HF} = ({\cal G}^{(0)\ }_{\sigma})^{-1} - ({\cal G}^{HF}_{\sigma})^{-1} = 
 \xi_{\sigma}^{HF} - \xi_{\sigma}^{(0)} = - n_{F}(\xi_{\sigma}^{HF}) \lambda \tilde{a}(0)\; .
\label{SC}
\end{equation}
On the other hand, the general expression for the temperature dependent energy \cite{FW71} 
for electrons in the LLL within the SHF is 
\begin{eqnarray}
E(T, \nu) & = & \frac{N_{\Phi}}{\beta} \sum_{\sigma, i \nu_{n}} e^{i \nu_{n} \eta}
\left [\epsilon^{(0)}_{\sigma} + \frac{1}{2} \Sigma_{\sigma}(i \nu_{n}) \right ]
{\cal G}_{\sigma}(i \nu_{n}) \nonumber \\
& = & \frac{N_{\Phi}}{2} \sum_{\sigma} n_{F}(\xi_{\sigma}^{HF}) 
(\epsilon_{\sigma}^{(0)} + \epsilon_{\sigma}^{HF}) \ .
\label{hartree-fock-energy}
\end{eqnarray}
At $\nu=1$, where particle-hole symmetry $\xi_{\uparrow}^{HF}=-\xi_{\downarrow}^{HF}$ holds, 
the chemical potential is $\mu=-\lambda \tilde a(0)/2$. The  
zero temperature SHF energies for the spin-directions are 
$\epsilon_{\uparrow}^{HF}= -\Delta_z/2 - \lambda \tilde a(0)$ and 
$\epsilon_{\downarrow}^{HF}= \Delta_z/2$, respectively.  Assuming $\Delta_z=0$,  
we recover the exact result for $E(T=0,\nu=1)/N=\epsilon_{Coul}(1)=
-\lambda \tilde a(0)/2 = - \sqrt{\pi/8}\, \lambda$ of Eq.~(\ref{exact-polar}). 
As usual, the consideration of the HF-diagrams gives for a uniform system at 
zero temperature a correction to the 
non-interacting result, which is half of the exchange energy 
contribution in the LLL $\epsilon_{ex}^{00}= -\lambda \tilde{a}(0)$. 
Therefore, the SHF yields the exact ground state energy. 

After the discussion of exactly filling factor one, we turn to the charged excitations, 
the quasiparticles. 
Similar to the case of the spin-polarized ground states at $\nu=1/q$, we can ask 
what happens if we remove one electron from, or add one electron to the spin-polarized 
ground state at $\nu=1$. However, in contrast to the spin-polarized case, where adding 
an electron at $\nu=1$ leads inevitably to the occupation of the second orbital Landau level 
$n=1$, and removing one electron creates $N+1$ quasiholes, which become degenerate in 
the thermodynamic limit, now, the emerging picture is much richer since 
we remain in the LLL utilizing the spin degree of freedom. 

\begin{figure}[h]
\centerline{\resizebox{6cm}{6cm}{\includegraphics{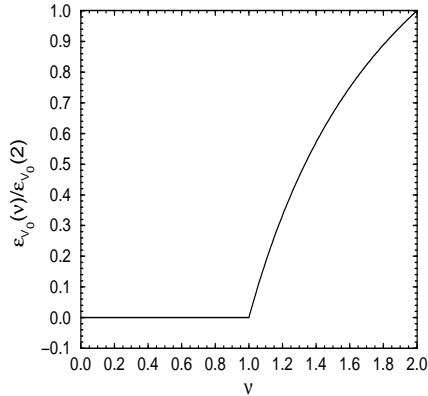}}}
\caption[]{\small
The ground state energy per particle $\epsilon_{V_{0}}(\nu)$
as a function of $\nu$ in units of 
$\epsilon_{V_{0}}(2)=V_{0} \lambda$ for the 
$V_{0}$ model covering the whole $\nu$ range $0 \le \nu \le 2$. 
Note the right side derivative at $\nu=1^{+}$ of value 2, which determines the 
QE energy in the model. In contrast to this 
example, for Coulomb interaction only single energy values are exactly known. 
In the case of spin-polarized electrons, approximate interpolation formulae  
for $0 < \nu \le 1$ exist \cite{HMM96}. 
}
\label{particle_hole_spin}
\end{figure}

We use again the $V_{0}$ model for a microscopic understanding of the quasiparticles.  
First, we discuss the quasihole charged excitations, which are also much more 
amenable to diagonalization in the disk geometry as the data are not spoiled by 
finite-size effects. Nevertheless, particle-hole symmetry about $\nu=1$ makes these 
considerations not specific to the quasiholes.
The neutral quasiholes are created by finding the ground state 
for $N$ particles moving in an area covered by $N_{\Phi}=N+1$ flux quanta. 
It is obvious that the various quasihole energies in the $V_{0}$ model equal zero, 
{\em i.~e.~}$\epsilon_{-}^{n}(1)=0$, and hence the ground state energy in the 
$V_{0}$ model ($\Delta_z=0$) is 
\[ \epsilon_{V_{0}}(\nu)/\lambda =  \left\{ \begin{array}{r@{\quad:\quad}l}
		  0 &   0 \le \nu \le 1  \\  2(1-\frac{1}{\nu}) V_{0} & 1 < \nu \le 2 
				    \end{array} \right. 	\]	
\label{ground state-hardcore-spin}
because of Eq.~(\ref{particle-hole-spin}), see Fig.~\ref{particle_hole_spin}. 

\begin{figure}[h]
\centerline{\resizebox{6cm}{6cm}{\includegraphics{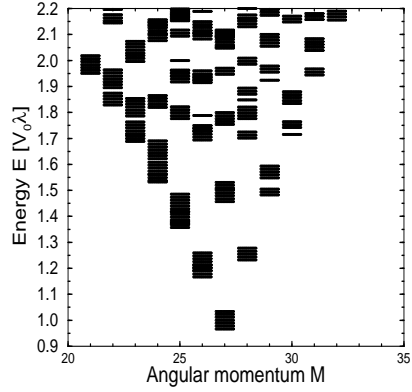}}}
\caption[]{\small
The quasielectron at filling factor $\nu=1$ for $N=8$ and $N_{\Phi}=7$ in 
a model with $V_{0}$-interaction and no background. The energy is given in 
units of $V_{0} \lambda$. The lowest lying state at $M=21$ has $S=3$ with 
7 particles spin up and one spin down. It has the lowest finite-size correction 
when compared with the thermodynamic limit of the energy of 
a quasielectron in the $V_{0}$ model, which is $\epsilon_{+}^{n}(1)=2V_{0} \lambda$. 
The energy correction is the smallest one if the quasielectron 
has its innermost position minimizing the influence of the edge of the system.
}
\label{quasielectron_spin}
\end{figure}

\begin{figure}[h]
\centerline{\resizebox{7cm}{5cm}{\includegraphics{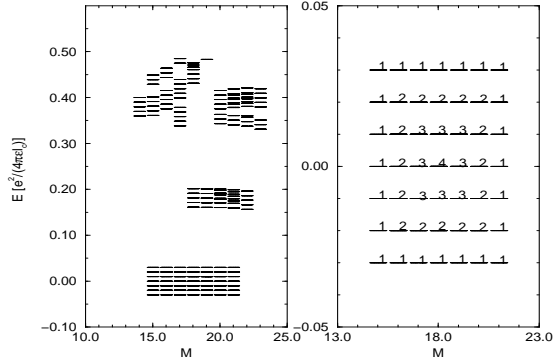}}}
\caption[]{\small
All skyrmionic states for six particles and $N_{\Phi}=7$ at $\Delta_z=0.01 \lambda$ in the $V_{0}$ model.
The right panel indicates the degeneracy of each of the skyrmionic states even 
after ordering w.~r.~t.~the angular momentum $M$ and in the presence of a non-vanishing Zeeman term. 
From \cite{Kas96}.
}
\label{quasihole_spin}
\end{figure}

On the other hand, the neutral quasielectron energy becomes in the thermodynamic limit 
\begin{eqnarray}
\epsilon_{+}^{n}(1)=\lim_{N_{\Phi} \to \infty} E_{0}(N,N_{\Phi}=N-1)-E_{0}(N,N_{\Phi}=N) \ ,
\nonumber \\
= - \lim_{N \to \infty} 
\left(\frac{\partial E_{0}}{\partial N_{\Phi}} \right)_{N_{\Phi}=N^+} 
= \nu^{2} \left ( \frac{\partial \epsilon}{\partial \nu} \right)_{\nu=1^+} \ ,
\end{eqnarray}
which leads to $\epsilon_{+}^{n}(1)= 2 V_{0} \lambda$\ ,  
see Fig.~\ref{particle_hole_spin}. In an alternate way, the finite-size 
quasiparticle energies can be also calculated directly from the energy of the 
quasielectron with a spin reversed particle in the state with angular momentum 
$m$, ($m=0,\ldots, N_{\Phi}-1$), {\em i.~e.~} 
$|\Psi_{1,m}^{(+)} \rangle = c^{\dagger}_{m \downarrow} c^{\dagger}_{0 \uparrow}
\ldots  c^{\dagger}_{N-2 \uparrow}|0 \rangle $.
In first quantization, the quasielectron state at the origin ($m=0$) reads \cite{MH86}
\begin{equation}
\Psi_{1,m=0}^{(+)}(z_{1},\ldots,z_{N}) = {\cal A} \left( \prod_{i>j=2}^{N} (z_{i}-z_{j})\  
|\downarrow_{1} \uparrow_{2} \ldots \uparrow_{N} \rangle \right ) 
e^{-\frac{1}{4}\sum_{i=1}^{N} |z_{i}|^2} \ 
\end{equation}
with ${\cal A}$ as antisymmetrization operator. 
The representation allows, on the one hand, to identify the quasielectron states 
from the numerical 
calculations, see Fig.~\ref{quasielectron_spin}, despite their strong finite-size effects, 
on the other hand, we find in 
the thermodynamic limit ($N_{\Phi} \to \infty$) that the quasielectrons are 
degenerate and have energy $2 V_{0} \lambda$. 
Similar calculations for the conventional quasiparticle states in 
the Coulomb model with background lead to results summarized in Table \ref{tab:energies-nu-1}.
Below, we shall see that these states are not the true 
elementary excitations for the Coulomb interaction, \ie~energetically 
lower excitations exist. 
Furthermore, it is interesting to note that the existence 
of a non-vanishing gap for particle-hole excitations far apart is necessary for the 
incompressibility of a state. 
However, this does not exclude the existence of gapless neutral excitations occurring as 
long-wavelength spin-waves above the polarized ground state at $\nu=1$ in the 
case of vanishing Zeeman energy, see Subsect.~\ref{subsect:spin-wave}. This is 
different from the situation at fractional $\nu$ for spin-polarized electrons 
in the LLL, where even in the long-wavelength limit $k \to 0$ of the 
single-mode approximation a gap exists \cite{GMP85}.  
\begin{table}[h]
{\small 
\begin{tabular}{lccccccc} 
\\ \hline \\[-3mm]
Model & $\epsilon(1)$	&  $\epsilon_{-}^{n}(1)$ & $\epsilon_{+}^{n}(1)$ & 
$\epsilon_{-}(1)$ & $\epsilon_{+}(1)$ & $\Delta_{qe-qh}$ & $\Delta_{sk-ask}$ \hspace*{0mm}
\\ \hline\hline \\[-3mm]
$V_{0}$ model &	$0$ & $0$ & $2V_{0}$ &  $0$ &  $2V_{0}$ & $2V_{0}$ & $ 2V_{0}$ \\ \hline
Coulomb & $-\frac{1}{2} \sqrt{\frac{\pi}{2}}$ & 
$\frac{1}{2}\sqrt{\frac{\pi}{2}}$ & 
$\frac{1}{2}\sqrt{\frac{\pi}{2}}$ & $\sqrt{\frac{\pi}{2}}$ & 0 &
 $\sqrt{\frac{\pi}{2}}$ & $\frac{1}{2}\sqrt{\frac{\pi}{2}}$ \\ \hline
\end{tabular}
}
\caption[]{ Exact results for the ground state energy per particle $\epsilon(1)$, 
for the neutral and gross quasihole energies $\epsilon_{-}^{n}(1)$ and $\epsilon_{-}(1)$, 
the neutral and 
gross quasielectron energies $\epsilon_{+}^{n}(1)$ and $\epsilon_{+}(1)$, the value 
of the quasielectron-quasihole gap $\Delta_{qe-qh}$, and the value of the 
skyrmion-antiskyrmion gap $\Delta_{sk-ask}$, see below, in units of $\lambda$ for 
the $V_0$ model and the Coulomb model with background at $\nu=1$.}
\label{tab:energies-nu-1}
\end{table}
It is remarkable that the case $\nu=1$ is the only example to date, where the 
incompressibility caused by interaction and {\em not} by a one-particle effect 
can rigorously be shown in the thermodynamic limit. 
From a microscopic point of view, the phenomenological distinction between integer QHE 
with a one-particle gap and the fractional QHE exhibiting a many-particle gap
loses its meaning at $\nu=1$ as in the LLL the one particle gap $\Delta_z$ is small in comparison 
with the enhanced spin-exchange gap due to many-particle effects. This becomes particularly obvious 
in experiment, where the exertion of hydrostatic pressure causing 
a vanishing $g$-factor does not lead to the destruction of the QHE \cite{MPP96}. 
The classification of the charged excitations of the quasihole type at $\nu=1$ for the 
$V_{0}$ model is explicitly possible and was done by MacDonald \ea~\cite{MFB96,Kas96}.
The QHs appear as zero energy eigenstates in Fig.~\ref{quasihole_spin}.  
The multiple degeneracy of most of the states for a given total angular momentum $M$ and 
$S_{z}$ in the right panel of Fig.~\ref{quasihole_spin} shows 
that not all states have $S=N/2$, {\em i.~e.~}not all states can 
be described as spin split polarized quasiholes known from the spin-polarized case 
discussion in Section \ref{sect:polar}. 
Again, all zero energy functions comply with Eq.~(\ref{zero-state}). 
Hence, only boson occupation numbers $n_{0\sigma}$ and $n_{1\sigma}$ 
can occur ensuring that the maximum power of $z_{i}$ is fixed due to $N=N_{\Phi}-1$. 
So-called seed states can be uniquely identified in Fig.~\ref{quasihole_spin}, which 
are eigenstates of the angular momentum, of $S_{z}$ and even of ${\bf S}^2$ in the 
thermodynamic limit. Hence, the application of the ladder operator 
$S^{+}$ on the state with $n_{0\downarrow}=K$, {\em i.~e.~}$|n_{0\downarrow}=
K, n_{0\uparrow}=0, n_{1\downarrow}=0 \rangle $ vanishes so that 
this state can be uniquely identified as an energy eigenstate 
with $S=S_{z}=M=N/2-K$. 
Then all states of the spin-split diagram can be identified since the 
degeneracy with respect to the center of mass motion leads to a degeneracy 
of states with given $S_{z}$ along the $M$ axis. 
Due to the mapping between bosons and fermions, the seed states 
can be translated into fermionic states with $K$ reversed spins \cite{APFGM97}
\begin{equation}
|\Psi_{K}^{0} \rangle  = \frac{1}{\sqrt{C(K)}} \sum_{m_{K} > \ldots > m_{1}=1}^{N_{\Phi}}
\frac{1}{\sqrt{m_{1} \ldots m_{K}}} 
c^{\dagger}_{m_{K}-1\downarrow} \dots c^{\dagger}_{m_{1}-1\downarrow}
c^{\dagger}_{l_{N-K}\uparrow} \ldots c^{\dagger}_{l_{1}\uparrow} |0 \rangle  \ ,
\end{equation}
where $l_{i} \in \{0,\ldots,N\} \backslash \{m_{1},\ldots,m_{K}\}$,  
$l_{i+1} > l_{i}$ and $C(K)$ is the normalization constant. 

\begin{figure}[h]
\centerline{\resizebox{8cm}{5cm}{\includegraphics{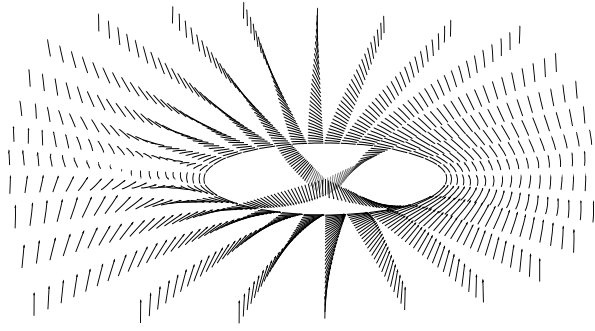}}}
\caption[]{\small
Illustration of the local magnetization ${\bf m}({\bf r})$ of a scale invariant 
skyrmion with a local magnetization at the origin 
pointing oppositely to the direction of the external magnetic field. 
The spin gradually turns up when departing from the origin. 
The extension of the skyrmion is 
defined by the radius $\lambda_{sk}$, for which $m_{z}(r)$ vanishes. From \cite{Gir99}.
}
\label{skyrmion}
\end{figure}

This somewhat idealized picture in the $V_{0}$ model is distorted when the interaction is of the 
Coulomb type so that the energy becomes lowest if the number of reversed spins $K$ increases.  
This behavior can be quantitatively described 
by the energy of the generalized quasiparticles with $K$ reversed spin 
$\epsilon(K)_{\pm}=E(K,N=N_{\Phi}\pm 1)-E_{0}(N=N_{\Phi})$, where 
$E(K,N=N_{\Phi}\pm 1)$ is the lowest energy state with $K$ flipped spins. 
Therefore, $\epsilon(K)$ is monotonically decreasing. 
In a sufficiently large finite system, this means $S=0$, see Fig.~\ref{skyrmion}.  
While in a microscopic quasiparticle 
state the additional spin has to change its direction on the microscopic length scale $\ell_c$, 
just the opposite happens for large $K$, where the spins turn smoothly 
in order to minimize the exchange energy. 
Thus, it is tempting to derive a classical field theory for 
the spin magnetization ${\bf m}({\bf r})$ with the 
local constraint ${\bf m}^{2}({\bf r})=1$ whose order parameter field changes slowly 
on the magnetic length scale $\ell_c$. The order parameter field ${\bf m}({\bf r})$ 
describes the deviation of the magnetization from the ferromagnetic ground state 
${\bf m}({\bf r})=(0,0,1)$. 
A gradient expansion of the energy functional $E[{\bf m}({\bf r})]$ 
yields in leading order two terms 
\cite{SKKR93,APFGM97,BMV96}. 
The first term corresponds to a classical ferromagnetic Heisenberg model and 
the second to a topological term, see below, 
\begin{equation}
E_{0}[{\bf m}] = \frac{\rho_{s}}{2} \int d^{2} r\, ({\bf \nabla} {\bf m})^2  
- 2 \rho_{s} \int d^{2} r\, ({\bf m} \cdot [ \partial_{x} {\bf m} 
\times \partial_{y} {\bf m}]\,) \ ,
\label{nlsigma}
\end{equation}
where $\rho_{s}$ is the spin-stiffness.  The next order of the gradient expansion is of 
the Hartree type
\begin{eqnarray}
E_{c}[{\bf m}] = \frac{\lambda}{2} \int d^2 r \int d^{2}r'\ 
\frac{\rho({\bf r}) \rho({\bf r'})} {|{\bf r}-{\bf r'}|} \ .
\label{coulomb}
\end{eqnarray}
It describes the direct interaction between the local charge densities $\rho({\bf r})$ 
and is given at $\nu=1$ by 
\begin{equation}
\rho({\bf r})= \frac{1}{8 \pi} \epsilon_{ab}\, {\bf m}({\bf r}) \cdot 
[\partial_{a} {\bf m}({\bf r}) \times  \partial_{b} {\bf m}({\bf r})] \ .
\label{topological-density}
\end{equation}
The mathematical structure of (\ref{topological-density}) ensures that the integration of  
the density $\rho({\bf r})$ over the entire two-dimensional space yields the 
topological charge 
\begin{equation}
Q=\int d^{2} r\, \rho({\bf r}) \ ,
\label{charge}
\end{equation}
which is always an integer number. Therefore, $\rho({\bf r})$ is called a topological charge density \cite{Raj87}.  
Magnetic field configurations of different $Q$ can be easily distinguished by mapping the 
magnetization field from the plane onto the surface of the Riemann sphere. 
The integer $Q$ counts how often the local field wraps the surface of the sphere 
showing that configurations with different topological charges cannot be reached by 
continuous deformations.

It has been known for a long time \cite{BP75} that local extrema of 
(\ref{nlsigma}) besides the global minimum energy field configuration 
${\bf m}({\bf r})=(0,0,1)$ 
characterized by the topological invariant $Q$ exist.  Such a classical field 
configuration, called either a skyrmion ($Q<0$) or an antiskyrmion ($Q>0$), 
is shown in Fig.~\ref{skyrmion}.
The size of a skyrmion $\ell_{sk}$ is defined as the distance $\ell_{sk}$ 
between the origin of the skyrmion, where the spin points down,  
($m_{z}(0)=-1$) to
a point, where the $z$-component of the local magnetization vanishes, 
{\em i.~e.~}$m_{z}(r=l_{sk})=0$. Here, we employ the radial symmetry so that 
the magnetization vector is in the $x$-$y$-plane. 
Note the symmetry of the skyrmions with respect to translations of the 
origin as well as to rotations about an axis along the $z$-direction 
through the origin. 
In case of a non-vanishing Zeeman term, we have to add the expression
\begin{eqnarray}
E_{z}[{\bf m}] = \frac{\Delta_z}{4 \pi \ell_c^2} \int d^{2}r\, [1 -  m_{z}({\bf r})] \ .
\label{zeeman}
\end{eqnarray}
The explicit solutions minimizing the functional (\ref{nlsigma}) are known for arbitrary 
$Q$, and a parameterization for the $|Q|=1$ skyrmion located at the origin is 
\begin{equation}
{\bf m}({\bf r}) = \left( \frac{4x \lambda_{sk}}{r^2+4\lambda_{sk}^2},
\frac{4y \lambda_{sk}}{r^2+4\lambda_{sk}^2}, 
\frac{r^2 -4\lambda_{sk}^2}{r^2 +4\lambda_{sk}^2} \right) \ ,
\label{skyrmion-param}
\end{equation}
which leads to the topological charge density 
\begin{equation}
\rho(r) = \frac{1}{\pi} \frac{4 \lambda_{sk}^2}{(r^2 + 4 \lambda_{sk}^2)^2} \ 
\end{equation}
and $Q=-1$ because of (\ref{charge}). Note that the topological charge is independent 
of the scale parameter $\lambda_{sk}$. 

This independence is equally true for the energy of an 
antiskyrmion with topological charge $Q>0$ or a skyrmion with $Q<0$,  
and the energy is given by 
\begin{equation}
E_{s}=4 \pi \rho_{s} |Q| \ .
\label{skyrmion-energy}
\end{equation}

If we take into account the Hartree-term (\ref{coulomb}), a skyrmion of maximum size is preferred,  
but the topological charge and energy of the skyrmions remain unchanged. 
Therefore, without a Zeeman term, a neutral excitation 
consisting of a skyrmion-antiskyrmion pair with $Q=\pm 1$ 
has an energy $\Delta_{sk-ask}=8 \pi \rho_{s} = \lambda \tilde a(0)/2 = 
\sqrt{\pi/8} \lambda$ in case of Coulomb interaction. Most 
remarkably, this is only half of the quasielectron-quasihole pair energy 
$\Delta_{qe-qh}=\lambda \tilde a(0)$, see Tab.~\ref{tab:energies-nu-1}. 
On the other hand, in the $V_{0}$ model, there is not such an energy gain for the 
skyrmion-antiskyrmion pair since $\rho_{s}= V_{0}/(4 \pi) \lambda$ and therefore 
$\Delta_{sk-ask}=\Delta_{qe-qh} = 2 V_{0} \lambda$. This is just what we found from the discussion 
of the highly degenerate ground states on the disk. In this respect, the $V_{0}$ model is only 
instructive to link the microscopic picture with the description in terms of a 
classical order parameter field.

Since the topological charge density and the electrical charge density are identical at $\nu=1$, 
the topological charge has direct experimental consequences. This is different 
from a Heisenberg ferromagnet, where only spin, but no charge degrees of freedom 
exist. The skyrmions can be observed in transport experiments. For example, 
characteristic features of skyrmions in tunneling are predicted in the 
current-voltage-characteristic, when electronic tunneling between two parallel 2DES near 
filling factor $\nu=1$ is performed \cite{PF98}. 

In general, the Zeeman term (\ref{zeeman}) cannot be neglected, and this causes 
competition between a smooth change of the local magnetization with $K \to \infty$ for 
vanishing Zeeman term and a single spin-flip when the Zeeman term dominates.  
The ground state is a function of the effective gyromagnetic factor 
$\tilde g =\Delta_z/\lambda$. The function $\epsilon(K(\tilde g))$ determines 
the number of flipped spins in the 
ground state as well as the gap for charged excitations. 
For increasing $\tilde g$, the classical model becomes less appropriate and the 
correct excitations can only be found from variational wavefunctions, from a Hartree-Fock 
variational Ansatz, and from exact diagonalizations with different accuracy \cite{APFGM97}.  
The appropriate generalization of the classical field-theory is a formulation that 
takes into account quantum fluctuations around the classical solution \cite{AB97,Abo98}. 

Moreover, there exists a critical effective gyromagnetic factor $\tilde g_{crit}$,  
above which the skyrmions have a higher energy than the quasiparticles. 
This value was found to be $\tilde g_{crit}=0.054 $ from diagonalization 
studies on the sphere \cite{SKKR93} and variational wavefunction 
calculations \cite{APFGM97}. From a diagonalization study in the disk geometry, 
we got the value $0.041$, which is presumably smaller due to the strong finite-size 
effects in this geometry \cite{Kas96}. 
In $GaAs$, the magnetic field dependent ratio is
$\tilde g=\Delta_{z}/\lambda=0.00574 \sqrt{B[T]}$.  In typical experiments, 
it is $\tilde g < 0.054$ as in Barrett's NMR-experiment, where 
$\tilde g \simeq 0.016$ \cite{BDPWT95}. Theory predicts for 
this value of $\tilde g$ a number of 
$K \simeq 3-4$ flipped spins \cite{APFGM97,FBCM94}.  
Note that although $K$ should be quantized, such a behavior has not been seen 
in experiment so far \cite{SEPW95}. 

\begin{figure}[h]
\centerline{\resizebox{6cm}{6cm}{\includegraphics{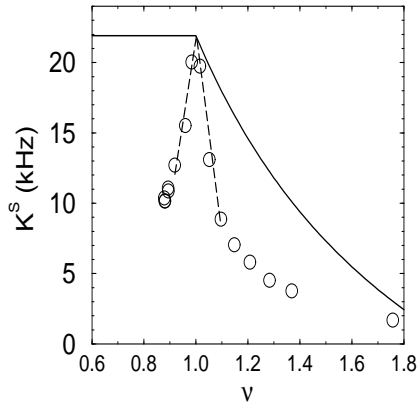}}}
\caption[]{\small
Knight-shift measurement data from an NMR-experiment by Barrett et al.~in a 
$GaAs$-$Al_{x}Ga_{1-x}As$ heterostructure at $T=1.55\ K$. 
The Knight-shift is proportional to the spin-polarization of the 2DES, and the maximum is 
due to saturated ferromagnetism at exactly $\nu=1$. The precipitous drop 
of the spin magnetization is attributed to antiskyrmions on the left and skyrmions on the 
right side. From \cite{BDPWT95}. 
}
\label{knight_shift}
\end{figure}

Besides these findings, the charge spin texture excitations 
have the following other important properties:
\begin{enumerate}
\item
The charge, as we have already seen, and statistics of skyrmions agree 
with those of the corresponding 
quasiparticles of the Laughlin wavefunction. The $\nu=1$ skyrmions 
are spin $1/2$ fermions and carry charge $\pm e$ \cite{AB97,YS96}. 
\item
The occurrence of skyrmions as the lowest charged excitations at larger, odd 
filling factors depends on the kind of interaction. For example, at $\nu=3$, $\tilde g=0$, 
and zero width $w$, skyrmions are not energetically favorable with a Coulomb interaction \cite{WS95}, 
but for a width larger than a critical value and a $\tilde g$ below a certain critical 
gyromagnetic factor, skyrmions occur as charged excitations \cite{Coo97}.
\item
There exists another critical gyromagnetic factor $\tilde {g}^{(2)}_{crit} \simeq 10^{-4}$. 
Below this value, two skyrmions can form a  bound state, but the smallness of 
$\tilde {g}^{(2)}_{crit}$ and disorder effects presumably prevent any experimental 
observation \cite{LKKS97}.  
\item
Skyrmions exist possibly also at the fractional filling factor $\nu=1/3$, 
however, $\tilde g$ is rather small \cite{KWJ96b}.   
\end{enumerate}
These theoretical results are corroborated by measurements utilizing different 
experimental techniques. 
Ground state spin magnetization measurements about $\nu=1$ were among the first 
experiments, which gave evidence of the existence of skyrmions \cite{BDPWT95}, see 
Fig.~\ref{knight_shift}.  
In a simple phenomenological model, the slope of the magnetization curve 
at $\nu=1$ can be related to the number of flipped spins $K$ of a skyrmion 
($\nu=1^{-}$) and an antiskyrmion ($\nu=1^{+}$), respectively, \cf~Eq.~(\ref{magn-skyrm}). 
Any $K>1$ leads to a drop of 
the spin magnetization that is stronger than that for simple quasiparticle 
excitations, see Subsect.~\ref{subsect:elementary}.
From the slope of the curves in Fig.~\ref{knight_shift}, an estimate of the number of flipped spins 
gives $K \simeq 3.6\pm0.3$ in good accord with theory. However, some 
uncertainty remains, in particular, the spin-polarization value at $T=1.55\ K$ 
is probably smaller than the assumed complete polarization for $T=0$ at $\nu=1$ \cite{KKB98a}. 
Second, magneto-absorption experiments, see Subsect.~\ref{subsect:comparison}, 
allow the determination of the spin-polarization as a function of $\nu$.  
Values of $K$ between $2.5$ and $3.7$ were found \cite{AGB96}. 
Third, as explained earlier, 
the standard method to determine the activation gap $\Delta$ for polarized 
FQHE states is based on the measurement of the longitudinal resistivity obeying the relation 
$\rho_{xx}=\rho^{0}_{xx} e^{-\Delta/(2 k_{B} T)}$ for sufficiently small temperatures 
in the QHE plateau region. The concomitant increase of 
$B$ while keeping $B_{z}=B_{\perp}$ fixed allows a continuous sweep of 
$\tilde g=\Delta_{z}/\lambda$ to larger values while staying at $\nu=1$ \cite{SEPW95}.   
If the number of reversed spins $K$ increases, the effective Zeeman contribution 
$|g \mu_{B}| {\bf B} {\bf S}$ is influenced via $S$, and the derivative of this 
expression with respect to $B$ allows the determination of $S$.
Schmeller et al.~found 7 reversed spins at $\tilde g \simeq 0.01$, which is close 
to the theoretical prediction. Another modification of this technique 
can be realized by exerting hydrostatic pressure, which allows to 
change continuously $\tilde g$ down to zero and to observe the 
increasing influence of spin-flips on the activation gap \cite{MPP96}.  
Despite these convincing experimental results, there is still some controversy about 
the influence of LL-mixing on the stability of these excitations \cite{KRL95}. 

So far, we reviewed the theory of the low-lying charged excitations at $\nu=1$.  
Each of these two elementary excitations forms by definition also the ground state of 
a system with an area that contains one more or one less flux quantum than at filling factor one. 
Since for small, but finite deviations from $\nu=1$ the density of skyrmions is 
small, one can assume that these particles arrange for energetic reasons in a crystal similar to the 
formation of a Wigner crystal in the low filling factor limit of a QH-system.  
There are a few predictions on this 
Skyrme crystal. Except for very small deviations from $|\nu - 1| \simeq 0.04$, where a 
triangular ferromagnetic arrangement of skyrmion centers is lower in energy, a HF-calculation 
favors a square-lattice arrangement of antiferromagnetically oriented spins at the skyrmion 
centers \cite{BFCM95,CMBFGS97}. Any of the Skyrme crystal ground states is characterized 
by the long-range order in the spatial arrangement of the skyrmion centers as well as 
by the azimuthal orientation of the in-plane component of the local spin magnetization 
near each skyrmion center.
Good agreement of a spin magnetization calculation in HF-approximation with the results 
from NMR-experiments by Barrett \ea~\cite{BDPWT95} supports 
such a picture up to deviations $|\nu-1| \simeq 0.2$, \cf~Fig.~\ref{magn_skyrm} \cite{BFCM95}.  
Moreover, a peak in the low-temperature specific heat can be related to the melting of 
the Skyrme lattice \cite{BGMSS96,BGVMS97}. 
If the skyrmion density further increases, a transition to a fluid state seems to be 
reasonable. Recently, the ground state energy of a spin-polarized and an unpolarized 
liquid was calculated in the filling factor range $2/3 \le \nu \le 1$ \cite{PP99}.  It turned out 
that below $\nu \simeq 0.92$ the spin-unpolarized liquid is favored over the square-lattice 
Skryme crystal, which questions the existence of a Skyrme crystal in such a wide range of 
filling factors.
In any case, a very rich $T-\nu - \Delta_z$ phase diagram is expected to occur, which 
comprises various quantum and thermodynamic phase transitions \cite{CMBFGS97,MM2000}. 

	\subsection{Exact neutral one-spin flip excitations at $\nu=1$}
	\label{subsect:spin-wave}

Thus far, we know the elementary charged excitations at fractional 
filling factors for spin-polarized systems and those for a model with 
spin at $\nu=1$. These charged excitations become the constituents of 
neutral excitations above the ground states. They are 
either a well separated neutral quasihole-quasielectron pair or 
a neutral skyrmion-antiskyrmion pair far apart. 
In the Landau gauge, the quasihole-quasielectron pairs are the $k \to \infty$ limit of
a whole class of excitations that describe one-spin-flip excitations 
above the ground state $|\Psi_{\nu=1} \rangle$ of a filled majority spin band at $ \nu = 1 $. 
In a finite system with $N=N_{\Phi}$ particles,  $N^2$ states   
\begin{eqnarray}
|{\bf k} \rangle = \frac{1}{\sqrt{N}} \sum_{q} e^{-iqk_x\ell_c^2} 
c_{q \downarrow}^{\dagger} c_{q+k_y \uparrow}^{\vphantom{\dagger}}|\Psi_{\nu=1} \rangle \ ,
\label{SW}
\end{eqnarray}
can be constructed.  
The summation over $q=2 \pi l/L_{y}$ ($l=0,\pm 1,\ldots$) comprises all 
momenta in $y$-direction.  
The operator acting on $|\Psi_{\nu=1} \rangle $ is proportional to the 
spin lowering operator 
$\bar{S}^{-}({\bf k})$, where the bar denotes the projection onto the LLL, and 
$\bar{S}^{-}({\bf k})= \int d^{2} r \; e^{-i {\bf r} {\bf k}}  
\bar{S}^{-}({\bf r})= \bar{S}_{x}({\bf k}) - i \bar{S}_{y}({\bf k})$ 
is the Fourier transformation of the projected spin density 
$\bar{S}^{-}({\bf r})= \Psi^{\dagger}_{\downarrow}({\bf r}) 
\Psi^{\vphantom{\dagger}}_{\uparrow}({\bf r})$. Then 
\begin{equation} 
\bar{S}^{-}({\bf k})= 
e^{-\frac{k^2}{4} \ell_{c}^{2}- \frac{i}{2} k_{x}k_{y} \ell_{c}^{2}}
\sum_{q} e^{-iq k_{x} \ell_{c}^{2}}\, c^{\dagger}_{q \downarrow} c_{q+k_{y} \uparrow}^{\vphantom{\dagger}} \ ,
\end{equation} 
and (\ref{SW}) can be written except for an irrelevant phase factor as 
\begin{equation} 
|{\bf k} \rangle = 
\frac{e^{\frac{k^2}{4} \ell_c^2}}{\sqrt{N}} \bar{S}^{-}({\bf k}) |\Psi_{\nu=1} \rangle \ .
\end{equation} 
The form of (\ref{SW}) ensures the normalization of the states $|{\bf k} \rangle$. 
One can check that these states $|{\bf k} \rangle $ are {\em exact} 
energy eigenfunctions of the LLL Hamiltonian in Landau gauge, \cf~Eq.~(\ref{ham-lll-landau}), 
\begin{eqnarray}
 H  =  - \frac{1}{2} \Delta_{z}(N_{\uparrow}-N_{\downarrow})
 +  \frac{\lambda}{2}
 \sum\limits_{p,p^{'},q \atop \sigma, \sigma^{'}}
 \tilde{W}(q,p-p^{'}) 
c^{\dagger}_{p^{\vphantom{'}},\sigma^{\vphantom{'}}}
c^{\dagger}_{p^{'}, \sigma^{'}} 
c^{\vphantom{\dagger}}_{p^{'}+q,\sigma^{'}}
c^{\vphantom{\dagger}}_{p-q, \sigma^{\vphantom{'}}} \ . 
\label{H-Landau}
\end{eqnarray}
Thus, the excitation energies $\epsilon_{SW}({\bf k})$ 
above the spin-polarized ground state $|\Psi_{\nu=1} \rangle $ are described in the 
thermodynamic limit by 
the continuum of excitations 
\begin{eqnarray}
\epsilon_{SW}({\bf k})  & = & \langle {\bf k} | H | {\bf k} \rangle 
- \langle \Psi_{\nu=1} | H | \Psi_{\nu=1} \rangle \nonumber \\
& = & \Delta_{z} + \lambda  (\tilde{a}(0) - \tilde{a}({\bf k}))\;.
\label{DISP}
\end{eqnarray}
The quantity 
\begin{equation}
\tilde{a}({\bf k})= \int \frac{d^{2}{\bf q}}{(2\pi)^2}
\tilde{V}({\bf q})e^{-\frac{q^2\ell_c^2}{2}} 
e^{i{\bf q} \cdot ({\bf e}_z \times {\bf k}) \ell_c^2}.
\label{aofk}
\end{equation}
is already known as the band formed by the eigenvalues of two attracting particles 
in the LLL, 
see Eq.~(\ref{attraction-two-particles}). 
The three terms can be easily interpreted. First, $\Delta_z$ is the one-particle 
energy that is necessary to promote an electron to the minority spin band. Second, 
$ \lambda \tilde{a}(0)$ is the exchange energy that is lost due to one electron less in 
the majority spin band. It is equivalent to the gross quasihole energy $\epsilon_{-}(1)$ 
in the majority spin band, see 
the Coulomb entry in Tab.~\ref{tab:energies-nu-1}. The 
third term, $ \lambda \tilde{a}({\bf k})$, describes the attractive interaction between the 
electron in the minority spin band and the hole in the majority spin band and the energy 
equals that of  two  spinless, but differently charged fermions attracting 
each other, {\em cf.~}Eq.~(\ref{energy-attractive}). 
The two-particle complex of an electron and a hole 
can also be viewed as a {\em spin-exciton} or {\em  magneto-exciton}, where 
the valence and conduction bands  
are identified as the dispersionless majority and minority spin bands of the LLL.  

Eq.~(\ref{DISP}) describes for $|{\bf k}| \ell_c \gg 1$ the single 
particle character of the excitations, namely the interaction between the minority electron 
and majority hole whose distance is given by $|{\bf k}| \ell_c^2$. This can either be seen 
from (\ref{attraction-two-particles}) or by expanding (\ref{DISP}) for large $k \ell_{c}$. For 
example, in case of an  
unscreened Coulomb interaction, this yields 
\begin{equation} 
\epsilon_{SW}({\bf k}) = \Delta_{z} + \lambda \left( a(0) - \frac{1}{k \ell_c} \right) + 
{\cal O} \left(\frac{1}{k \ell_c^{3}} \right) \ .
\label{asymptotic}
\end{equation}
The second term is the energy of the particle-hole pair excitation $\Delta_{qe-qh}$ 
at infinite distance and the lowest order correction due to the 
attractive Coulomb interaction 
between charges $\pm e$ of different sign at dimensionless distance $k \ell_c$.
It is $\Delta_{qe-qh}= \lambda a(0) = \lambda \sqrt{\pi/2}$.  
This behavior is analogous to that for the large distance limit of attractive 
quasiparticle interaction at $\nu=1/q$, where the leading correction reflects the 
fractional charge of the quasiparticles by a prefactor $\nu^{2}$. 
Below, in our diagrammatic analysis, we will see that it is necessary to work with a 
screened interaction $\tilde a(k)$ instead of $a(k)$.  

Of course, corrections to the spin-gap enhancement occur when one includes higher 
Landau level contributions. Then, the expansion parameter is $\lambda/(\hbar \omega_{c})
\propto 1/\sqrt{B}$, which vanishes in the LLL approximation.  
Besides analytical investigations, which are based on the leading order result of a 
perturbation theory \cite{SKKR93} and an RPA-calculation \cite{SMG92},  
the reliability was checked by means of a variational quantum Monte Carlo 
(VQMC) approach \cite{KRL95}.  In that case, variational wavefunctions for the ground state 
as well as for the quasiparticles with a spin flip were constructed 
by multiplying these wavefunctions at $\nu=1$ by a Jastrow function  
$\prod_{i<j} e^{u(r_{ij})}$ mixing in higher LLs. 
The perturbation theory by Sondhi et al.~\cite{SKKR93} leads in zeroth order of 
$1/\sqrt{B}$, which can be written as the linear contribution in the 
dimensionless parameter $r_{s}=r_{0}/a_{B}$, to a particle-hole gap $\Delta_{qe-qh}$ with 
\begin{equation} 
\Delta_{qe-qh} = \Delta_{z} + 
\lambda ( \frac{\sqrt{\pi}}{2} - \frac{0.58}{\sqrt{2}} r_{s}) \ .
\label{spin-gap-higher-LL}
\end{equation}
In the high field limit $r_{s} \to 0$, we recover the 
LLL result, but for 
realistic values at $\nu=1$ and an electron density $n \sim 10^{11}\ cm^{-2}$,  
it is in $GaAs$ $r_{s} \sim 1.8$.  The Monte Carlo approach results in a reduction 
of the LLL gap of about 45\% to a value of approximately $0.7 \lambda$ 
compared with the LLL value $1.253 \lambda$. 
Thus, in this parameter range the MC results 
are more reliable than the perturbation theory, which  
overestimates the influence of mixing for values larger than $r_{s} \simeq 1.2$ 
leading to a value $\Delta_{qe-qh}=0.515 \lambda$. 
In summary, the VQMC interpolates between the perturbation results reliable at 
small $r_{s}$ and the RPA \cite{SMG92}, which is particularly suited to large 
values of the electron parameter $r_{s}$.

The experimental determination of the spin gap at $\nu=1$ 
by transport measurements of the activation gap 
\cite{UNHF90}, magneto-capacitance measurements \cite{DSA97}, and 
optical methods \cite{KT96} 
do not support this ideal picture: first, the gap experimentally found, is much 
smaller, and, secondly, a linear instead of a square root dependence on $B$ is 
measured. The main effect is attributed to the influence of disorder, 
but theoretical investigations could not support such a conjecture so far.  
Note, however, that these studies did not aim at the detection of 
skyrmion-antiskyrmion pairs.

For small values $|{\bf k}| \ell_c \ll 1$, {\em i.~e.~}in 
the long-wavelength limit, the collective character of 
the excitations are of spin-wave character, and an expansion of 
(\ref{DISP}) up to $k^2$ yields 
\begin{equation} 
\epsilon_{SW}({\bf k}) = \Delta_z + 4 \pi \rho_s \ell_c^2 k^2 \ .
\label{longwavelengths}
\end{equation}
The spin-stiffness $\rho_s = \lim_{k \to 0} \lambda 
(\tilde{a}(0) - \tilde{a}({\bf k}))/(4 \pi \ell_c^2 k^2 )$, which 
appears as a phenomenological constant in field-theories, \cf~Eq.~(\ref{nlsigma}),  
\cite{LK90,SKKR93}, has its microscopic origin in the electron-electron interaction. 
Note that in the limit ${\bf k} \to 0$ the excitation energy is that of the 
non-interacting system in accordance with Kohn's theorem \cite{Koh61}.  
\begin{figure}[t]
\centerline{\resizebox{6cm}{6cm}{\includegraphics{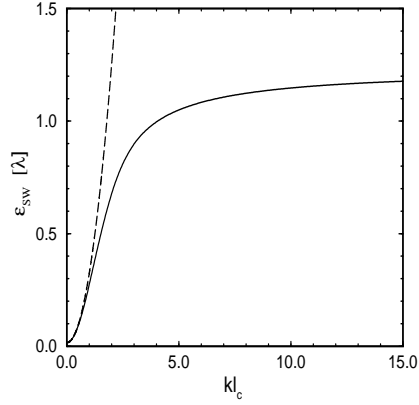}}}
\caption[]{\small
Comparison of the exact dispersion curve (solid line) of Eq.~(\ref{DISP}) with 
the spin-wave approximation (long dashed curve) 
in (\ref{longwavelengths}). Note the small shift at $k=0$ due to the non-zero 
Zeeman term $\Delta_z= 0.016 \lambda$ in comparison with the spin-wave 
bandwidth $\lambda \tilde{a}(0)$. 
A static screening wavevector $k_{sc}=0.01 \ell_{c}^{-1}$ is used. 
Around $k \ell_{c} \sim 1$, the two curves bifurcate indicating the 
diminishing collective character of the one-spin flip excitations for 
$k \ell_c > 1$ and the emergence of the minority spin-down 
electron-majority spin-up hole pair of single particle character. From \cite{KPM2000}. 
}
\label{disp}
\end{figure}
There are various other ways to derive the dispersion relation (\ref{DISP}). 
Among these is the thermodynamic Green's function formalism where one has to 
calculate the spin-spin correlation function in the LLL. The identification 
of the GF's pole in the limit $T \to 0$ leads again to the dispersion 
relation of Eq.~(\ref{DISP}). 

The excitations of Eq.~(\ref{SW}) describe for each ${\bf k}$ 
bosonic excitations above the spin-polarized ground state since $S=S_{z}=N/2-1$. 
In systems exhibiting saturated ferromagnetism, such 
low-lying excitations are known as magnons \cite{Mat85a}. Therefore one is inclined  
to explain the thermodynamics of the QHF at $\nu=1$ on the basis of non-interacting 
bosons, see Subsect.~\ref{subsect:elementary}. 
However, such an approximation becomes more questionable if the number 
of these excitations is too large and the interactions between the bosonic spin-waves 
has to be taken into account. Such an effect can be seen from the diagonalization of 
a finite number of electrons on a sphere at $\nu=1$ identifying one and 
two spin-wave states \cite{KPM2000,Kas2000}. 
Consequently, spin-wave spin-wave interaction plays a role in thermodynamics, especially, 
when the temperature increases and therefore limits the validity of a free magnon theory.

	\subsection{Elementary theories of spin magnetization around $\nu=1$}
	\label{subsect:elementary}

The knowledge of the ground state of a QH many-particle system allows the calculation 
of the spin magnetization $M(\nu,T=0))$ in dependence on the filling factor. But 
the investigation of the thermodynamic behavior of $M(\nu,T)$ over the whole 
temperature range makes information about all excited states necessary. In general, 
such a task is too ambitious to be rigorously solved,  
and one needs approximative schemes to calculate 
thermodynamic quantities in a certain temperature range. 
 
\begin{figure}[t]
\centerline{\resizebox{6cm}{6cm}{\includegraphics{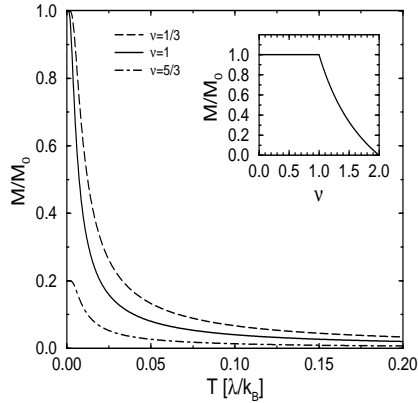}}}
\caption[]{\small 
The temperature dependence of the spin magnetization of non-interacting 
electrons in the lowest Landau level at filling factors $1/3, 1$, and $5/3$ 
for $\Delta_z=0.016 \lambda$. The inset shows the ground state magnetization as 
a function of $\nu$, {\em cf.~}Eq.~(\ref{magn-free-nu-0}). From \cite{Kas2000}.
}
\label{magn_free}
\end{figure}

Due to the two-dimensionality, spontaneous magnetization at vanishing 
Zeeman term can only occur in the ground state, and we saw that the 
existence of a QHFM at $\nu=1$ is caused by interaction.
On the other hand, a non-zero Zeeman term destroys 
the $SU(2)$ symmetry and allows also for finite temperatures a non-vanishing 
magnetization. Before we study the influence of the interaction on the 
temperature dependent spin magnetization, we set out  from 
the temperature and filling factor dependent spin magnetization of 
non-interacting electrons, which is given by   
\begin{equation}
M^{(0)}(\nu,T=0) = M_{0} \left (\Theta(1-\nu) + \Theta(\nu-1)(\frac{2}{\nu}-1) \right ) \ .
\label{magn-free-nu-0}
\end{equation}
Eq.~(\ref{magn-free-nu-0}) shows constant magnetization as long as the 
majority spin band is filled up, but starts to decrease with the occupation of the minority band and 
vanishes for the completely filled lowest orbital LL at $\nu=2$. 
The generalization to arbitrary temperature yields 
\begin{equation}
M^{(0)}(T,\nu)= \frac{N g \mu_{B}}{2}\; 
\frac{\sinh(\beta \Delta_z/2)}{z+\cosh(\beta \Delta_z/2)} \; , 
\label{magn-free-nu-T}
\end{equation}
where the fugacity $z=e^{\beta \mu}$ ($\mu$ - chemical 
potential) is related to the filling factor $\nu$ by 
\begin{equation}
z(\nu,T,\Delta_z) = 
\frac{1}{(2-\nu)} \left (\sqrt{(1-\nu)^2 \cosh^2\frac{\beta \Delta_z}{2}+\nu(2-\nu)}
 - (1-\nu) \cosh\frac{\beta \Delta_z}{2} \right ) \ ,
\label{fugacity}
\end{equation}
see Fig.~\ref{magn_free}.
For filling factor $\nu=1$, we get a closed expression due to the 
{\em independence} of the chemical potential of temperature and Zeeman energy so that 
$\mu=0$ and 
\begin{equation}
M^{(0)}(\nu=1,T)= M_{0} \tanh(\beta \Delta_z/4) \ .
\label{magn-free-nu=1}
\end{equation}
Note that the magnetization in a QH-system drops stronger than for a localized 
spin $1/2$ system corresponding to a Heisenberg model on a lattice with zero 
coupling ($J=0$), where $M/M_{0}=\tanh(\beta \Delta_z/2)$. This is a 
consequence of the existence of spin as well as charge degrees of freedom, 
{\em i.~e.~}of the itinerant character of the QH-system. 
Indeed, an electron excited into the minority spin band 
can concomitantly reverse its spin and occupy any of the $N_{\Phi}=N$ orbital 
one-particle states. In total, in a finite system $N^{2}$ excitations above the 
ground state are possible. 

From our previous discussion, we know that the zero temperature magnetization at $\nu=1$ 
is exact since all spins are polarized and show therefore saturated ferromagnetism. 
However, the non-interacting result of (\ref{magn-free-nu-0}) is in 
obvious contradiction to our findings on the influence of the charged spin texture excitations, 
the skyrmions, on the ground state magnetization vs.~$\nu$ shown 
in Fig.~\ref{knight_shift}. 
From a phenomenological point of 
view, the creation of one neutral quasihole at filling factor one changing the filling 
factor infinitesimally to $1-\delta \nu$ leads to a rearrangement 
of the ground state, which leaves in average 
${\cal A}$ holes in the majority spin band, {\em i.~e.~}$K={\cal A}-1$ spins are 
flipped in average. Analogously, 
the creation of a neutral quasielectron leads to $K={\cal S}$ reversed spins in 
the minority spin band. The values 
${\cal A}$ and ${\cal S}$ depend on the effective gyromagnetic factor 
$\tilde g=\Delta_{z}/\lambda$. 
Thus, the spin magnetization about $\nu=1$ can be expressed as  
\begin{equation}
M(\nu,T=0) = M_{0} \left[ \Theta(1 - \nu) (\frac{2}{\nu}(1-{\cal A}) - (1-2{\cal A})) + 
\Theta(\nu-1) (\frac{2{\cal S}}{\nu}+1-2 {\cal S}) \right ].
\label{magn-skyrm}
\end{equation}
This relation breaks down when the underlying assumption that independent skyrmions and antiskyrmions, 
respectively, form the 
ground state becomes invalid. 
Eq.~(\ref{magn-free-nu-0}), which is actually only true if $\Delta_z \gg \lambda$, 
{\em i.~e.~}when the one-particle character dominates the magnetization curve, 
can be recovered when setting ${\cal A}={\cal S}=1$ in (\ref{magn-skyrm}). 
For any value ${\cal A}$ and ${\cal S}$ larger than one, the magnetization drops 
precipitously from the 
maximum polarization at $\nu=1$. The deviation of the relative magnetization $M/M_{0}$ from the 
maximum value one is therefore up to second order in small values of $|\nu-1|$:  
$2({\cal A}-1)(|\nu-1| + |\nu-1|^2 + \ldots)$ for $\nu<1$ and 
$2 {\cal S}(|\nu-1| - |\nu-1|^2 + \ldots)$ for $\nu>1$, respectively. 
Recent NMR-experiments by Barrett {\em et al.~}\cite{BDPWT95} 
measure the Knight shift of $^{71}Ga$, which is due to the additional magnetic field caused by 
the spins of the 2DES. This shift corresponds to the spin-polarization per particle around $\nu=1$ at 
low temperatures, see Subsect.~\ref{subsect:comparison}. Comparison with 
the value of the left and right first derivative of the experimentally found 
spin magnetization curve at $\nu=1$ (\ref{magn-skyrm}) 
allows the determination of ${\cal A}$ and ${\cal S}$, respectively, see Fig.~\ref{knight_shift}.
Moreover, even the stronger drop in Barrett's curve 
for $\nu<1$ breaking the symmetry of the magnetization 
about $\nu=1$ in leading order of $|\nu-1|$ can be explained within this simple theory. 
The determined value from the curve are ${\cal A}={\cal S}=3.6 \pm 0.3$ for 
an experimental ratio $\Delta_z/\lambda=0.016$. 
\begin{figure}[t]
\centerline{\resizebox{6cm}{6cm}{\includegraphics{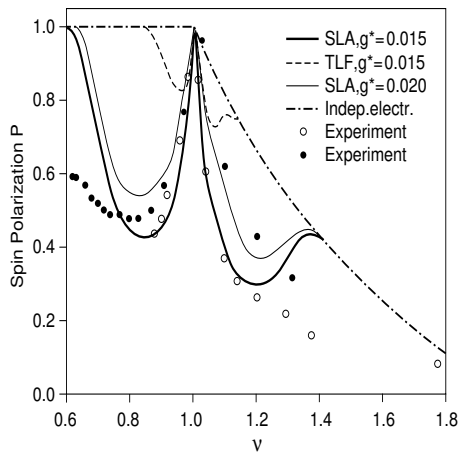}}}
\caption[]{\small
The ground state magnetization near $\nu=1$ for Skyrme lattice states at typical 
values of $\tilde g=\Delta_z/\lambda$. The open ($7.05\ T$) and closed ($B=9.39\ T$) 
circles are the experimental results of 
Barrett et al.~\cite{BDPWT95}. The legend indicates two different Skyrme lattice 
states: the SLA state is a square lattice state with opposing skyrmion orientations;
the TLF state is a triangular lattice with aligned skyrmion orientation. From \cite{BFCM95}.
}
\label{magn_skyrm}
\end{figure}
These results corroborate strongly the existence of finite-size skyrmions in a QH-system near $\nu=1$. 
The found values are in fair agreement with theoretical 
estimates that predicted an effective spin of about 3.5 for this value of $\tilde g$ \cite{FBCM94}.  
The experimental values can be fitted within this simple theory for $0.9 < \nu < 1.1$ 
and give a hint of the range of validity of the independent skyrmion assumption. 

A more ambitious theory to explain the ground state magnetization near $\nu=1$ 
is based on the assumption of a Skyrme lattice, which we mentioned in conjunction with the 
discussion of skyrmions \cite{BFCM95}.  
Results from HF-calculations are shown for a square and a triangular lattice in Fig.~\ref{magn_skyrm}.
In the following, we will focus on the spin magnetization at exactly $\nu=1$ whose study is 
essentially facilitated by the knowledge of the spin-polarized ground state. This 
is the reason why more ambitious theories on the temperature-dependence of the magnetization 
for filling factors unequal one, except for some FQHE filling factors, are still missing.

Before we start to develop a many-particle theory of the magnetization 
at $\nu=1$, let us write down the spin magnetization for the case 
that the only excitations are those non-interacting bosons, which we  
described as one-spin-flip excitations in the last Subsection.
Starting from the expression for the temperature dependent spin magnetization whose saturated magnetization 
value at zero temperature $M_{0}$ is diminished by the number of occupied 
bosonic spin-waves with dispersion $\epsilon_{SW}({\bf k})$ 
\begin{equation}
M(T)=M_{0} - g \mu_B \sum_{{\bf k}} n_{B}(\epsilon_{SW}({\bf k})) \ ,
\end{equation}
we get after approximating the dispersion relation by Eq.~(\ref{longwavelengths}) 
\begin{equation}
M(T)=M_{0} \left [1 - \frac{ k_{B} T}{4 \pi \rho_{s} }
\; \ln \left (\frac{1}{(1 - e^{-\beta \Delta_z})} \right ) \right ] \;  .
\label{magn-magnons}
\end{equation} 
This result is formally equivalent to the expression one gets for a localized 
spin system described by a Heisenberg model with spin $S$ in an external 
magnetic field of strength $\Delta_{z}/|g \mu_{B}|$ when $k_{B} T \ll 2 \pi S J $, 
{\em i.~e.~}for 
low temperatures and spin $S=1/2$. Hence, (\ref{magn-magnons}) represents the 
two-dimensional (linear $T$-term) analog in an external magnetic field 
of the $T^{3/2}$--Bloch law in three dimensions and 
suggests that for low temperatures the thermodynamics is governed by independent 
spin-waves \cite{Mat85a}. The paramagnetic behavior for temperatures greater than zero in the 
limit of $\Delta_z=0$ is signaled by the divergence of the logarithm in the expression. 

\begin{figure}[t]
\centerline{\resizebox{6cm}{6cm}{\includegraphics{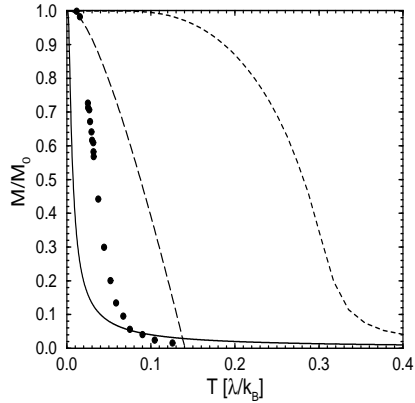}}}
\caption[]{\small
The temperature dependence of the spin magnetization at $\nu=1$ in 
the self-consistent Hartree-Fock approximation (dashed line), see below, and the 
result assuming free magnons (long-dashed curve). It is 
$\Delta_z=0.016 \lambda$. For comparison, the results for non-interacting 
electrons (solid line, see also Fig.~\ref{magn_free}) and of 
Barrett's experiment \cite{BDPWT95} (filled circles) are added. From \cite{Kas2000}.
}
\label{magn_magnons}
\end{figure}

For two temperature regions, (\ref{magn-magnons}) simplifies to 
\begin{eqnarray}
M(T) =   M_{0} \left [1 - \frac{k_{B} T}{4 \pi \rho_{s}} e^{- \beta \Delta_z} \right ]
& & k_{B} T \ll  \Delta_{z}  \nonumber\\
M(T)  =  M_{0} \left [1 - \frac{k_{B} T}{4 \pi \rho_{s}} 
\ln(\frac{k_{B} T}{\Delta_{z}}) \right]   & &
 \Delta_z \ll k_{B} T \ll 4 \pi \rho_{s}  \ .
\label{magn-low-temperature}
\end{eqnarray}
In the first case, the magnetization is governed by activated behavior for 
temperatures much smaller than the activation threshold $\Delta_z$ due to the non-zero 
Zeeman term, the second case describes an approximately linear regime.
The curve for the independent magnon approximation is shown in Fig.~\ref{magn_magnons}.
At temperatures above $0.14 \lambda/k_{B}$, the magnetization becomes 
negative and the approximation breaks down. There are some obvious deficiencies, which  
are well known from the study of localized spin systems in three 
dimensions without magnetic field \cite{Dys56b}. First, the number of fermionic excitations is 
restricted by the Pauli principle, while the occupation number of a 
bosonic state $|{\bf k} \rangle $ can exceed one.
Second, the interaction of spin-waves, whose existence was shown above, becomes  
at higher temperatures more important as the number of excited spin-waves strongly increases. 
This leads inevitably to an overestimate of the decrease of the magnetization and to negative values of $M$.   
Despite the restricted value of this simple theory, it will give us some guidance 
for the low-temperature behavior of our many-particle approach discussed in 
Subsect.~\ref{subsect:scattering}.

	\subsection{Hartree-Fock approximation and beyond: electron scattering on spin-waves}
	\label{subsect:scattering}

\begin{figure}[t]
\centerline{\resizebox{6cm}{6cm}{\includegraphics{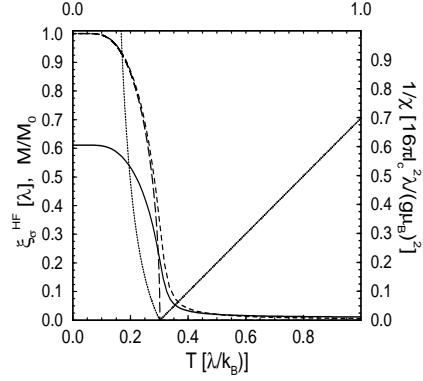}}}
\caption[]{\small
The self-consistent Hartree-Fock eigenenergy $\xi_{\downarrow}^{HF}$
(solid line) as a function of temperature at $\nu =1$;
$\xi_{\uparrow}^{HF}= -\xi_{\downarrow}^{HF}$ because of particle-hole
symmetry at $\nu=1$. The magnetization $M=M_0
(\nu_{\uparrow}-\nu_{\downarrow})$
within the SHF is depicted for
$\Delta_z=0.016 \lambda$ (dashed curve) and $\Delta_z=0.0 \lambda$
(long dashed curve),
respectively.
Note the finite magnetization at low $T$ in the latter case
incorrectly indicating the existence of an ordered phase for $T$ below
$T_c=\tilde{a}(0) \lambda/(4 k_B)$. The uniform static inverse susceptibility
is plotted as a dotted line in units of $(16 \pi l_c^2 \lambda)/(g \mu_B)^2$ for
$T \ge T_c$ as well as for $T < T_c$. From \cite{KPM2000}.
}
\label{schf}
\end{figure}

Before we outline our many-particle theory taking into account the spin-wave excitations,  
let us analyze the consequences of the self-consistent Hartree-Fock approximation (SHF) at 
finite temperatures. Obviously, the SHF of Eq.~(\ref{SC}) reduces 
to an algebraic equation for $\xi^{HF}_{\sigma}$ to be solved numerically for a 
given filling factor $\nu=\nu_{\uparrow} + \nu_{\downarrow}$. With the particle-hole symmetry 
at $\nu=1$, the temperature dependent SHF energies $\xi^{HF}_{\sigma} = \epsilon^{HF}_{\sigma} 
- \mu$ satisfy the relation 
$\xi^{HF}_{\uparrow}(T) = -\xi^{HF}_{\downarrow}(T) < 0$, see Fig.~\ref{schf}. Note that 
the chemical potential in the SHF at $\nu=1$ remains 
{\em independent} of temperature {\em and} Zeeman energy 
and is given by $\mu= -\lambda \tilde{a}(0)/2$. The temperature dependent energy difference 
$\xi^{HF}_{\uparrow}-\xi^{HF}_{\downarrow}$ is the exchange splitting 
that shows oscillating behavior in dependence on the filling factor \cite{AU74}.  
Here, we are rather interested in its temperature dependence for fixed $\nu=1$. 
The gap starts at 
zero temperature from its maximum value $2 \xi^{HF}_{\downarrow} 
= \Delta_z + \lambda \tilde{a}(0)$ that is clearly dominated by interaction 
and reaches for $T \to \infty$ the bare gap value $\Delta_z$ reflecting 
the non-interacting situation. Experimentalists relate this enhanced spin gap to an effective 
$g$-factor, which can be determined from activation measurements \cite{UNHF90}.  
However, comparison shows that the values found theoretically are too large and 
need to be corrected by incorporating the influence of 
the finite width of the sample, screening effects, Landau level mixing as well as 
disorder, \cf~Subsect.~\ref{subsect:comparison}.

The spin magnetization in the SHF yields 
\begin{equation}
M^{HF}(T,\nu=1)=M_{0}(\nu^{HF}_{\uparrow} -\nu^{HF}_{\downarrow}) 
= M_{0}  \tanh(\beta \xi_{\downarrow}^{HF}/2) 
\label{magn-HF}
\end{equation}
and 
is shown in Fig.~\ref{schf}. Not surprisingly for a mean-field theory, the 
SHF exhibits in the limit $\Delta_z=0$ {\em incorrectly} 
spontaneous magnetization due to a continuously broken symmetry 
in contradiction to the Mermin-Wagner theorem prohibiting this in dimension two 
for temperatures larger than zero.
For non-zero $\Delta_z$, this leads to a softening of the magnetization curve, and 
the inclination point is a remnant of the mean-field phase-transition. 
In any case, also for non-zero $\Delta_z$, the magnetization is much too large 
when compared with experiment, see Fig.~\ref{magn_magnons}, 
and such a behavior will prevail as long as the interaction is the dominating 
energy scale.
Since the temperature dependent $\xi_{\sigma}^{HF}$ is proportional to the order parameter, 
the spin magnetization, 
the mean-field critical exponent $\beta=1/2$ as well as all the other mean-field values can be found: 
$\alpha=0$ for the specific heat, $\gamma=1$ for the divergence of the spin-susceptibility and 
$\delta=3$ for the vanishing of the symmetry breaking field at the mean-field critical temperature. 
Formally, the discussion is analogous to the Weiss theory of ferromagnetism, where 
one finds a similar equation for the order parameter and whose results are 
independent of the spatial dimension. At 
temperatures above the mean-field critical temperature $T^{HF}_{c}=\lambda \tilde{a}(0)/(4 k_{B})$,  
the vanishing magnetization leads to paramagnetic behavior, while for $T<T^{HF}_{c}$ 
a thermodynamically stable solution unequal zero signals ferromagnetism. 
Note that the role of the symmetry breaking field is played by the Zeeman energy $\Delta_z$ and not by 
the external magnetic field.

We see that HF theories are useful in predicting the ground state 
and the spin-flip excitations from the filled majority spin band to the 
minority spin band at $\nu=1$. However, they yield rather poor results for 
the thermodynamic behavior. In particular, simple theories 
for the temperature dependent spin magnetization at $\nu=1$ are insufficient: 
neither the non-interacting picture with its too strong drop of the magnetization at temperatures larger than the 
bare Zeeman energy $\Delta_z$ of about $2\ K$ nor the SHF picture provides a trustworthy result when compared 
with experiment. 
In the latter case the exchange enhanced one-particle gap $2 \xi_{\downarrow}^{HF}$ 
that only slowly decreases with temperature grossly overestimates the 
magnetization at temperatures below $T_{c}^{HF}$. 
Therefore, the magnetization curves of these two theories represent only 
lower and upper bounds for any improved theory as well as 
for the experimental data. The latter is even true if we keep in mind that in 
experiments additional corrections have to be taken into account, 
{\em e.~g.~}the finite width in $z$-direction. 
The SHF theory applied to the quantum Hall system at $\nu=1$ has much in common with 
mean-field theories of itinerant magnetism like the Stoner theory in 
the Hubbard model \cite{Mor85}. 

The improvement we propose accounts for the fact that spin-wave excitations are 
important for the suppression of the magnetization with increasing temperature \cite{KPM2000}.  
We will systematically incorporate this mechanism in our diagrammatic treatment 
going beyond the SHF, 
{\em i.~e.~}we will identify those diagrams in an approximation of 
the self-energy $\Sigma(\omega)$ 
that are responsible for such a mechanism. This approach is similar to 
an approximation that was applied to a single-band Hubbard model with strongly 
ferromagnetic ground states at zero temperature by Hertz and Edwards \cite{HE73}. 

Technically, we calculate the correction 
$\tilde{\Sigma}_{\sigma}(\omega)$ to the SHF self-energy $\Sigma^{HF}_{\sigma}$ so that 
$\Sigma_{\sigma}(\omega) = \Sigma^{HF}_{\sigma} + \tilde{\Sigma}_{\sigma}(\omega)$.  
We express the self-energy by the four-scattering vertex 
$\Gamma_{\sigma,\sigma'}^{(4)}(1,2,3,4)$, which is calculated in a ladder approximation 
exactly describing the interaction between a single electron in 
the minority spin-down band and a single hole in the majority spin-up band and vice versa.  
Since the pair-propagator $\bar{\chi}_{\sigma,\sigma'}$ enters the four-scattering vertex 
as a product of two GFs, the four quantities, \ie~the GF ${\cal G}_{\sigma}$, 
the corrected self-energy $\tilde{\Sigma}_{\sigma}$, the 
four-scattering-vertex $\Gamma^{(4)}_{\sigma, \sigma'}$ and 
the pair-propagator $\bar{\chi}_{\sigma,\sigma'}$  form a closed set of 
equations, \cf~Eq.~(\ref{system}).
Note that the diagrammatic treatment calculates corrections to the SHF-GF and not 
as usually to the bare GF. 
\begin{figure}[h]
\centerline{\resizebox{10cm}{3cm}{\includegraphics{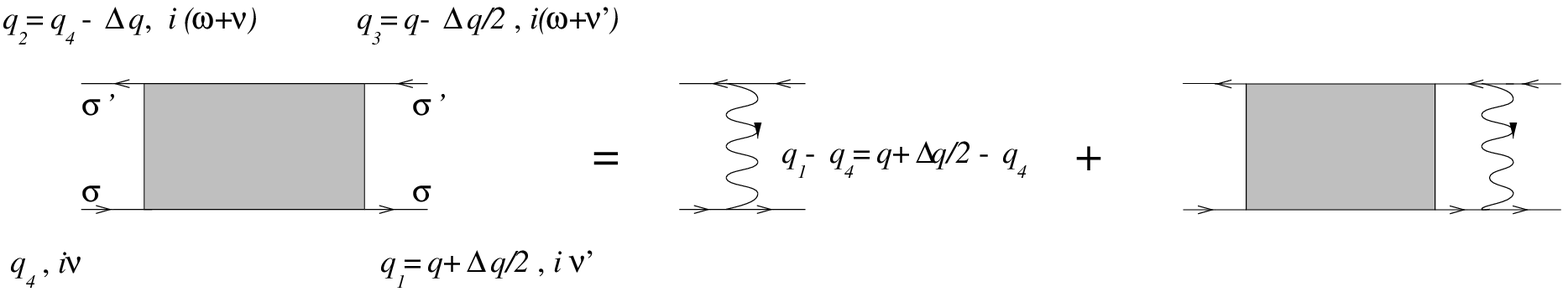}}}
\caption[]{
The self-consistent integral equation for the scattering vertex 
$\Gamma^{(4)}_{\sigma,\sigma'}$ evaluated 
in the particle-hole ladder approximation where $\sigma=\uparrow$ and 
$\sigma'=\downarrow$. The first order term, the interaction vertex,  
has to be subtracted since it leads to a tadpole diagram for the self-energy, which has to be 
omitted due to the neutralizing background charge. 
}
\label{leit}
\end{figure}
The essential step is the summation of the particle-hole ladder by 
solving the Bethe-Salpeter integral equation for the scattering vertex shown 
in Fig.~\ref{leit} 
\noindent
\begin{eqnarray}
\Gamma_{\sigma,\sigma'}^{(4)}(q + \Delta q/2 - q_4,\Delta q; i \omega_n)  = 
\lambda \tilde{W}(q+ \Delta q/2 - q_4,\Delta q)  \nonumber \\
  +  \lambda \bar{ \chi}_{\sigma,\sigma'}(i \omega_n)
\ell_c \int_{-\infty}^{\infty} dq'\; \tilde{W}(q-q',\Delta q)
\Gamma_{\sigma,\sigma'}^{(4)}(q'+ \Delta q/2 - q_4,\Delta q; i \omega_n) \; ,
\label{BS}
\end{eqnarray}
where the momentum part of the scattering vertex solely depends on 
the transfer of momentum $q_1-q_4=q+\Delta q/2 -q_4$ of one particle 
($q \equiv (q_1 + q_3)/2, \Delta q \equiv q_1-q_3$)  
and the difference of momentum between one outgoing and the other ingoing particle $\Delta q$. 

Denoting $\tilde{\Gamma}_{\sigma,\sigma'}^{(4)}(i\omega_{n})$ as the four scattering vertex integrated 
over the two-dimensional momentum dependence of ${\bf k}$ our approximation can be summarized in 
the following set of four coupled equations 
\begin{eqnarray}
({\cal G}_{\sigma}(i\nu_{n}))^{-1} - ({\cal G}^{HF}_{\sigma}(i\nu_{n}))^{-1}
& = & - \tilde{\Sigma}_{\sigma}(i \nu_{n}) \nonumber \\
\tilde{\Sigma}_{\sigma}(i \nu_n) & = & \frac{1}{\beta} \sum_{\omega_n} 
\tilde{\Gamma}_{\sigma, \sigma'}^{(4)}(i \omega_n) {\cal G}_{\sigma'}(i(\nu_n 
+ \omega_n)) \nonumber \\ 
\tilde{\Gamma}_{\sigma,\sigma'}^{(4)}(i \omega_n) & = &
2 \pi \ell_c^2\int \frac{d^2 {\bf k}}{(2 \pi)^2} \biggl\lbrace 
\frac{\tilde{a}^{2}({\bf k})\lambda^2 \bar{\chi}_{\sigma,\sigma'}(i \omega_n)}
{(1 - \tilde{ a}({\bf k})\lambda \bar{\chi}_{\sigma,\sigma'}(i \omega_n))} 
\biggr\rbrace \nonumber \\ 
\bar{\chi}_{\sigma,\sigma'}(i \omega_n) & = & - \frac{1}{\beta}
\sum_{i \nu_{n}} {\cal G}_{\sigma}(i \nu_{n}) {\cal G}_{\sigma'}(i (\nu_{n} + \omega_{n})).
\label{system}
\end{eqnarray}
The self-consistent solution of this equation does not have stable solutions at low 
temperatures as was shown by Haussmann in conjunction with the problem of 
tunneling between two layers \cite{Hau96,Hau97b}.  
Our treatment is based on an approximation, where the propagator in the 
pair-propagator and the self-energy is evaluated with the SHF Green's function 
(\ref{GHF}). This leads to the correct low-temperature behavior as we see below. 
For example, the resulting self-energy correction for spin up 
${\tilde \Sigma}_{\uparrow}(i \nu_n)$ is 
\begin{eqnarray}
\tilde{\Sigma}_{\uparrow}(i\nu_{n}) = \lambda^{2}(\nu_{\uparrow}^{HF} - 
\nu_{\downarrow}^{HF})
\int_{0}^{\infty} d(\frac{k^2\ell_c^2}{2}) \tilde{a}^{2}(k)
\frac{\lbrace n_{B}(\tilde{\epsilon}_{SW}(k)) + n_{F}(\xi_{\downarrow}^{HF})\rbrace }
{(i \hbar \nu_{n} + \tilde{\epsilon}_{SW}(k) - \xi_{\downarrow}^{HF})} \; .
\label{SE1}
\end{eqnarray}
Here, the Bose-Einstein distribution function $n_{B}({\tilde \epsilon}_{SW}(k))$ 
of the temperature-dependent spin-wave dispersion 
\begin{equation}
\tilde{\epsilon}_{SW}(k) = \Delta_z + \lambda (\nu_{\uparrow}^{HF} - \nu_{\downarrow}^{HF}) 
(\tilde{a}(0) - \tilde{a}({k}))
\end{equation}
occurs. The expression for the minority spin can be found from the 
relation $\tilde{\Sigma}_{\downarrow}(i \nu_n) = - \tilde{\Sigma}_{\uparrow}( -i \nu_n)$, 
which holds for $\nu=1$ due to the particle-hole symmetry.
At zero temperature, the correctness of the SHF is recovered since 
the self-energy expression equals zero due to a vanishing numerator.  

Eq.~(\ref{SE1}) and the corresponding minority spin expression are formally 
equivalent to the second order electronic self-energy expression due to virtual 
phonon-exchange in a model of free band electrons coupled to phonons \cite{Mah90}. 
Eq.~(\ref{SE1}) describes the scattering of a majority spin electron 
into a minority spin state upon absorption of a spin-wave playing the role of the phonons. 
However, there is an important difference: the bosonic spin-waves carry spin $S_{z}=-1$ 
and not spin zero like the phonons. Therefore, it is not possible for the 
majority spin electron to have simultaneously 
scattering in the minority spin state $S_{z}=-1/2$ and emission of a spin-wave. 
Hence, an emission term as in the phonon-case is missing.  
Reversely, in the minority spin case 
scattering into majority spin states is only permitted when an emission of a 
spin-wave, but not of an absorption, occurs.  
In our case, the role of the electron-phonon coupling constant is played 
by the electron-spin-wave coupling, which is proportional to $\lambda \tilde{a}(k)$. 
Note that, unlike the case of deformation potential phonon-coupling, the 
spin-wave coupling approaches for $k=0$ a non-zero constant. 
This picture can be extended to the region of $k \ell_c \gg 1$, where 
magneto-excitons with {\em real space} distance $k \ell_c^2$ can be absorbed and 
emitted, respectively. 

It is interesting to note that an unscreened Coulomb interaction leads to diverging 
self-energy expressions, which causes the slow asymptotic fall-off of $a(k) \sim 1/k$, 
{\em i.~e.~}the long-range behavior at $k \ell_c^2$. 
This can already be seen from the second order ladder diagram for the self-energy. 
Although screening does not play an important role in the physics of 
quantum Hall systems with gapped ground states at low temperatures, we are forced to introduce 
a screening mechanism. This is accomplished in our approach by replacing 
the unscreened interaction ${\tilde V}(k)
=2\pi/k$ by ${\tilde V}(k) = 2 \pi/(k+k_{sc})$, where we introduced the 
static screening wavevector $k_{sc}$. 
Below, we will discuss the approximate temperature dependence of $k_{sc}$
accounting for the stronger screening with increasing temperature.
Formally, any non-zero screening wavevector removes for real space distances $k \ell_{c}^2$ 
larger than $1/k_{sc}$ the divergence of the self-energy due to a quicker decrease of 
${\tilde a}(k)$.  

In summary, our partial summation of diagrammatic electron-electron contributions 
leads effectively to a model of SHF-electrons and spin-waves, which are coupled 
via an electron-spin-wave constant $\lambda {\tilde a}(k)$.  
It is obvious that such a model goes beyond the naive picture of free magnons without 
coupling as described in Subsect.~\ref{subsect:elementary}. 

The analytical structure of the retarded GF can be studied after analytical 
continuation to the real frequency axis $(i \hbar \nu_{n} \to E+i \eta), \eta \to 0^{+}$ 
\begin{equation}
G_{\sigma}^{ret}(E) = \frac{1}{(E+i \eta -\xi_{\sigma}^{HF} - 
{\tilde \Sigma}_{\sigma}^{ret}(E))} \ .
\end{equation}
We are particularly interested in the properties of the spectral function  
$A_{\sigma}(E)$, which is diagonal in the spin index 
\begin{equation}
A_{\sigma}(E) = -2 Im G^{ret}_{\sigma}(E)
\label{imaginary}
\end{equation}
with the normalization condition
\begin{equation}
\int_{-\infty}^{\infty} \frac{dE}{2\pi} A_{\sigma}(E) = 1 \ .
\label{normalization}
\end{equation}
The particle-hole symmetry at $\nu=1$ results in the relation 
$A_{\uparrow}(E)=A_{\downarrow}(-E)$ for the spin-dependent spectral functions. 

From the analytical structure of $\Sigma^{ret}_{\uparrow}(E)$ we can infer 
that in the temperature dependent interval 
$(\xi^{HF}_{\uparrow},\xi_{\downarrow}^{HF}-\Delta_z)$, the imaginary part 
of the self-energy is non-zero, and, hence, the retarded GF exhibits a branch cut 
in this energy range leading to a continuous contribution to the spectral function. This 
is due to the incoherent band contribution whose bandwidth 
$(\xi_{\downarrow}^{HF}-\Delta_z) - \xi^{HF}_{\uparrow} = 
(\nu_{\uparrow}^{HF} - \nu_{\downarrow}^{HF}) \lambda {\tilde a}(0) $
agrees with the width of the spin-wave band. Outside this region, the zeros of the equation 
\begin{equation}
E-\xi_{\uparrow}^{HF} = Re {\tilde \Sigma}_{\uparrow}^{ret}(E)
\label{poles}
\end{equation}
define the quasiparticle poles in the spectral function. The self-energy 
has for $T>0$ always two quasiparticles poles, $E_{\uparrow}^{-}$ and 
$E_{\uparrow}^{+}$ outside the spin-wave excitation interval. The former one is located 
below the spin-up SHF pole $\xi_{\uparrow}^{HF}$,  
and  the latter one in the positive energy region 
satisfying the relation $E_{\uparrow}^{+} > \xi^{HF}_{\downarrow} - \Delta_z$. 

This behavior is summarized in the explicit expression for the spectral function 
with the two quasiparticle terms and the incoherent band contribution 
\begin{eqnarray}
A_{\sigma}(E) & = & \frac{2\pi \delta(E-E_{\sigma}^{-})}
{|1 -  \frac{\partial \tilde{\Sigma}_{\sigma}^{ret}(E)}{ \partial E}_{|E =
E_{\sigma}^{-}}|} +
\frac{2\pi \delta(E-E_{\sigma}^{+})}
{|1 -  \frac{\partial \tilde{\Sigma}_{\sigma}^{ret}(E)}{ \partial E}_{|E =
E_{\sigma}^{+}}|}
\nonumber \\
&+& \theta(E-\xi^{-}_{\sigma})\theta(\xi^{+}_
{\sigma}-E)
\frac{(-2 Im \tilde{\Sigma}_{\sigma} ^{ret} (E))}{((E-\xi_{\sigma}^{HF}
-Re \tilde{\Sigma}_{\sigma}^{ret}(E))^2 +
(Im \tilde{\Sigma}_{\sigma}^{ret} (E))^2)} \; .
\label{Spectral}
\end{eqnarray}

In order to get a qualitative feel for the low-temperature region, we discuss 
the behavior of the residues of the lower and upper quasiparticle poles 
for spin up
\begin{equation}
z_{\uparrow}^{-,+}= \frac{1}{|1 -  \frac{\partial \tilde{\Sigma}_{\uparrow}^{ret}(E)}
{ \partial E}_{|E = E_{\uparrow}^{-,+}}|} \ .
\label{residue}
\end{equation}
Dividing the self-energy integral (\ref{SE1}) for 
${\tilde \Sigma}_{\uparrow}(E)$ into a collective contribution for $k \ell_c <1$ 
and a single-particle contribution when $k \ell_c>1$ shows the importance 
of the collective contribution for small $k$ at low temperatures. 
The weight and location of the 
low-energy pole found by means of (\ref{poles}), (\ref{residue}) are 
\begin{eqnarray}
z_{\uparrow}^{-}
& = & [ 1 + \ell_{c}^{2}
\int_{0}^{l_{c}^{-1}} dk\, k\, n_{B}(\epsilon_{SW}(k)) ]^{-1}
\simeq 1 - \ell_{c}^{2} \int_{0}^{l_{c}^{-1}} dk\, k\, n_{B}(\epsilon_{SW}(k)) \nonumber \\
E_{\uparrow}^{-} & = & \xi_{\uparrow}^{HF} - \lambda \tilde{a}(0)[1 - z_{\uparrow}^{-} ] \;,
\end{eqnarray}
where we require $4 \pi \rho_{s} \gg k_{B}T$ in order to keep the weight factor positive.
It is remarkable that the coupling constant $\tilde{a}(k)$ drops out of the expression 
for $z_{\uparrow}^{-}$. 
The weight of the low-energy pole, which is one at zero temperature,   
is diminished by the number of excited spin-waves and its shift toward 
smaller values is proportional to the decrease of the weight.
In the first line, we have assumed that the occupation number of bosonic spin-waves 
is small, an approximation that is in accordance with the omission of spin-wave 
interaction in our diagrammatic approach. 
A similar treatment for the upper spin-up pole yields 
\begin{eqnarray}
z_{\uparrow}^{+}  & = &  \frac{1}{z_{\uparrow}^{-}}-1 = 
\ell_{c}^{2} \int_{0}^{l_{c}^{-1}} dk\, k\, n_{B}(\epsilon_{SW}(k)) \ll 1 \nonumber \\
E_{\uparrow}^{+} & = &  \xi_{\downarrow}^{HF} - \Delta_z + 
\lambda \tilde{a}(0) z_{\uparrow}^{+} \ .
\end{eqnarray}
This consideration shows how the high-energy pole is shifted from the upper limit of the 
incoherent contribution to higher energies.  Thus, there exists repulsion of the 
two quasiparticle poles with increasing temperature. 
In the temperature range $4 \pi \rho_s \gg k_{B} T \gg \Delta_z$, it follows that 
\begin{equation}
z_{\uparrow}^{+} = \frac{k_{B} T}{8 \pi \rho_s} ln \left(\frac{k_{B}T}{\Delta_z}\right) \ ,
\end{equation}
while we get in the activated temperature region $k_{B} T \ll \Delta_z$ 
\begin{equation}
 z_{\uparrow}^{+} = \frac{k_{B} T}{8 \pi \rho_s} e^{-\frac{\Delta_z}{k_{B} T}} \ .
\end{equation} 
These dependences lead to the same temperature behavior of the magnetization as 
described in (\ref{magn-low-temperature}). 

\begin{figure}[h]
\centerline{\resizebox{7cm}{7cm}{\includegraphics{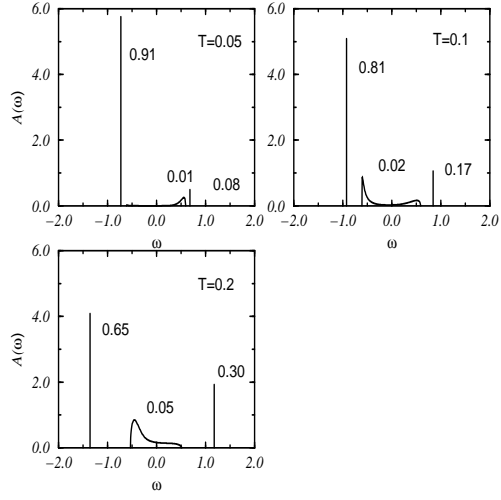}}}
\caption[]{\small
The spectral function $A_{\uparrow}(\omega)$ at 
$\Delta_z = 0.016 \lambda$ and screening wavevector $k_{sc} \ell_c = 0.01$ for temperatures 
$k_{B} T/\lambda=0.05, 0.1$ and $0.2$. 
The numbers indicate the fraction of the total spectral weight from the 
two poles and from the branch cut. The energy $\omega$ is given in 
units of $\lambda/\hbar$. From \cite{KM96}.
}
\label{spectral-function}
\end{figure}

The discussion shows that in the low-temperature region the weights of the 
two quasiparticle poles exhaust almost the entire spectral weight. This 
remains even true at higher temperatures, where 
a transfer of weight from the lower to the higher pole occurs. 
This behavior is depicted in Fig.~\ref{spectral-function} 
with results from numerical calculations of the spin-up spectral function for three different 
temperatures.
For temperatures larger than $k_{B} T \sim 0.35 \lambda$, the weight factors tend to 
$0.5$ from above and below, respectively \cite{KPM2000}.  
The same happens for the Fermi factors at much higher temperatures 
$k_{B} T \gg \lambda \tilde{a}(0) $
so that the magnetization vanishes for $T \to \infty$ as it should be. 
Although the incoherent spectral function contribution has only a tiny influence 
on the spin magnetization, it becomes important when studying the nuclear 
spin-lattice relaxation rate \cite{KPM2000}. 
 
Before discussing the magnetization, let us 
comment on the role of screening \cite{Kas2000}. As already mentioned, it is 
necessary to introduce an {\em ad hoc} screening mechanism.  This adds to the 
bare Coulomb interaction 
$\tilde V_{c}(k)$ a static, but temperature dependent screening wavevector 
$k_{sc}$ entering the Fourier transformation of the interaction as $\tilde{V}=2 \pi/(k+k_{sc})$. 
At zero temperature, screening does not occur at any integer $\nu$ like $\nu=1$ 
due to the one-particle gap, but its importance grows with increasing temperature 
larger than zero. 
A dynamical screening theory in random-phase approximation down to intermediate magnetic 
field strengths was developed by Smith et al.~\cite{SMG92}. 
Here, we restrict ourselves to the LLL and start from a Thomas-Fermi theory expression 
\cite{AFS82} 
\begin{equation}
k_{sc}(T) = 2 \pi \ell_c \lambda n^2 \kappa
= 2 \pi \ell_c \lambda \biggl( \frac{\partial n}{\partial \mu} \biggr)_{\scriptstyle T,V,N}
=\frac{\lambda}{\ell_c} \biggl( \frac{d \nu}{d \mu} \biggr)_{\scriptstyle T,V,N}\; .
\label{scr4}
\end{equation}
The existence of incompressible states at integer filling and zero temperature means 
$\kappa=0$ and ensures the vanishing of $k_{sc}$. Eq.~(\ref{scr4}) can be written as 
\begin{eqnarray}
k_{sc} = \frac{2 \beta \lambda}{\ell_c} \int_{- \infty}^{\infty} \frac{d E}
         {(2 \pi)} A_{\uparrow}(E) n_{F}(E)(1-n_{F}(E)) \ ,
\label{scr5}
\end{eqnarray}
which allows to use the calculated spin-wave spectral function. 
This is an implicit equation for $k_{sc}$ since the spectral function depends on $k_{sc}$.
Explicit determination of $k_{sc}$ was carried out by evaluation of the spectral 
function entering the r.~h.~s.~of (\ref{scr5}), 
which depends on $k_{sc}$ via $\tilde{V}({\bf k})$.  
We found for the two temperatures $k_{B}T=0.09 \lambda$ and 
$k_{B}T=0.18 \lambda$ wavevector values $k_{sc} \ell_c = 0.01$ and 
$k_{sc} \ell_c = 0.1$, respectively, which shows that $k_{sc}$ is more than 
a simple fit parameter. 

The knowledge of the spectral functions $A_{\sigma}(E)$, which is the 
many-particle analog of 
the one-particle density of states, allows the calculation of thermodynamic quantities. 
The temperature-dependent spin magnetization Eq.~(\ref{magn-formula}) can be written 
for $\nu=1$ as
\begin{equation}
M(T) = - M_{0} \int_{- \infty}^{\infty} \frac{d E} {2 \pi} tanh(\frac{\beta E}{2}) 
A_{\uparrow}(E) \ .
\label{magn-general}
\end{equation}
The one-particle results at $\nu=1$ for non-interacting electrons and within the SHF can 
be easily derived. The usage of the spectral functions 
$A_{\uparrow}(E)=2 \pi \delta(E-\Delta_{z}/2)$ and 
$A_{\uparrow}(E)=2 \pi \delta(E-\xi_{\uparrow}^{HF})$, respectively, yields the 
expressions for the magnetization of Eqs.~(\ref{magn-free-nu=1}) and (\ref{magn-HF}), see 
Fig.~\ref{magn_magnons}. 
The temperature dependence of the magnetization over the entire temperature range 
can only be found by numerically 
evaluating (\ref{magn-general}) on the basis of the spectral function (\ref{Spectral}). 

	\subsection{Comparison of theory and experiment: the spin magnetization at $\nu=1$}
	\label{subsect:comparison}

In this Subsection we describe shortly two other theoretical approaches 
that also aim at the understanding of the physics at filling factor $\nu=1$, 
before we introduce experimental techniques allowing the study of the spin magnetization. 
Eventually, we will extend our theoretical results by taking into account deviations 
from our model of an ideal 2DES and compare these results with those from experiment. 

To date, there exist two strategies to attack the problem of calculating the 
magnetization at $\nu=1$: first, by starting from the electronic many-particle 
system emphasizing the charge {\em and} spin 
degrees of freedom, and second, by mapping the original problem onto 
a ferromagnetic lattice Heisenberg model and deriving an effective continuum quantum 
field theory (CQFT).
Examples of the former approach are a work by Haussmann \cite{Hau97b} and 
our calculation discussed at length before. 
The latter was put forward by Read and Sachdev \cite{RS95} on a mean-field level 
and later extended by Timm et al.~\cite{TGH98} calculating corrections. 
Numerical studies like exact diagonalizations in 
various geometries \cite{CPS97,KPM2000,Kas96} or MC calculations \cite{HSTG2000}  
play rather an intermediate role. Although they  
struggle in general with finite-size effects, they serve as a check of analytical theories. 

The theory developed by Haussmann is particularly successful in predicting 
the spin magnetization in the high-temperature 
range down to temperatures, at which the relative spin magnetization does not exceed a value 
of about $0.4$ \cite{Hau97b}.  
It is a so-called modified self-consistent random-phase approximation and  
was originally developed in order to explain tunneling experiments \cite{EPW92b,Hau96}.  
The idea is quite general and has its origin in the treatment of the independent 
boson model, an exactly solvable model that describes the interaction between {\em one electron 
of fixed energy} and independent bosons linearly coupled to the single electron \cite{Mah90}. 
This set of ideas can be transferred to the problem of interacting electrons in the LLL
using a diagrammatic formulation. Breaking the Coulomb interaction in two 
parts, where one part can be viewed as bosonic, allows to transform the 
original interacting problem into 
an interacting electronic problem with a residual interaction. 
Applying a modified self-consistent RPA he was able to determine the self-energy of the 
transformed problem, but also the 
spectral function $A_{\sigma}(E)$ of the original problem, which 
consists of two Gaussian peaks each of them weighted by Fermi factors. 
This reminds of the similar structure of the spectral function (\ref{Spectral}), where,  
however, delta peaks occur. Although Haussmann's theory becomes instable when 
calculating $\xi_{\sigma}$ at low temperatures, the 
form of the spectral function seems to be rather independent of temperature. 
At zero temperature neglecting the Fermi factors, two delta-peaks of a distance 
that agrees with the exchange energy in the LLL were found. 
Due to the temperature suppression of the upper pole, only one pole as in the SHF occurs.

Unfortunately, this theory pretends also a phase transition when $\Delta_z \to 0$ 
at $k_{B}T_{c}=0.041 \lambda$. This is just the temperature region, where 
this theory breaks down. We think that this fact is due to the neglect of 
collective excitations as in the SHF-approximation.

The field theoretical description is based on the observation that 
the quantum Hall ferromagnet at $\nu=1$ 
and a ferromagnetic quantum-mechanical Heisenberg model for spin $S=1/2$ on 
a square lattice in an external magnetic field are phenomenologically  
equivalent because of a ferromagnetic ground state and the existence of spin-waves  
as low-lying excitations. The Heisenberg Hamiltonian is given by 
\begin{equation}
H = -J \sum_{<{\bf ij}>} {\bf S}_{\bf i} {\bf S}_{\bf j} - \Delta_{z} \sum_{\bf i} S_{z,{\bf i}} \ ,
\label{Heisenberg}
\end{equation}
where $<\bf ij>$ means summation over nearest neighbor sites of a quadratic lattice.
The one-spin flip excitations 
\begin{equation}
|{\bf k}> = \frac{1}{\sqrt{N}} \sum_{j} e^{i {\bf k} {\bf r}_{j}} S^{-}_{j}|0> 
\end{equation}
$N$ - number of lattice sites,  
describe spin-waves of energy $\epsilon^{Hei}_{SW}= -J(cos(k_{x}a) + cos(k_{y}a))$ 
above the spin-polarized ground state $|0 \rangle$.
The mapping to the original problem is accomplished by matching the  
coupling constant of the Heisenberg model $J = 4 \rho_{s} > 0$ to the spin-stiffness found in 
the long-wavelength limit (\ref{longwavelengths}) of the dispersion relation of the microscopic 
Hamiltonian, and which is proportional to the exchange energy $\lambda \tilde a(0)$.
For an unscreened Coulomb interaction, we find $J/\lambda = a(0)/(4 \pi) 
= 1/(4 \sqrt{2 \pi}) \simeq 0.0997$, while screening and finite width effects result 
in a reduction of the spin-stiffness. 
It is obvious that such a mapping neglects the charge degree of 
freedom and reduces the problem to a localized spin problem, 
whose quantum number ${\bf k}$ is restricted to the first Brillouin zone. 
Moreover, the neglect of itineracy leads to differences in the physically less interesting 
high-temperature region. The leading order in a high-temperature expansion 
of the Heisenberg model is proportional to $\beta \Delta_z/2$ and is twice of 
the value for the QHFM, see Eq.~(\ref{magn-free-nu=1}) \cite{Kas97}. 
Another basic question concerning the completeness of the model 
was raised when the influence of the Hartree term (\ref{coulomb}) on the 
magnetization was studied \cite{Bre2000}. As we have already seen, a systematic gradient 
expansion of the order parameter field ${\bf m}({\bf r})$ contains additional terms beyond 
the leading non-zero Heisenberg model contribution.
Brey showed that the next order contribution, the Hartree term, 
does have a non-negligible influence on the magnetization in the temperature 
range $0.05 \le k_{B}T/\lambda \le 0.15$ leading to an increase of 
the magnetization by approximately thirty per cent. 
This is due to the  preference of a more homogeneous topological density in the Hartree term 
suppressing spin-fluctuations.
This conclusion was drawn from classical MC results, which should agree quite well in this 
temperature range with the quantum-mechanical result.
Hence some doubt is left whether the Heisenberg model (\ref{Heisenberg}) can 
really serve as an appropriate model in such a wide temperature range. 

To get immediately numerical results, one can evaluate (\ref{Heisenberg}) by means 
of quantum Monte Carlo calculations on a lattice as it was done by Henelius et al.~obtaining 
very accurate results \cite{HSTG2000}.  
An analytical treatment of the Heisenberg model is possible by generalizing the 
original model with  $SU(2) \cong O(3)$  symmetry in spin space to  
the cases of $SU(N)$ and $O(N)$ symmetry in a Schwinger boson representation. 
That means, when treating the general case of spin $S$, {\em i.~e.~}${\bf S}^2=S(S+1)$,
that the entries of the $N \times N$ $SU(N)$-matrix $S_{\alpha}^{\beta}=
b_{\alpha}^{\dagger} b_{\beta}$ are composed out of $N$ bosonic 
operators (Schwinger bosons). Additionally, they have to satisfy the constraint 
$\sum_{\alpha} b^{\dagger}_{\alpha} b_{\alpha}= NS$. 
Performing the continuum approximation leads to a CQFT whose partition 
function is evaluated in the $N \to \infty$ limit, which is known as the 
large $N$ mean-field theory. Improvements are expected from the calculation of 
$1/N$-corrections for $N=2$ and $3$, respectively, which are not really small. 
In total, this approach provides four different results, depending on the symmetry 
and on the accuracy of the large $N$ calculation. 

The most accessible observable reflecting the peculiar features of this 
correlated electron system at $\nu=1$ is the spin magnetization. 
So far, three methods were applied for a more or less direct measurement of 
this observable at $\nu=1$.  
These are optically pumped nuclear magnetic resonance (OPNMR) measurements, 
the magneto-optical absorption technique and photo-luminescence experiments.  \\
Information about the spin magnetization of a 2DES in an electron doped 
$GaAs$-$Al_{x}Ga_{1-x}As$ quantum well can be extracted from NMR 
experiments. Essentially, a nucleus with spin ${\bf I}$ and magnetic moment 
$\mu_{N}$ is not only Zeeman coupled to the external magnetic field ${\bf B}=B {\bf e}_{z}$, 
but also to the electron's spin $\bf S$ of the 2DES via the hyperfine interaction, \ie  
\begin{equation}
H_{N} = - g_{N} \mu_{N} {\bf B} {\bf I} + A {\bf I} {\bf S},
\label{hyperfine}
\end{equation}
where $A$ is the hyperfine coupling constant. Rewriting the coupling between 
nuclear and electron spin-operators 
as ${\bf I} {\bf S} = (I^{+}S^{-} + I^{-}S^{+})/2 + I_{z} S_{z}$  
allows to distinguish between the influence of static and dynamical mechanism
on the NMR-spectra. The 
expectation value of the transversal part of the hyperfine interaction 
vanishes for an external field in $z$-direction. The additional magnetic 
field at the location of the nucleus, which is proportional 
to $\langle S_{z} \rangle$, causes a shift of the resonance NMR-signal with respect to 
the resonance in a $GaAs$-sample without hyperfine coupling. 
This shift of the NMR-resonance, the Knight shift, can 
be used to extract the spin magnetization of the 2DES from 
the measured NMR-spectra \cite{Sli90}. Unfortunately, the shift does not allow the 
determination of the absolute value for the spin-polarization, however, there exist often 
reference points.
For example, the maximum Knight shift at the lowest available temperature can be 
identified with the occurrence of saturated ferromagnetism at exactly $\nu=1$.
Deviations from this ideal behavior appear as measurements at zero temperature 
are not possible,  
and the influence of disorder diminishing the spin-polarization is often unknown \cite{SMG2000}.  
Hence, some source of uncertainty remains when evaluating experimental data.
Such an NMR-technique was developed and applied to the 2DES in a strong magnetic field by 
Barrett and collaborators \cite{BTPW94,BDPWT95}. 
One of the main obstacles was to reach a sufficiently large NMR-signal 
from the $^{71}Ga$-nuclei. An NMR-signal enhancement of $\sim 100$ could be obtained by means 
of an optically pumped NMR-technique (OPNMR). Moreover the technique 
allows to distinguish the signal from the nuclei within the well from that of the nuclei 
in the barrier region \cite{BTPW94}. 
More information can be obtained from the time-dependence of the Knight-shift, 
which is related to the spin-lattice relaxation rate $1/T_{1}$ of the 
out-of-equilibrium nuclear spins due 
to the coupling via the transversal terms to the electrons. 
The measurements were done with $GaAs$-$Al_{0.1}Ga_{0.9}As$ multiple quantum wells 
of width $300$ {\AA} separated by $1800$ {\AA} wide barriers exhibiting  
typical electron densities of around $10^{11} cm^{-2}$ for the 2DES in the well 
and a high mobility larger than $10^{6}\,cm^2/Vs$. The strength of the magnetic 
fields was between $3$ and $12\ T$ in dependence on the investigated filling factor.  
Initially, the lowest available temperatures were about $1.55\ K$, but meanwhile values 
smaller than $0.3\ K$ are possible \cite{KKB98a}.  

The magneto-optical method initiated by Goldberg's group \cite{AGB96,MAGBPW96} utilizes 
the temperature dependence of the absorption coefficient $\alpha(\omega,T)$ 
upon shining light onto a $GaAs$-heterostructure. 
The peaks in the absorption spectrum can be associated with optical 
transitions of electrons from an initial state $i$ in one of the 
valence bands into a final state $j$ in the LLL with spin $\sigma$. 
Integrating over the entire peak w.~r.~t.~the frequency 
of the experimentally determined absorption curve assuming some frequency cut-off yields the 
intensity $I_{ij}= \int d\omega \alpha_{ij}(\omega)$, which is proportional to 
the optical matrix element $f_{ij}(\omega)$ for this interband transition as well as to 
the total number $N_{A_{\sigma}}=N_{\Phi}-N_{\sigma}$ of available or non-occupied states 
in the LLL with spin $\sigma$, \ie~  
\begin{equation}
I_{ij} =  C f_{ij} N_{A_{\sigma}} \ . 
\end{equation}
A theoretical calculation of the $f_{ij}$ and the 
determination of the factor $C$, which is approximately a constant independent of the 
transition and temperature, yield the quantities $N_{A_{\uparrow}}$ and 
$N_{A_{\downarrow}}$. These are related to the spin magnetization by 
\begin{equation}
M(T) = M_0 \frac{N_{\uparrow}-N_{\downarrow}}{N} = 
M_0 \frac{N_{A_{\downarrow}}-N_{A_{\uparrow}}}{N} \ .
\end{equation}
The unknown total particle number $N$ in the LLL can be inferred from the relation 
$(N_{A_{\uparrow}}+N_{A_{\downarrow}})/N = (2/\nu) - 1$.   
The advantage of this method is the determination of the absolute value of 
the spin magnetization, while, on the other hand, additional theoretical calculations 
of the optical matrix elements are necessary. 

Theoretical simplifications made so far comprise mainly the omission of the 
finite width $w$ and the finite barrier 
height of the quantum well containing the 2DES, the neglect of 
higher orbital Landau levels as well as of disorder.

The first issue can be taken into account by a form factor $F({\bf k},w)$ leading to 
an effective interaction $\tilde{V}_{eff}({\bf k},w)= F({\bf k},w) \tilde{V}({\bf k})$,  
which is due to the extension of the envelope wave function $\phi_{0}(z)$ in $z$-direction 
instead of a delta-function for zero width. 
Assuming an infinite barrier well height and a symmetric charge distribution 
$\rho(z)=|\phi_{0}(z)|^2=\theta(w-z) (2/w) sin^{2}(\pi z/w)$ as in Barrett's 
experiment \cite{BDPWT95}, we obtain for the form factor 
\begin{equation}
F(k,w)=\frac{32 \pi^{4}(e^{-kw}-1)}{(kw(4\pi^2 + k^2w^2))^2} + 
\frac{8 \pi^2}{(kw(4\pi^2 + k^2w^2))} + \frac{3kw}{(4\pi^2+k^2w^2)} \  
\label{form}
\end{equation}
altering the short-range behavior for $kw \gg 1$.
In the phenomenological field-theory, the spin-stiffness $\rho_{s}$ becomes 
\begin{equation}
\rho_s(w)  = \frac{\lambda \ell^2_c}{32 \pi^2}  
\int_{0} ^{\infty} dk k^3 \tilde V_{eff}(k,w) \exp ( - k^2 \ell^2_c /2 ).
\label{rhoswidth}
\end{equation}
For a typical width of $30\ nm$ and fields of $B=7.05\ T$ as in Barrett's experiment,  
the width in units of the cyclotron length is $w=3.11 \ell_c$. The spin-stiffness 
becomes then approximately half of its zero-width value \cite{KPM2000} and causes 
a considerable decrease of the energy scale.  

Since the ratio of Landau level distance and interaction energy scales with the 
square root of $B$, see (\ref{scales}), the correction due to higher Landau levels 
depends on the magnitude of the magnetic field used in experiment. In Barrett's
experiment, there is $\lambda/(\hbar \omega_c)=0.966$, which is not really much 
smaller than one as it should be. 
In fact, allowing arbitrary values for the magnetic field, but still keeping $\nu=1$, 
the task becomes much more complicated as even saturated ferromagnetism cannot 
be simply assumed, \cf~an example at $\nu=2$ with a first-order phase transition 
from paramagnetism to ferromagnetism upon lowering the magnetic field \cite{GQ85}. 
The Monte Carlo results \cite{KRL95} for 
the one-spin flip gap indicate a considerable decrease in the $k \to \infty$-limit 
of the spin-wave dispersion. Nevertheless, it remains difficult to estimate 
the quantitative influence of a renormalized dispersion relation on the temperature
dependent magnetization. In any case, the magnetization should be diminished due to 
the consideration of LL mixing. 

The largest uncertainty arises from the influence of disorder. Almost all 
theories aiming at the calculation of thermodynamic properties neglect it. 
In spite of the high-mobility of the samples, disorder occurs, which leads 
in general to a broadening of the Landau levels. This can be taken into account in a 
crude approximation by the self-consistent Born approximation (SCBA) resulting in the 
somewhat unphysical semi-elliptical broadening of the one-particle density of states \cite{AU74}.  
Similarly, the many-particle gap of the FQHE is reduced, what was quantitatively shown 
for short-range impurities at $\nu=1/3$ \cite{RH85}, as well as the disappearance of the FQHE 
when decreasing the mobility of the samples below a certain value \cite{KT85}.  
Qualitatively, disorder should decrease the magnetization of the ground state 
due to reversed spins in regions of strong potential fluctuations.

\begin{figure}[t]
\centerline{\resizebox{7cm}{7cm}{\includegraphics{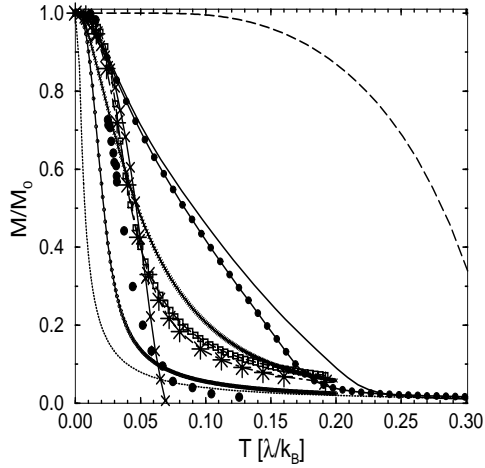}}}
\caption[]{\small
Results for $M(T)$ at $\Delta_z=0.016 \lambda $ for various values of 
the screening vector $k_{sc}$ and the widths $w$ and $\alpha$, respectively. 
The quantity $\alpha$ used in finite-size numerical calculations describes the width
of a Gaussian charge distribution in $z$-direction instead of the width $w$ of a
hard-wall quantum well used elsewhere.
1. free electrons: $w=0$ - dotted curve.
2. SHF: $k_{sc}=0.01 \ell_c^{-1}, w=0$ - long-dashed line.
3. our theory: $k_{sc}=0.01 \ell_c^{-1}, w=3.11 \ell_c$ - solid line.
4. our theory: $k_{sc}=0.1 \ell_c^{-1}, w=3.11 \ell_c$ - solid line with dots.
5. exact diagonalization on the sphere: $N=9, k_{sc}=0,\alpha=0$ solid line with crosses. 
6. exact diagonalization on the sphere: $N=9, k_{sc}=0, \alpha=2 \ell_c$ - solid line
 with circles. 
7. Barrett's experimental data \cite{BDPWT95}: $w=3.11 \ell_c$ - filled points.
8. $O(N)$-field theory with $1/N$-corrections \cite{TGH98}: $w=3.11 \ell_c$ - 
long dashed line with squares.
9. $SU(N)$-field theory with $1/N$-corrections \cite{TGH98}: $w=3.11 \ell_c$ - 
long dashed line with crosses.
10. Monte Carlo results for the Heisenberg model \cite{TGH98}: $w=3.11 \ell_c$ - 
dot-dashed line with stars. From \cite{KPM2000}.
}
\label{magn_comparison}
\end{figure}

\begin{figure}[t]
\centerline{\resizebox{6cm}{6cm}{\includegraphics{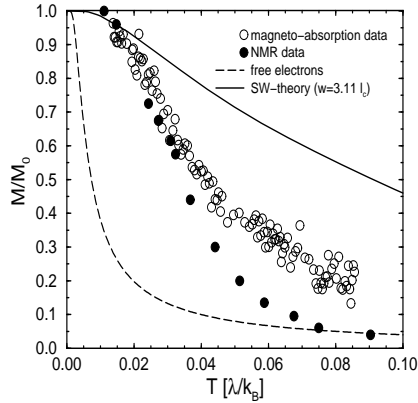}}}
\caption[]{\small
Comparison of the temperature dependent spin magnetization $M(T)$ from 
magneto-absorption data \cite{MAGBPW96} with Barrett's data \cite{TBDPW95}. 
The results for non-interacting particles with Zeeman energy $\Delta_z=0.016 \lambda$ 
and our many-particle result are also shown. Note the higher magnetization at 
higher temperature from magneto-absorption experiments, which do not fall below 
the free particle result as Barrett's data do. 
}
\label{magn_exp}
\end{figure}

Comparison of theoretical and experimental results is for many 
reasons not easy. Simplifications and approximations 
in theory as well as non-unique experimental conditions influence the results. 
Nevertheless, the 2DES in the QH-regime offers outstanding conditions 
for comparison of theory with experiment in a highly-correlated system.
Therefore, let us quickly compare various theoretical and experimental 
magnetization results, which are depicted in 
Figs.~\ref{magn_comparison} and \ref{magn_exp}, 
for a more thorough discussion, see \cite{KPM2000}.
Our comparison with experiment is mainly based on Barrett's data, while results 
from magneto-absorption experiments are only shown in Fig.~\ref{magn_exp}. 
The SHF and the free particle results serve as upper 
and lower bound for any magnetization data. 
Barrett's NMR data are in between these two curves except for temperatures below 
$0.02$, where they seem to be too high, and for 
$k_{B}T/\lambda \ge 0.09$, where they even fall below the free particle curve. 
A possible explanation for the former behavior is the measurement at $\nu=0.98$ instead 
of $\nu=1$ as well as disorder preventing full polarization 
at zero temperature and therefore keeping the curve smoother when approaching zero temperature. 
Not surprisingly, Goldberg's data whose technique allows the determination of the 
absolute value of the relative spin-polarization are smaller in 
this temperature range, see Fig.~\ref{magn_exp}.

Comparison of the numerical data from the MC calculation and the exact 
diagonalization results on the sphere 
for nine particles assuming a finite width gives additional 
important information. The MC data are 
always larger than the diagonalization data. While this has to be true at larger 
temperatures due to the omission of the itinerant character of the electrons 
in the Heisenberg model, the discrepancy at lower temperatures is mainly due to the strong 
finite-size corrections of the diagonalization data at small $\Delta_z$. 
The temperature, at which these 
finite-size corrections become important, can be estimated from numerical 
results for the susceptibility on the sphere \cite{KPM2000}.  
Therefore, the MC data seem to be more trustworthy at low temperatures.
The results of the CQFT partially differ in dependence on the approximation used. 
At low temperatures, the $SU(N)$ mean-field 
theory is most appropriate as the free magnon scenario (\ref{magn-magnons}) can be recovered. 
Except for tiny deviations, this is also true for the $SU(N)$-theory with $1/N$-corrections. 
At temperatures above $0.4$\,, the curves start to diverge. The mean-field curve is 
above the MC data while the $1/N$-correction curve drops so strongly that it becomes 
negative similar to the result of a free magnon theory of Fig.~\ref{magn_magnons}.
On the other hand, the $O(N)$ results do not yield the correct free magnon behavior,  
but agree well with the MC data. Here, the $1/N$-corrections represent a notable 
improvement over the mean-field $O(N)$-theory. In spite of doubts regarding 
the completeness of the model, the results are impressive for the experimental 
value of $\tilde g$. These findings within the CQFT corroborate the picture that 
the collective behavior dominates in the considered temperature range for 
$4 \pi \rho_s \gg \Delta_z$. Such an interpretation gets further support when comparing 
the dispersion curves for the QHFM and the 
Heisenberg model, which start to diverge in a system of finite width 
only at one-spin flip excitation energies that correspond to temperatures of more than 
$50 K$, \cf~Fig.~\ref{disp}. 

Among the many-particle approaches, our theory is evidently an essential improvement over 
the SHF-theory. The correct low- and high-temperature behaviors are additional 
pleasing features. Moreover, the discussion of the low-temperature excitations 
showed that in the case of vanishing Zeeman term the spin-wave excitations 
suppress any spontaneous magnetization at temperatures larger zero.
On the other hand, the deviations from experiment are still large at moderate temperatures 
despite improvement when accounting for the finite width of the quantum well. 
The theory of Haussmann suggests that at higher temperatures screening plays an important 
role. In Fig.~\ref{magn_comparison}, two curves with different screening 
wavevectors $k_{sc}=0.01\ell_{c}^{-1}$ and $0.1 \ell_{c}^{-1}$ are shown. 
They correspond to those two temperatures 
$0.09 \lambda/k_{B}$ and $0.18 \lambda/k_{B}$, respectively, that we found from 
the self-consistent determination within the RPA-calculation, \cf~Eq.~(\ref{scr5}).
Therefore, we expect the curve with the correct temperature dependence to interpolate between the 
two curves in the temperature range  between $0.09\ \lambda/k_B$ and $0.18\ \lambda/k_B$. 

Within our diagrammatic approach, skyrmions cannot be taken into account, but 
their influence on the spin magnetization is presumably not very strong. 
Although skyrmion-antiskyrmion neutral excitations are more effective 
in decreasing the magnetization due to the larger number of flipped spins,  
their thermodynamic weight is rather small because of their 
comparably large energies. Moreover, the experimental Zeeman term 
value $\Delta_z/\lambda=0.016$ lets dramatically shrink the energetic preference 
for these excitations over the neutral quasielectron-quasihole excitations. 

The knowledge of the spectral function allows also comparison with other 
experimental observables at filling factor one. For example, 
the filling factor dependence of the spin-lattice relaxation rate 
$1/T_1$, which was systematically studied at temperature $T=2.2K$ \cite{TBDPW95} 
shows a deep dip at $\nu=1$ due to the lack of low-lying excitations in contrast to 
those existing above the ground state of a Skyrme crystal. 
Unfortunately, only single values for the temperature dependence of $1/T_1$ at 
exactly $\nu=1$ are available \cite{TBDPW95,BDP96}. On the basis of a Korringa theory of nuclear 
magnetic resonance in metals \cite{Sli90} employing an approximation without 
vertex corrections we predict a Gaussian like maximum in the temperature dependence of the 
relaxation rate \cite{KPM2000}. 
The rate was also discussed in the framework of the CQFT \cite{RS95,HSTG2000}.\\
A traditional method to measure spectral functions is tunneling \cite{Schrieffer64}. In our case,   
a current can flow between two parallel 2DES layers in a magnetic field when 
a voltage $V$ between the layers is applied. 
In the case of two weakly coupled layers, the current 
can be expressed as the convolution of the energy dependent spectral functions.  
We predict pronounced peaks in the $I$-$V$-characteristic 
if the voltage $V$ equals the temperature dependent difference 
of the quasiparticle peaks, \ie~, $eV=E^{+}_{\uparrow}-E^{-}_{\uparrow}
=E^{+}_{\downarrow}-E^{-}_{\downarrow}$ \cite{KPM2000}. 
Although tunneling experiments were performed in the range $0.2 < \nu < 0.9$ for 
spin-polarized electrons \cite{EPW92b}, experimental data at filling factor one are 
not available to date.

\section{Summary and outlook}
\label{sect:summary}

This paper is concerned with the physics of interacting electrons in  
two-dimensions in a strong magnetic field neglecting the influence of 
disorder. 
Such systems can be realized in certain semiconductor structures of sufficiently high mobility,  
where the electrons form a 2DES. The electronic degrees of freedom can be well observed 
at low temperatures. 
There are some unique properties of such a 2DES. Among them are: the position 
of the chemical potential can be tuned, the band structure 
of a clean sample in a magnetic field is exactly known, and 
the coupling of the electron's spin to an external magnetic field can be changed. 
These properties allow a much easier comparison of theory with experiment  
than in other strongly correlated systems. 
In the strong magnetic field limit, however, we encounter the common problem 
that any naive perturbation theory fails. 
This fact as well as the many unpredicted experimental observations in the 
quantum Hall regime have been a challenge to theory for more than twenty years. 

Our theoretical treatment is restricted to the physics of the lowest orbital Landau level 
based on a microscopic electronic Hamiltonian. 
First, we study the extreme case of an infinitely strong magnetic field, where the physics 
is solely determined by interacting spinless electrons. Although this case is rather artificial,  
it provides important information, which can 
explain the incompressibility of the many-particle ground state at certain filling factors. 
This is a necessary ingredient for the understanding of the fractional quantum Hall effect. 
In particular, we investigate 
the properties and quality of various quasiparticle trial wavefunctions.
The inclusion of the spin degree of freedom is essential at and near 
filling factor one. 
We show that at $\nu=1$ the ground state exhibits saturated ferromagnetism, even 
in the absence of a symmetry breaking field, 
while novel spin-structures appear near $\nu=1$ if the effective gyromagnetic factor 
$\tilde g$ is small enough. 
The thermodynamics of such a quantum Hall ferromagnet is studied at exactly filling factor $\nu=1$ 
in the framework of a many-particle theory emphasizing 
the role of spin-wave excitations and screening. We investigate the temperature 
dependence of the spin magnetization in the presence of a small Zeeman term and compare 
with other theoretical approaches and with experimental data. 

Although predictions of future research are rather limited as many of the unprecedented 
experimental observations show, let us shortly speculate, what will be of particular 
theoretical interest in the next time.  
So far, we know quite a lot about the physics at single filling factors, where experimental 
effects are most pronounced. On the other hand, there remain large regions along the 
filling factor line, where much less is known. Incompressible, compressible metallic-like 
and insulating ground states are known at certain filling factors, 
but the transition between them is hardly understood. 
Support from the experimental side is rather restricted due to the occurrence  
of disorder. Thus, the consideration of disorder is an interesting and promising field. 
Unfortunately, quantitative control of disorder in experiment is limited. 
Moreover, such investigations could 
also diminish the gap between the rather loosely connected fields of interacting electrons 
in a magnetic  field and non-interacting electrons in a random potential. 

Another interesting field seems to be the systematic study of higher Landau levels. 
Recent experiments suggest that 
some of the spectacular effects of the lowest orbital Landau level disappear 
when the electrons reside in higher Landau levels. Nevertheless, our knowledge about this 
region is rather sparse. In all, the study of the transition from the high-field region 
down to the low-field region is still almost unexplored. 

In closing, the two-dimensional electron system in a strong magnetic field 
exhibits a lot of different physical phases, and an astonishing number of 
theoretical concepts finds their application. 
In view of the progress of the last twenty years and the number of unsolved questions, 
future experiments and theories will certainly deepen and enlarge our current understanding. 

\vspace*{0.25cm} \baselineskip=10pt{\small \noindent
The present work is a shortened version of the Habilitation thesis of the 
author submitted to the Otto-von-Guericke-Universit\"at Magdeburg. 
Discussion and collaboration with W.~Apel, A.~H.~MacDonald, J.~J.~Palacios, 
S.~Barrett, S.~M.~Girvin, B.~Goldberg, R.~Haussmann, P.~Henelius, C.~Timm as well as the 
support by H.~B\"ottger, P.~Fulde, B.~Kramer, and U.~R\"o{\ss}ler 
are gratefully acknowledged. 
The work was partially supported by a grant of the DAAD. 
}


\begin{thebibliography}{300}
\bibitem{Rem1}
It is worth to remark that except for corrections due to quantum
  electrodynamical effects, the Zeeman splitting energy $\Delta_z$ equals the
  distance between adjacent Landau levels $\hbar \omega_c$ in vacuum. This
  leads to the same energy for particles in Landau level $n$ and minority spin
  $\downarrow$ and in the adjacent Landau level $n+1$ and majority spin
  $\uparrow$. However, this relation breaks down in a semiconductor environment
  as we will see below.
\bibitem{Rem6}
In this work, we use the term strong magnetic field and quantum Hall regime as
  synonyms, although the second notion is more precise. It indicates that
  transport measurements show at low filling factors the typical features of
  the quantum Hall effects. This can be accomplished, at least in principle,
  even at moderate magnetic fields, but at sufficiently small particle density
  $n$. On the other hand, it implies a sufficiently clean sample.
\bibitem{Rem5}
The largest energy gap can be found for the fraction $1/3$, whereas the other
  fractions have smaller gaps and are more difficult to observe.
\bibitem{Rem2}
For sake of simplicity, we choose the $z$-direction as the direction of the
  external magnetic field. Tilting the sample by an angle $\theta$ changes the
  effective field for the orbital motion to $B_{\perp}=B cos \theta$, but keeps
  the isotropic Zeeman term constant. Therefore tilting opens the opportunity
  to increase the ratio of the Zeeman energy to the Landau level distance as
  well as to the interaction energy.
\bibitem{symbol}
Here, the bracket $[x]$ means the greatest integer that is not greater than $x$.
\bibitem{Rem7}
The role of the neutralizing background potential becomes unimportant in the
  thermodynamic limit because the particle-background term acts like a constant
  due to the constant single angular momentum distribution in the Laughlin
  wavefunction.
\bibitem{Rem8}
Technically, the projection of a one-particle term onto the LLL can easily be
  done employing the formula: ${\cal P}_{0}(z^{*k} z^{l} e^{-|z|^{2}/4}) =
  2^{k} (l!/(l-k)!) z^{(l-k)} e^{-|z|^{2}/4}\;$.
\end{thebibliography}
\end{document}